\documentclass[a4paper,11pt]{article}
\pdfoutput=1 

\usepackage{jheppub} 

\usepackage{graphicx}
\usepackage{amsmath}
\usepackage{tensor}
\usepackage{hyperref}
\usepackage{comment}
\usepackage{pdfpages}
\usepackage[section]{placeins}
\usepackage{epstopdf}
\usepackage[dvipsnames]{xcolor}
\usepackage[colorinlistoftodos]{todonotes}
\usepackage{epsfig}
\usepackage{graphicx,color}
\usepackage{cancel}
\usepackage[utf8]{inputenc}
\usepackage{simpler-wick}
\usepackage{mathtools}

\DeclareUnicodeCharacter{2212}{-}

\newcommand{\SCETI}{\text{SCET}_{\text{I}}}
\newcommand{\SCETII}{\text{SCET}_{\text{II}}}

\newcommand{\mbf}[1]{\mathbf{#1}}
\newcommand{\tv}{\text{v}}
\newcommand{\textoverline}[1]{$\overline{\mbox{#1}}$}
 
\newcommand{\eps}{\varepsilon}

\newcommand{\la}{\lambda}

\newcommand{\nn}{\nonumber}

\newcommand{\veir}{\varepsilon_{\rm {IR}}}
\newcommand{\veuv}{\varepsilon_{\rm {UV}}}

\newcommand{\be}{\begin{equation}}
\newcommand{\ee}{\end{equation}}
\newcommand{\bea}{\begin{eqnarray}}
\newcommand{\eea}{\end{eqnarray}}
\newcommand{\balign}{\begin{align}}
\newcommand{\ealign}{\end{align}}

\newcommand{\bra}[1]{\left< #1 \right |}
\newcommand{\ket}[1]{\left | #1 \right >}
\newcommand{\braket}[1]{\left< #1 \right>}

\newcommand{\sandwich}[3]{\left< #1 \right | #2 \left | #3 \right >}

\newcommand{\bg}{\begin{gather}}
\newcommand{\foma}{\end{gather}}

\newcommand{\noopsort}[1]{}

\newcommand{\vecb}[1]{\mbox{\boldmath $#1$}}

\def\e{\epsilon}

\def\ve{\varepsilon}

\def\<{\langle}
\def\>{\rangle}

\def\a{\alpha}

\def\d{\delta}  
\def\s{\sigma}
\def\r{\rho}  

\def\m{\mu}
\def\n{\nu}
\def\t{\tau}

\def\({\left(}
\def\[{\left[}
\def\){\right)}
\def\]{\right]}

\def\ln{\hbox{ln}}

\def\Slash#1{{#1\!\!\!\slash}}

\def\Dslash{D\!\!\!\!\slash}

\def\le{\left }
\def\ri{\right}

\newcommand{\ben}{\begin{eqnarray}}
\newcommand{\een}{\end{eqnarray}}

\newcommand{\bef}{\begin{figure}[htb]\centering}
\newcommand{\eef}{\end{figure}}

\usepackage[normalem]{ulem} 
\newcommand{\old}[1]{{\color[rgb]{1,0,0} {\sout{#1}}}} 

\newcommand{\eq}[1]{eq.~\eqref{#1}}

\reversemarginpar

\title{Factorization for $J/\psi$ leptoproduction at small transverse momentum}

\author[a]{Miguel G. Echevarria,}
\author[a]{Samuel F. Romera,}
\author[b]{and Pieter Taels}

\affiliation[a]{Department of Physics \& EHU Quantum Center, University of the Basque Country,\\ Barrio Sarriena s/n, 48940 Leioa, Spain}
\affiliation[b]{Department  of Physics, University of Antwerp, \\ Groenenborgerlaan
171, 2020 Antwerpen, Belgium}

\emailAdd{miguel.garciae@ehu.eus}
\emailAdd{samuel.fernandez@ehu.eus}
\emailAdd{pieter.taels@uantwerpen.be}

\abstract{
Nonrelativistic Quantum Chromodynamics (NRQCD) breaks down in the region of low transverse momentum, where the transverse momentum of the produced quarkonium state is sensitive to multiple scattering with the incoming hadron and to soft gluon radiation. In this kinematic regime, the transverse-momentum-dependent (TMD) factorization framework is required, promoting the long-distance matrix elements (LDMEs) of NRQCD to the so-called TMD shape functions (TMDShFs), which encode both the soft gluon radiation and the formation of the heavy-quark bound state. 
In this work, we apply an effective-field theory approach (combining NRQCD and SCET) to the photon-gluon fusion process in inclusive $J/\psi$ leptoproduction.
We derive a factorization theorem for the cross section in terms of TMDShFs, compute these quantities at next-to-leading order, establish their evolution, and study their matching onto the corresponding LDMEs in the high-transverse-momentum region. 
Our results are particularly relevant to the Electron-Ion Collider, where $J/\psi$ leptoproduction can be used to probe gluon transverse-momentum-dependent parton distribution functions (gluon TMDPDFs).
}

\begin{document}
\maketitle

\section{Introduction} \label{sec:1}


Quarkonium production in semi-inclusive deep inelastic scattering (SIDIS) is a particularly valuable process for probing transverse-momentum-dependent (TMD) gluon distributions. Indeed, since heavy quarks couple to the gluon distribution already at leading order in the strong coupling, measuring the transverse momentum of the produced quarkonium state can give access to the incoming gluon's transverse momentum.
Therefore, particularly in the context of the future Electron-Ion Collider (EIC)~\cite{AbdulKhalek:2021gbh}, quarkonium production in SIDIS has already been studied extensively~\cite{Fleming:1997fq,Kniehl:2001tk,Godbole:2013bca,Mukherjee:2016qxa,Rajesh:2018qks,Bacchetta:2018ivt,DAlesio:2019qpk,Boer:2020bbd,DAlesio:2021yws,ColpaniSerri:2021bla,Kishore:2022ddb,Bor:2022fga,DAlesio:2023qyo,Boer:2023zit,Kang:2023doo,Maxia:2024cjh,Kato:2024vzt} (see also the recent review on quarkonium physics at the EIC~\cite{Boer:2024ylx}).
In addition, quarkonium production in \textit{pp} collisions at the Large Hadron Collider (LHC) has been studied in, e.g.~\cite{denDunnen:2014kjo,Lansberg:2017tlc,Lansberg:2017dzg,Scarpa:2020sdy,Scarpa:2019fol,Wei:2024ksb}, in
$e^+ e^-$ annihilation at B factories~\cite{Jia:2024cvv,Lee:2020dza} and at Belle II~\cite{Garg:2024rkt}, in
Z-boson decays at LHC~\cite{Wang:2023ssg,Bodwin:2017pzj},
and in Higgs decay at HL-LHC~\cite{Martynenko:2023xyx,Dong:2022bkd} (see also~\cite{Chapon:2020heu} for the prospects at HL-LHC), as well as many others (see \cite{
Lansberg:2019adr} for a comprehensive review). Moreover, recent work regarding quarkonium in fragmentation functions can be found, e.g., in~\cite{Echevarria:2020qjk,Celiberto:2023fzz,Dai:2023rvd,Copeland:2023qed,Echevarria:2023dme,Copeland:2023wbu}, as well as heavy-quark TMD distributions and fragmentations in~\cite{vonKuk:2023jfd,vonKuk:2024uxe}.

The main goal of this paper is to establish the factorization for $J/\psi$ leptoproduction in terms of TMD shape functions\footnote{The shape function was first introduced in \cite{Beneke:1997qw,Neubert:1993ch} to describe the redistribution of the endpoint region by the non-perturbative effects in inclusive weak decays of hadrons containing a heavy quark.
In \cite{Fleming:2006cd,Fleming:2003gt}, photoproduction and $e^+ e^-$ annihilation were considered in the endpoint region promoting the LDMEs to the shape functions.
Lastly, the factorization in terms of TMDShFs was firstly introduced in~\cite{Echevarria:2019ynx} for $\eta_c$ production and in \cite{Fleming:2019pzj} for $\chi_Q$ decay to light quarks.} (TMDShFs) and gluon TMD parton distribution functions~\cite{Mulders:2000sh,Meissner:2007rx,Echevarria:2015uaa} (gluon TMDPDFs). TMDShFs describe the formation of the quarkonium bound state, while gluon TMDPDFs parameterize the structure of the incoming hadron in terms of gluons. Both TMDShFs and gluon TMDPDFs are transverse-momentum dependent and highly sensitive to the soft gluon radiation involved in the entire process. Our aim is to develop a complete and consistent theoretical framework for quarkonium leptoproduction at small transverse momentum, describing its factorization in terms of TMDShFs and TMDPDFs, attributing the different modes of gluon radiation to the proper objects. Afterwards, we study the scale evolution of the TMDShFs as well as their matching onto collinear functions at high transverse momentum.


Nonrelativistic Quantum Chromodynamics (NRQCD)~\cite{Bodwin:1994jh} is an effective field theory (EFT) for the description of heavy quarkonium, using the relative velocity $\tv$ of the heavy quark in the rest frame of the bound state as the power counting parameter.
Quarkonium production is studied within the framework of the NRQCD factorization conjecture~\cite{Nayak:2005rt,Nayak:2005rw,Kang:2011zza,Kang:2014tta}, where the cross section can be separated into two parts: the short-distance coefficients (SDCs) and the long-distance matrix elements (LDMEs).
The SDCs encode the heavy-quark pair production and are calculable in perturbative QCD. The LDMEs, on the other hand, parameterize the probability that the heavy-quark pair with a specific spin $S$, angular momentum $L$, and color configuration (octet $[8]$ or singlet $[1]$), denoted as $[n] = \, ^{2S+1}L_J^{[1,8]}$, hadronizes into the bound state. 
Therefore, LDMEs are inherently nonperturbative and need to be extracted from experiments, see, e.g.,~\cite{Butenschoen:2011yh,Shao:2014yta,Bodwin:2015iua,Feng:2018ukp,Brambilla:2022ayc}. 
Not all the possible heavy-quark configurations contribute to the final quarkonium state, since LDMEs scale with powers of the relative velocity~\cite{Braaten:1996ix} and calculations are performed up to a fixed order in the v-expansion. 
In the case of the production of a $J/\psi$ meson, the leading power (LP) in the v-expansion corresponds to the color-singlet $[n] =\!\,^3S_1^{[1]}$ and the next-to-leading power (NLP) to the color-octet states $^1 S_0^{[8]}$, $^3S_1^{[8]}$, and $^3P_J^{[8]}$, see, e.g.,~\cite{Schuler:1997is}.
In our work, we perform the calculations in the so-called vNRQCD framework~\cite{Rothstein:2018dzq} (briefly reviewed in \hyperref[sec:A-vNRQCD]{appendix A}), which is a formulation of NRQCD in which there is a clear separation between potential, soft, and ultrasoft (usoft) modes through the label-momentum notation. This separation is crucial in our analysis, since soft and usoft gluon radiation plays a central role in the process we are interested in, and heavy quarks couple to both.

\begin{figure}
    \centering
    \includegraphics[scale=0.47]{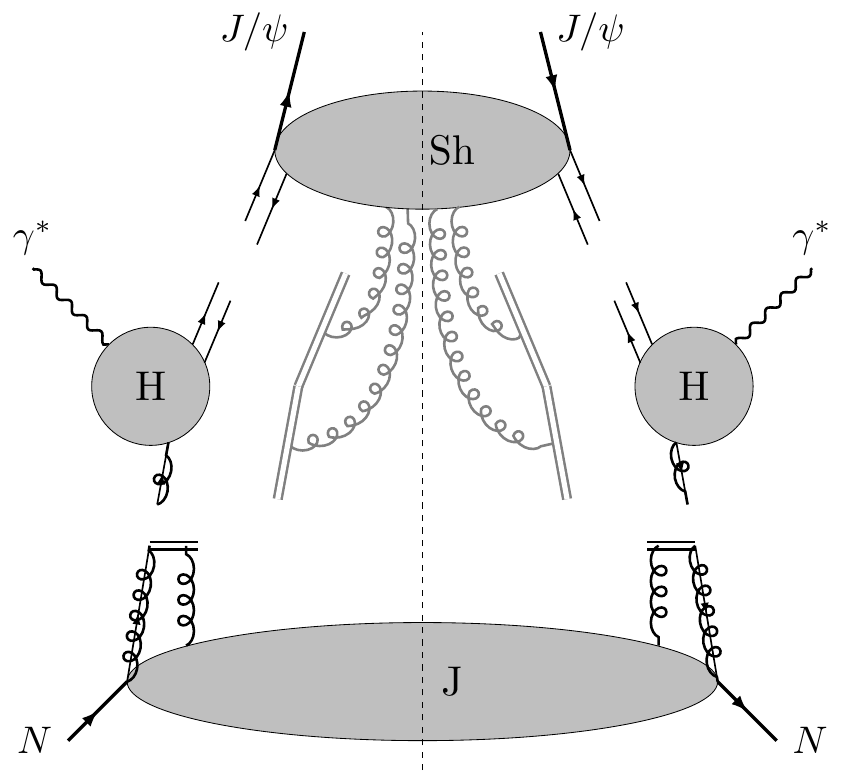}
    \caption{Diagrammatic representation of the factorization theorem for $\gamma^* \! +\! N \!\to\! J/\psi \!+\! X$ at small transverse momentum, which states that the cross section can be written as $d \sigma \!\sim\! \text{H} \!\otimes\! \text{J}\! \otimes \text{Sh}$. H describes the hard collision $\gamma^*\! +\! g\!\to\! c\bar{c}$, with the heavy-quark pair being created in a particular state $[n]\!= \!\, ^{2S\!+\!1}L_J^{[1,8]}$. J stands for the gluon TMDPDF, where we have used a full line with arrow with a coil overlay to indicate the incoming collinear gluon. Finally, Sh stands for the TMDShF. The black and gray double lines indicate the collinear and soft Wilson lines, respectively (usoft Wilson lines are not drawn). The gray lines connecting Sh with the soft Wilson lines indicate the contribution of soft radiation in the two directions involved in the process.}
    \label{fig:Factorization}
\end{figure}

Indeed, NRQCD factorization in terms of LDMEs is only well-defined when the heavy-quark pair is produced in the hard scattering with a large transverse momentum that remains largely unaffected by initial and final state radiation (see~\cite{Braaten:1994kd,Braaten:1994vv,Ma:1995vi,Braaten:1996rp,Petrelli:1997ge,Bodwin:2015iua}).
In contrast, when the heavy-quark pair is produced with low transverse momentum\footnote{\label{footnote:matching}In recent studies \cite{Boer:2020bbd,DAlesio:2021yws,Boer:2023zit}, the interplay between the three regions: high ($\Lambda_{\text{QCD}} \! \ll\! p_T \!\sim \!M$), intermediate ($\Lambda_{\text{QCD}}\! \ll\! p_T \!\ll \!M$) and low ($p_T \!\sim \!\Lambda_{\text{QCD}}\! \ll \!M$) transverse momentum has been considered, where $M$ is the mass of the bound state.} $p_T \ll Q_H$
(with $Q_H$ the hard scale of the process), radiative corrections due to abundant gluon emissions become crucial.
These emissions manifest as collinear, soft, and usoft modes, organized in the momentum region with the help of the scaling parameter $\lambda \equiv p_T/Q_H$, and are extensively studied within the framework of soft-collinear effective theory (SCET)
\cite{Bauer:2000ew,Bauer:2001yt,Bauer:2001ct,Beneke:2002ph}, in particular $\text{SCET}_{\text{II}}$ \cite{Bauer:2002aj} which includes soft particles with momentum scaling as $k^\mu \sim Q_H \lambda$.
More specifically, as illustrated in \hyperref[fig:Factorization]{figure 1}, soft radiation at low transverse momentum becomes entangled with the (u)soft radiation responsible for the binding mechanism of the quarkonium pair, since the heavy quarks interact with both gluon modes. Therefore, in the low-transverse momentum region, the LDMEs must be promoted to TMDShFs which incorporate this soft+usoft radiation. In consequence, we need to work in the vNRQCD+SCET framework where the calculations can be organized according to two power-counting parameters in addition to the QCD coupling constant: v and $\lambda$, which determine the scaling in vNRQCD and SCET, respectively.

The remainder of this paper is structured as follows: 
in \hyperref[sec:2]{section 2}, we establish the factorization of the hadronic tensor in SIDIS in terms of the TMDShFs. We define the effective operators within the vNRQCD+SCET framework and present the results of the matching coefficients derived from the tree-level matching onto QCD, as calculated in \hyperref[sec:A-MatTensors]{appendix B}. 
In \hyperref[sec:3]{section 3}, we perform the calculation of the LDMEs and TMDShFs at next-to-leading order (NLO), after which we study their scale evolution. Furthermore, we examine the perturbative behavior of the TMDShFs and their matching on the LDMEs at high transverse momentum through the operator product expansion (OPE). Finally, we briefly discuss the hard function. Our conclusions and outlook are presented in \hyperref[sec:4]{section 4}.


\section{Factorization}
\label{sec:2}
We study the inclusive production of an unpolarized $J/\psi$ meson in the collision of a lepton $\ell$ with a hadron $N$:\footnote{For conciseness, we denote $J/\psi$ as $\psi$ in the equations.}
\begin{equation}
\ell (k) + N (P_N)  \to \ell'(k') + J/\psi (P_\psi) + X(P_X) \; .
\end{equation} 
Here the particle momenta are given in brackets and the virtual photon momentum is given by $q = k - k'$. 
The mass of the $J/\psi$: $M_\psi$, and the photon virtuality $Q^2 \! = \!- q^2 \!>\!0$, are the two large scales in the reaction: $Q, M_\psi \! \gg \! \Lambda_{\mathrm{QCD}}$. For definiteness, in what follows we will denote the hard scale of the process as $Q_H$ (which is in general a function of $Q$ and $M_\psi$), where it is understood that $Q_H\!\sim\!M_{\psi}\!\sim\!Q$.
For our purposes, it is not necessary to make a more precise choice for $Q_H$.

The underlying partonic hard scattering process is the production of a heavy-quark pair with small relative momentum $\mbf q = m_c \mbf v$.\footnote{As mentioned above, the creation of the $c \bar c$ pair requires at least energy larger than $2 m_c$, and therefore it is created at time $ \sim \! 1/m_c$. However, the effects that bind the pair into quarkonium ensue at time $ \sim \! 1/m_c \tv^2$, so one can deal with the pair creation with not take into account those binding effects.} At leading order in the strong coupling, the heavy quarks are produced through photon-gluon fusion:  
\begin{equation}
\gamma^*(q) + g(p)  \to  c \bar c (P) \; ,
\end{equation}
which is the channel we will consider in this work. 
The momentum of the pair is given by $P^\mu \!=\! M v^\mu \!+\! r^\mu$, where $M \!=\! 2 m_c$, $r$ is the so-called residual momentum, and $v^\mu$ with $v^2 \!= \!1$ is the time-like velocity of the quarkonium state, which should not be confused with v: the relative momentum of the heavy quarks in the rest frame of the $J/\psi$. 
In this frame, where $v^\mu = (1, \mbf 0)$, interactions of the pair with the (u)soft gluons in the bound state change the residual momentum by an amount of $\Delta r \! \sim \! m_c \tv^2$ (note that the energy of the heavy quarks in the bound state scales as $m_c \tv^2$), but quarkonium velocity remains unchanged.
On the other hand, the momentum of the $J/\psi$ will be $P_\psi^\mu \!= \! M_\psi v^\mu$, so we can write the heavy-quark pair momentum as $P^\mu\! =\! M/ M_\psi \, P_\psi^\mu + r^\mu$ ($M/M_\psi\! <\! 1$).
Since the $J/\psi$ is on-shell: $P_\psi^2 \!=\! M_\psi^2$, but $P^2 \!=\! M^2 \left(1 \!+\! \mathcal O ( \tv^2) \right)$, where the velocity corrections are due to interactions between the heavy-quark pair and gluons in the interior of the meson.\footnote{
This is the same as for heavy-quark effective theory (HQET) \cite{Georgi:1990um}, where the heavy quark is moving with momentum $p_Q^\mu = m v^\mu + k^\mu$. Here $v$ is the hadron's velocity and $k$ is the residual momentum which is much smaller than $m$ and which interacts with the light degrees of freedom in the interior of the meson (e.g., $H \sim Q\bar q$). In this picture the mass of the hadron will be the mass of the heavy quark up to corrections: $m_H = m + \mathcal O (1/m)$.}
Notice residual momentum is only well defined in the rest frame of the $J/\psi$, so actually $r^\mu$ is the boosted residual momentum from rest frame to an arbitrary frame.
We have not defined any specific frame yet, the four-velocity $v^\mu$ merely indicates the direction of propagation of the $J/\psi$.

In order to set up the kinematics, we will make use of the light-like vectors $n$, $\bar n$ with $n^2 = \bar n^2 = 0$ and $n \cdot \bar n = 2$. The components of a generic vector $u$ are then defined as $u^{+}\equiv u\cdot\bar{n}$ and $u^{-}\equiv u\cdot n$, such that $u^2 = u^+ u^- + u_\perp^2$ and $u = (u^+,u^-,\mbf u _\perp)$. Note that we will denote vectors in two-dimensional Euclidean space in boldface, such that, e.g., $u_\perp^2 = - \mbf u_\perp^2 < 0$. We also use $u_T = |\mbf u_\perp|$, so that $u_\perp^2 = - b_T^2 < 0$.

Let us now take a closer look at the kinematics of the process. Throughout this work, we will occasionally go back and forth between two different frames. In both frames, the proton does not have any transverse momentum, and its momentum can be parameterized as:
\begin{equation} \label{eq:proton_momentum}
P_N^{\mu}  = P_N^{+}\frac{n^{\mu}}{2}+\frac{M_N^2}{P_N^{+}}\frac{\bar{n}^\mu}{2}\;.
\end{equation}
The physical picture is arguably most clear in a \emph{photon frame}~\cite{Boglione:2019nwk}, where both the proton and the virtual photon have vanishing transverse momentum. In such a frame:
\begin{equation}
    \begin{aligned}
        q^\mu &=\frac{-x_{B}P_{N}^{+}}{\gamma_{q}}\frac{n^{\mu}}{2}+\frac{Q^{2}\gamma_{q}}{x_{B}P_{N}^{+}}\frac{\bar{n}^{\mu}}{2}\;,\\
        P_{\psi}^{\mu}&=\frac{x_{B}P_N^{+}M_{\psi\perp}^{2}}{zQ^{2}\gamma_{\psi}}\frac{n^{\mu}}{2}+\frac{zQ^{2}\gamma_{\psi}}{x_{B}P_N^{+}}\frac{\bar{n}^{\mu}}{2}+P_{\psi\perp}^{\mu}\;.
    \end{aligned}
    \label{eq:Breitframe}
\end{equation}
where $M_{\psi\perp}^{2}\equiv M_{\psi}^{2}+\mathbf{P}_{\psi\perp}^{2}$ and where we have introduced the usual SIDIS invariants: 
\begin{equation} \label{eq:invariantsSIDIS}
\begin{aligned}
x_B = \frac{Q^2 }{2 P_N \cdot q} \; ,
\qquad y = \frac{P_N \cdot q}{P_N \cdot k} \;,
\qquad z = \frac{P_N \cdot P_\psi}{P_N \cdot q} \;,
\end{aligned}
\end{equation}
which are related to the center-of-mass energy squared $s=(k + P_N)^2$ via the relation $Q^2=x_B y (s-M_N^2)$. 
On the other hand, we will set up the factorization in the \emph{hadron frame}~\cite{Scimemi:2019cmh}, where all the transverse momentum is carried by the virtual photon:
\begin{equation}
    \begin{aligned}
        q^\mu &=\frac{-x_B P^+_N Q_\perp^2}{Q^2 \gamma_q}\frac{n^{\mu}}{2}+\frac{Q^{2}\gamma_{q}}{x_{B}P_{N}^{+}} \frac{\bar{n}^{\mu}}{2}+q_\perp^\mu\;,\\
        P_{\psi}^{\mu}&=\frac{x_{B}P_N^{+}M_{\psi}^{2}}{zQ^{2}\gamma_{\psi}}\frac{n^{\mu}}{2}+\frac{zQ^{2}\gamma_{\psi}}{x_{B}P_N^{+}}\frac{\bar{n}^{\mu}}{2}\;,
    \end{aligned}
    \label{eq:hadronframe}
\end{equation}
with $Q_\perp^2\equiv Q^2 - \mathbf{q}_\perp^2$. The quantities $\gamma_q$ and $\gamma_\psi$ in eqs.~\eqref{eq:Breitframe} and~\eqref{eq:hadronframe} encode proton-mass corrections; they are defined by:
\begin{equation} \label{eq:targetmass}
\begin{aligned}
\gamma_{q } \equiv \frac{1}{2} \left(1  +\sqrt{1+\frac{4x_{B}^{2}M_{N}^{2}Q_\perp^2}{Q^{4}}} \right) \quad\mathrm{and}\quad
\gamma_{\psi } \equiv \frac{1}{2} \left( 1 +\sqrt{1-\frac{4x_{B}^{2}M_{N}^{2}M_{\psi\perp}^{2}}{z^{2}Q^{4}}} \right) \;.
\end{aligned}
\end{equation}
We note that, in a photon frame, $Q_\perp \!=\! Q$, while in a hadron frame $M_{\psi\perp}\!=\!M_{\psi}$. The photon transverse momentum in the hadron frame is related to the quarkonium transverse momentum in a photon frame through the following relation:
\begin{equation}
    q_\perp^2 = \frac{P_{\psi \perp}^2}{z^2} \kappa \quad\mathrm{with}\quad\kappa=\frac{z^2 Q^2\big(Q^2+4x_B^2 M_N^2\big)}{z^2 Q^4-4x_B^2 M_\psi^2 M_N^2}\;,
\end{equation}
where target-mass corrections are encoded in the term $\kappa$, which is equal to one when the proton mass is neglected.
The phase space elements of the outgoing lepton and of the outgoing quarkonium can be written in function of the SIDIS invariants and the azimuthal angle $\varphi$ of the outgoing lepton:
\begin{equation}
\begin{aligned}
  \int\! d^{4}k'\delta(k'{}^{2})&=\int\!\frac{y(s-M_N^2)}{4}dx_{B}dyd\varphi \;,\\
\int\! d^{4}P_{\psi}\delta(P_{\psi}^{2}-M_{\psi}^{2})&=\int\!\frac{1}{2\gamma_{\psi}-1}\frac{dzd^{2}P_{\psi\perp}}{2z}\;.  
  \end{aligned}
\end{equation}
The first of the above identities is obtained from an easy calculation in the rest frame of the proton, while the second follows directly from eq.~\eqref{eq:Breitframe}.
In the approximation of a single virtual photon exchange, and integrating over $\varphi$, the differential cross section for SIDIS in a photon frame can be written as~\cite{Boglione:2019nwk}:
\begin{equation}
    \begin{aligned}
\frac{d^5\s}{dx_B\,dy\,dz\,d^2\vecb P_{\psi \perp}} &=
\frac{\pi \a_{em}^2e_c^2}{2Q^4} \frac{y}{z}\frac{1}{2\gamma_{\psi}-1} L_{\m\n} W^{\m\n}\;,
\end{aligned}
\label{eq:crosssectiondef}
\end{equation}
where $e_c=2/3$ is the fractional electric charge of the charm quark.
The cross section in eq.~\eqref{eq:crosssectiondef} depends on the leptonic tensor $L_{\m\n}$, given by:
\begin{align} \label{eq:leptonictensor}
L_{\m\n} &= 
2\le(
k_\m k'_\n + k_\n k'_\m - g_{\m\n} k\cdot k'
\ri)
+ 2i\la_l \e_{\m\n\r\s}l^\r q^\s
\; ,
\end{align}
where we have summed over the spin of the final lepton, and where $\lambda_\ell = \pm 1$ is twice the helicity of the incoming lepton.
Moreover, $W^{\m\n}$ is the hadronic tensor, defined as:
\begin{align} \label{eq:hadronictensor}
W^{\m\n} & =
\frac{1}{(2\pi)^4} \sum_X
\int \frac{d^3 \mathbf P_X}{(2\pi)^3 2E_X}
(2\pi)^4 \d^{4}(P_N+q-P_\psi-P_X)
\nn\\
& \times
\sandwich{N}{ J^{\mu \dagger}(0)}{J/\psi,X}
\sandwich{J/\psi,X}{J^\nu(0)}{N}
\nn\\
& =	
\sum_{\,\,X}\!\!\!\!\!\!\!\!\int
\int \frac{d^4b}{(2\pi)^4} e^{i q \cdot b}
\sandwich{N}{J^{ \mu\dagger}(b)}{J/\psi,X}
\sandwich{J/\psi,X}{J^\nu(0)}{N}
\; .
\end{align}
The sum over the undetected hadrons in the final state, $X$, includes as well the integration over their momenta, $P_X$.
In the second line, we have used momentum conservation to shift the position of the first current.

The current $J^\nu$ is the electromagnetic current for photon-gluon fusion in QCD.
It is matched onto the corresponding effective current in SCET, which is built from operators in the vNRQCD+SCET framework. This effective current can be written as the following sum:
\begin{equation}
J^\nu = \sum_{[n]} \sum_p J^{\nu (p)}_{[n]} \; ,
\label{eq:eff_current}
\end{equation}
where the superscript $(p)$ indicates the particular scaling with $\lambda$ in the SCET-expansion, and where $[n]= \, ^{2S+1}L_J^{[1,8]}$ denotes the color- and angular-momentum state of the heavy-quark pair. The open index $\nu$ couples to the leptonic tensor in~\eq{eq:leptonictensor}, which has a trivial matching in the effective field theory, since we work at tree level in the electroweak theory.

In this work, we limit ourselves to a study of TMD factorization at leading power.
In the SIDIS process under consideration, this also implies that only the photon-gluon fusion channel plays a role, where the heavy-quark pair is produced in a color-octet configuration.
Any other channel, be it with an incoming quark instead of a gluon, or with the heavy quarks being in a color-singlet state, would require an extra hard emission in the final state to carry away the remaining color charge.
However, any such emission not only adds an additional power in the strong coupling, but also in the power counting parameter $\lambda \sim P_{\psi T}/Q_H$.
Notice the choice of working at LP in SCET forces us to work at NLP in NRQCD, as the color-octet states are suppressed by a power of $\tv$ with respect to the color-singlet state.\footnote{Indeed,
the scaling of a particular configuration $[n]$ of the heavy-quark pair is determined by the scaling with $\tv$ of the corresponding LDME. As mentioned in the introduction, color-singlet states are leading power in the heavy-quark velocity ($\tv^3$), while color-octet states are next-to-leading power ($\tv^7$).}

Thus, for our purposes, the sum~\eqref{eq:eff_current} runs over $[n]= \left\{ ^1S_0^{[8]},\, ^3S_1^{[8]} , \, ^3P_{0}^{[8]}, \, ^3P_{1}^{[8]}, \, ^3P_{2}^{[8]} \right\}$ with the corresponding operators given by:
\begin{equation} \label{eq:PowerCountOpe}
\left\{ \mathcal O_{^1S_0^{[8]}}^\nu \; , \mathcal O_{^3S_1^{[8]}}^\nu \; , \mathcal O_{^3P_{0}^{[8]}}^\nu \; , \mathcal O_{^3P_{1}^{[8]}}^\nu \; , \mathcal O_{^3P_{2}^{[8]}}^\nu   \right\} \; .
\end{equation}
These operators are built from the heavy-quark fields (defined in vNRQCD), the incoming collinear gluon field, and the corresponding soft and usoft Wilson lines (defined in SCET).
The procedure to construct SCET operators at any power in $\lambda$ is well known.
For a collinear gluon field the building block is simply $\mathcal B_{n \perp}^\mu$, which is defined in the text below, and scales as $\mathcal B_{n \perp}^\mu \sim \lambda$. 
Moreover, we know that the spin states are described from the Pauli matrices $\boldsymbol \sigma$, and the angular momentum states by powers of the relative momentum, $\mbf q = m_c \mbf v$, of the heavy-quark pair in the quarkonium rest frame (the $L=0$ contribution is given by order $\mbf q^0$ and $L= 1$ by order $\mbf q^1$ in the $\mbf q$-expansion).

We write the effective $[n]$-operators as follows:
\begin{equation}
\begin{aligned} \label{eq:operators}
\mathcal O^\nu_{^1S_0^{[8]}} & = \Gamma_{^1S_0^{[8]}}^{\nu \alpha} \left( \mathcal S_v^{cd} \, \psi^\dagger_{\mbf{p}}  \, T^d \, \chi_{\bar{\mbf p}} \right) \times  \left( \mathcal S_n^{ce} \, \mathcal{B}^{e}_{n \perp \alpha} \right) \;, \\
\mathcal O^{\nu}_{^3S_1^{[8]}} & = \Gamma_{^3S_1^{[8]}}^{\nu \alpha \rho} \left( \mathcal S_v^{cd} \, \psi^\dagger_{\mbf{p}}  \, T^d \left( \boldsymbol \Lambda \cdot \boldsymbol \sigma \right)_\rho \chi_{\bar{\mbf p}} \right) \times  \left( \mathcal S_n^{ce} \, \mathcal{B}^{e}_{n \perp \alpha} \right) \;, \\
\mathcal O^{\nu}_{^3P_{0}^{[8]}} & = \frac{1}{M} \Gamma_{^3P^{[8]}_{0}}^{\nu \alpha}  \left( \mathcal S_v^{cd} \,  \psi^\dagger_{\mbf{p}} \, T^d  \left( \mbf q \cdot \boldsymbol \sigma \right)  \chi_{\bar{\mbf{p}}} \right) \times  \left( \mathcal S_n^{ce} \, \mathcal{B}^{e}_{n \perp \alpha} \right) \;, \\
\mathcal O^{\nu}_{^3P_{1}^{[8]}} & = \frac{1}{M} \Gamma_{^3P^{[8]}_{1}}^{\nu \alpha, k}  \left( \mathcal S_v^{cd}  \,  \psi^\dagger_{\mbf{p}} \, T^d  \left( \mbf q \times \boldsymbol \sigma \right)^k \chi_{\bar{\mbf{p}}} \right) \times  \left( \mathcal S_n^{ce} \, \mathcal{B}^{e}_{n \perp \alpha} \right) \;, \\
\mathcal O^{\nu}_{^3P_{2}^{[8]}} & = \frac{1}{M} \Gamma_{^3P^{[8]}_{2}}^{\nu \alpha, ij}  \left( \mathcal S_v^{cd} \,  \psi^\dagger_{\mbf{p}} \, T^d \,  q^{(i} \sigma^{j)} \,  \chi_{\bar{\mbf{p}}} \right) \times  \left( \mathcal S_n^{ce} \, \mathcal{B}^{e}_{n \perp \alpha} \right) \;.
\end{aligned}
\end{equation}
In the above expressions, the fields $\psi_{\mbf{p}}$ and $\chi_{\bar{\mbf{p}}}$ are the Dirac spinors of the heavy quark and antiquark, respectively, defined in  \hyperref[sec:A-vNRQCD]{appendix A}. 
The matrices $T^a$ are the generators of the Lie algebra of $SU(N_c)$ in the fundamental representation: they contain two fundamental indices that are suppressed for brevity. 
The factor $1/M$ (with $M=2 m_c$) in the third line of~\eq{eq:operators} is a normalization factor from NRQCD: $ \sandwich{c \bar c}{\psi^\dagger \chi}{0} = M \, \xi^\dagger \eta $, where $\xi$ and $\eta$ are the Pauli spinors. 
Moreover, the tensors $\Lambda^i_\mu$ are Lorentz boosts from the rest frame of the heavy-quark pair to the frame in which the three-momentum of the quarkonium coincides with the total three-momentum of the pair (see \hyperref[sec:A-MatTensors]{appendix B}).
The structures $\mathcal{S}$ are soft Wilson lines in the adjoint representation, parameterizing the soft radiation off the incoming proton and the outgoing heavy meson.\footnote{Under soft and collinear gauge transformations in $\SCETII$ the heavy-quark fields transform as $\psi \to V_{s} \psi$, $\psi \to \psi$, respectively.
Therefore to construct a gauge invariant operator requires the addition to soft Wilson lines, i.e., soft gauge invariance determines how $S$ appears \cite{Bauer:2001yt}.
Moreover, a nice way of reaching $\SCETII$ is carrying out the matching into $\text{QCD} \to \SCETI \to \SCETII$, where $\SCETII$ has usoft modes \cite{Bauer:2002aj}. In this way, a BPS field redefinition~\cite{Bauer:2001yt} is needed to decouple the usoft gluons from the leading power collinear $\SCETI$ Lagrangian and induces usoft Wilson lines. The definitions of those are like in~\eq{eq:SoftFunctions} but with usoft fields $A_{us}$. They are needed to get an $\SCETII$ operator which is manifestly invariant under collinear gauge transformation, as well as under soft and usoft gauge transformations.
Additionally, in \cite{Fleming:2019pzj}, a detailed discussion on the gauge invariance of the operators of interest was presented. They performed a tree-level matching of the QCD diagrams with an arbitrary number of (u)soft gluon emissions from the heavy quark and antiquark lines, in particular for quarkonium production in the light quark pair annihilation channel.}
In the former case, they are directed along the collinear vector $n^\mu$, while in the latter they point along the time-like vector $v^\mu = (n^\mu + \bar n^\mu)/2$.
Note that, for ease of notation, we keep the ultrasoft Wilson lines for the proton and the quarkonium state implicit inside the soft Wilson lines.
In the fundamental representation, which is related with the adjoint as $\mathcal{S}^{ba} T^a = S^\dagger T^b S $, the soft Wilson lines are defined as the path-ordered (denoted by $P$) exponentials of gauge fields along the light-cone and the time-like directions:
\begin{equation}
\begin{aligned} \label{eq:SoftFunctions}
S_n (x) & = 
P \exp \left[ ig \int_{- \infty}^0 d\t \,  n \cdot A_{s} (x +  n\t) \, e^{+ \t  \delta^-}  \right] \; , \\
S_v (x) & = P \exp \left[ ig \int_{- \infty}^0 d\t \,  v \cdot A_{s} (x + v \t)  \right] \; .
\end{aligned}
\end{equation}
In the light-cone direction $n$, the soft Wilson line contains rapidity divergences which we regularize using the $\delta$-regulator~\cite{Echevarria:2016scs}.
Moreover, $\mathcal{B}_{n \perp}^{\mu}$ is the gluon field strength defined in SCET as
\begin{equation} \label{eq:Bfield}
\mathcal{B}_{n \perp}^\mu (x) = \frac{1}{g} \left[ W_{n}^\dagger  \, i D_{n \perp}^{ \mu} W_{n} \right](x) \; ,
\end{equation}
with the covariant derivative $iD_{n \perp}^\mu = i\partial^{\mu}_{n \perp} +  g A_{n \perp}^\mu$, and
where $W_n$ is the $n$-collinear Wilson line~\cite{Echevarria:2016scs}:
\begin{equation}
\begin{aligned} \label{eq:WilsonLine}
W_n (x) & =   P \exp 
\left[ ig \int_{- \infty}^0 d\t \, \bar n \cdot A_n (x + \bar n \t) \, 
e^{+ \t  x \delta^+}\right] \; .
\end{aligned}
\end{equation}
Finally, in eq.~\eqref{eq:operators} we have used the notation $q^{(i} \sigma^{j)} = (q^i \sigma^j + q^j \sigma^i)/2 - \mbf q \cdot \boldsymbol \sigma \delta^{ij}/3$.

We now derive the factorization formula for SIDIS at small transverse momentum in terms of the effective $[n]$-operators.
We start from the hadronic tensor in~\eq{eq:hadronictensor}:
\begin{equation} \label{eq:Htensor1}
\begin{aligned}
W^{\mu\nu}&=\sum_N \sum_{\,\,X}\!\!\!\!\!\!\!\!\int
\int \frac{d^4b}{(2\pi)^4} e^{i q \cdot b}
\sandwich{N}{J_{[n]}^{ \mu\dagger}(b)}{J/\psi,X}
\sandwich{J/\psi,X}{J_{[n]}^\nu(0)}{N}
\; ,
\end{aligned}
\end{equation}
where we define the currents $J_{[n]}$ as follows:
\begin{equation} \label{eq:LPcurrents}
\begin{aligned}
J_{[n]}^{\nu}(0) & = C_{[n]}(Q,M, \mu) \left[  \mathcal{S}_v^{cd} \mathcal{S}_n^{ce} \psi_{\mbf p}^\dagger \left( \Gamma \cdot {\mathcal K} \right)_{[n]}^{\nu \alpha} T^d \mathcal B_{n \perp \alpha}^{e} \chi_{\bar{\mbf{p}}} \right](0) \; ,\\
J_{[n]}^{\dagger \mu}(0) & =  C^\dagger_{[n]}(Q,M, \mu) \left[  \left( \mathcal{S}_v^{c'd'} \mathcal{S}_n^{c'e'} \right)^\dagger \chi_{\bar{\mbf{p}}}^\dagger \left( \Gamma \cdot {\mathcal K} \right)_{[n]}^{\dagger \mu \alpha'} T^{d'} \mathcal B_{n \perp \alpha'}^{e'} \psi_{\mbf{p}} \right](0) \; ,
\end{aligned}
\end{equation}
where $\mathcal{B}_{n \perp}^\mu = \mathcal{B}_{n \perp}^{a,\mu}\, T^a$ ($T^a$ in the fundamental representation). It is understood that the summation runs over $ [n] = \left\{\, ^1S_0^{[8]}, \, ^3P_{0}^{[8]}, \, ^3P_{1}^{[8]}, \, ^3P_{2}^{[8]} \right\}$, since the matching tensor for $\, ^3S_1^{[8]}$ is zero (see~\eq{eq:EffOpLO}) and
\begin{equation}
\begin{gathered}\label{eq:GammadotK}
    \left( \Gamma \cdot \mathcal K \right)^{\mu \alpha}_{^1S_0^{[8]}} = \Gamma_{^1S_0^{[8]}}^{\mu \alpha} \, , 
    \quad
    \left( \Gamma \cdot \mathcal K \right)^{\mu \alpha}_{^3S_1^{[8]}} = \Gamma_{^3S_1^{[8]}}^{\mu \alpha \rho} \Lambda^i_{\rho}\sigma_i \, ,\\
    \left( \Gamma \cdot \mathcal K \right)^{\mu \alpha}_{^3P_{0}^{[8]}}  =   \Gamma^{\mu \alpha}_{^3P_{0}^{[8]}} \frac{\mbf q \cdot \boldsymbol{\sigma}}{M}   ,
    \quad
    \left( \Gamma \cdot \mathcal K \right)^{\mu \alpha}_{^3P_{1}^{[8]}}  = \Gamma^{\mu \alpha}_{ ^3P_{1}^{[8]},k} \frac{(\mbf q \times \boldsymbol{\sigma})^k}{M}   ,
    \quad
    \left( \Gamma \cdot \mathcal K \right)^{\mu \alpha}_{^3P_{2}^{[8]}}  = \Gamma^{\mu \alpha}_{ ^3P_{2}^{[8]},ij} \frac{q^{(i} \sigma^{j)}}{M} .
 \end{gathered}
\end{equation}
The so-called matching tensors $\Gamma_{[n]}$ and Wilson coefficients $C_{[n]}$ in eqs.~\eqref{eq:operators} and~\eqref{eq:LPcurrents} are obtained by requiring the $[n]$-operators to agree with the QCD calculation (in the nonrelativistic limit) for the partonic process $g+\gamma^*\to c\bar c$. 
At LP in the $\lambda$-expansion, we only need to match onto the tree-level result and virtual higher-order QCD corrections. Matching onto the former will yield the matching tensors. Virtual higher-order corrections cannot change the structure of these tensors but will contribute to the Wilson coefficients, which are pure numbers, functions of $Q$ and $M$, and equal to one at tree level: $C_{[n]} = 1+\mathcal{O}(\alpha_s)$.
Thus the so-called hard function in the factorization theorem will be given by:
\begin{equation}
H_{[n]}  = |C_{[n]}(Q,M,\mu)|^2=1+\mathcal{O}(\alpha_s) \; ,
\end{equation}
with $\mu$ the renormalization scale.
The contribution $\mathcal O (\alpha_s)$ is different for each $[n]$.
Matching onto the QCD calculation at tree level gives the matching tensors. As seen in \hyperref[sec:A-MatTensors]{appendix B} and mentioned above, since the effective operators scale as $\lambda$, we perform the tree-level QCD calculation at order $\lambda$. Noting that the soft Wilson lines are identically one at tree level, we get the following results in a photon frame:
\begin{equation}
\begin{aligned}
& \Gamma^{\mu \nu}_{^1S_0^{[8]}}  = \frac{ ge}{M} \gamma_q \, \epsilon_\perp^{\mu \nu} \; , \\
& \Gamma^{\mu \nu}_{^3S_1^{[8]}}  = 0 \; , \\
& \Gamma^{\mu \nu}_{^3 P_0^{[8]}}  = - \frac{i 2 \, ge }{3M} \left( \frac{M^2 (\gamma_q^2 + 2) + Q^2 \gamma_q^2}{M^2 + Q^2} \right) g_\perp^{\mu \nu} \; , \\
& \Gamma^{\mu \nu, k}_{ ^3 P_1^{[8]}}  = \frac{i  \, ge}{M} \left( \frac{x_B^2 P_N^{+2} \left(M^2 \gamma_q + Q^2 \left(\gamma_q - 1 \right)\right) (n \cdot \Lambda)_k - \gamma _q^2 Q^4 ( \bar{n} \cdot \Lambda)_k}{\gamma_q x_B P_N^+ M \left(M^2+Q^2\right)} \right) \epsilon_\perp^{\mu \nu} \; ,\\
& \Gamma^{\mu \nu, ij}_{ ^3 P_2^{[8]}}  = - \frac{i2\, ge}{M} \left( \frac{x_B^2 P_N^{+2} M^2}{Q^4} \right) (n \cdot \Lambda)_i (n \cdot \Lambda)_j \, g_\perp^{\mu \nu} \; ,
\end{aligned}
\label{eq:EffOpLO}
\end{equation}
with $\epsilon_\perp^{\mu \nu} = \tensor{\epsilon}{^\mu^\nu_\alpha_\beta} \bar n^\alpha n^\beta/2$ and $g_\perp^{\mu \nu} = g^{\mu \nu} - (n^\mu \bar n ^\nu + \bar n^\mu n^\nu)/2$.
The explicit form of $(n \cdot \Lambda)_i$ and $(\bar n \cdot \Lambda)_i$ can be seen in the \hyperref[sec:A-MatTensors]{appendix B}, but as we will see later, it is convenient to write the matching tensors as in \eq{eq:EffOpLO}.

A first important step in the factorization procedure is to identify the sources of the outgoing radiation $X$. The incoming gluon and the proton remnants generate collinearly enhanced radiation in the $n$-collinear direction. Moreover, together with the quarkonium state, they will emit soft (and ultrasoft) radiation. Being a bound state of heavy quarks, the $J/\psi$ cannot emit collinearly enhanced radiation, and neither can the virtual photon. Therefore, there is no collinear radiation in the anticollinear direction, and we can write $X=X_n+X_s$. Moreover, since in the SCET framework the collinear mode is decoupled from the soft one, we can separate the Hilbert space of the quarkonium and the outgoing radiation into two distinct ones:
\begin{equation}
\begin{aligned} \label{eq:StatesDeco}
\ket{J/\psi,X} = \ket{X_n}  \otimes \ket{J/\psi, X_{s}} \; .
\end{aligned}
\end{equation} 
Note that it is not possible to further factorize the Hilbert space $\ket{J/\psi, X_{s}}$, since soft SCET scale is equivalent with the soft vNRQCD scale, and therefore soft radiation is entangled with the interactions responsible for the formation of the bound state (see \hyperref[sec:A-vNRQCD]{appendix A}).
Eq.~\eqref{eq:StatesDeco} allows us to rewrite the hadronic tensor in \eq{eq:Htensor1} as follows:
\begin{equation}
\begin{aligned}
W^{\mu\nu} &=\sum_{[n]}|C_{[n]}|^{2}
\sum_{\,\,X_{n}}\!\!\!\!\!\!\!\!\int \int\frac{d^{4}b}{(2\pi)^{4}}\,e^{i q \cdot b}\langle N\big|\mathcal B_{n\perp\alpha'}^{\dagger e'}(b)\big|X_{n}\rangle\langle X_{n}\big|\mathcal B_{n\perp\alpha}^e(0)\big|N\rangle\\
& \times\sum_{\,\,X_{s}}\!\!\!\!\!\!\!\!\int\langle0\big|\big[ \mathcal{S}_v^{^\dagger  c'd'} \mathcal{S}_n^{^\dagger c'e'}  \chi_{\bar{\mbf{p}}}^\dagger \left( \Gamma \cdot \mathcal K \right)_{[n]}^{\dagger \mu \alpha'} T^{d'} \psi_{\mbf{p}}\big](b)\big|J/\psi,X_{s}\rangle
\\
&\times
\langle J/\psi,X_{s}\big|\big[\mathcal{S}_v^{cd} \mathcal{S}_n^{ce} \psi_{\mbf p}^\dagger \left( \Gamma \cdot \mathcal K \right)_{[n]}^{ \nu \alpha} T^d \chi_{\bar{\mbf{p}}}\big]\!(0)\big|0\rangle \;.
\end{aligned}
\label{eq:hadronictensor3}
\end{equation}
The next step is to consider the scaling of the different momenta, which is the same in the different frames in eqs.~\eqref{eq:Breitframe} and \eqref{eq:hadronframe}. 
The virtual photon always scales as a hard mode $q^\mu \!\sim\! Q_H(1, 1, \lambda)$, such that the position coordinate $b$ in the Fourier transform of eq.~\eqref{eq:hadronictensor3} scales as $b^\mu \!\sim \!Q_H^{-1}(1,1,\lambda^{-1})$. Since the covariant gluon field, \eq{eq:Bfield}, depends on the collinear momentum $\sim \! Q_H(1,\lambda^2,\lambda)$, a multipole expansion then allows us to neglect the $b^+$-dependence of the field. Likewise, the operators in the second and last line of eq.~\eqref{eq:hadronictensor3} only depend on soft momenta $\sim \!Q_H(\lambda,\lambda,\lambda)$, since the hard components of the heavy-quark momenta have been integrated out.\footnote{On one hand, after integrating out the label momentum in vNRQCD, the dynamical momentum of the heavy-quark fields scales as usoft mode, $k \sim m_c \tv^2$, and therefore $\partial^\mu \psi_{\mbf p} (k) \sim \tv^2 \psi_{\mbf p} (k) \sim \lambda^2 \psi_{\mbf p} (k)$ (we assume $\tv \sim \lambda$). On the other hand, regarding the soft Wilson lines in those operators, it is true that $\partial^\mu \mathcal S_n = i g \mathcal S_n \partial^\mu (n \cdot A_s (x + n \tau) ) \sim \lambda A_s^-(x+n \tau)$ because $\mathcal S_n \sim \lambda^0$. Thus the soft scaling of those operators comes from both soft Wilson lines, $\mathcal S_n$ and $\mathcal S_v$, and the interaction among them.} The multipole expansion of these operators then implies that only the dependence on the transverse coordinate should be kept, to the LP in $\lambda$. We conclude that:
\begin{equation}
\begin{aligned}
W^{\mu\nu} &=\sum_{[n]}|C_{[n]}|^{2}
\sum_{\,\,X_{n}}\!\!\!\!\!\!\!\!\int \int\frac{d^{4}b}{(2\pi)^{4}}\,e^{i q \cdot b}\langle N\big|\mathcal B_{n\perp\alpha'}^{\dagger e'}(b^-,b_\perp)\big|X_{n}\rangle\langle X_{n}\big|\mathcal B_{n\perp\alpha}^{e}(0)\big|N\rangle\\
& \times\sum_{\,\,X_{s}}\!\!\!\!\!\!\!\!\int\langle0\big|\left[ \mathcal{S}_v^{^\dagger  c'd'} \mathcal{S}_n^{^\dagger c'e'}\chi_{\bar{\mbf{p}}}^\dagger \left( \Gamma \cdot \mathcal K \right)_{[n]}^{\dagger \mu \alpha'} T^{d'} \psi_{\mbf{p}}\right](b_\perp)\big|J/\psi,X_{s}\rangle
\\
&\times
\langle J/\psi,X_{s}\big|\left[\mathcal{S}_v^{cd} \mathcal{S}_n^{ce} \psi_{\mbf p}^\dagger \left( \Gamma \cdot \mathcal K \right)_{[n]}^{ \nu \alpha} T^d \chi_{\bar{\mbf{p}}}\right](0)\big|0\rangle \;,
\end{aligned}
\end{equation}
Now, the fact that the $b^+$-dependence of the above operators is suppressed by powers of $\lambda$ allows us to shift it back to the phase. In other words, in the LP of SCET, we can partially undo the second step of eq.~\eqref{eq:hadronictensor} since 
\begin{equation}
\begin{aligned}
e^{(1/2)i\hat{P}^-b^+}O^\dagger(0^+)e^{-(1/2)i\hat{P}^-b^+}&=O^\dagger(b^+)=O^\dagger(0^+)+\mathcal{O}(\lambda)\;.
\end{aligned}
\end{equation}
We thus obtain:
\begin{equation}
\begin{aligned}
W^{\mu\nu} & =\sum_{[n]}|C_{[n]}|^{2} \int\frac{d^{4}b}{(2\pi)^{4}}\,e^{\frac{1}{2}ib^+ (q^-+P^-_N-P_{\psi}^--P_X^-)}e^{\frac{1}{2}ib^-q^+}e^{ib_\perp\cdot q_\perp}\\
 &\times \sum_{\,\,X_{n}}\!\!\!\!\!\!\!\!\int \langle N\big|\mathcal B_{n\perp\alpha'}^{\dagger e'}(b^-,b_\perp)\big|X_{n}\rangle\langle X_{n}\big|\mathcal B_{n\perp\alpha}^{e}(0)\big|N\rangle\\
 & \times\sum_{\,\,X_{s}}\!\!\!\!\!\!\!\!\int\langle0\big|\left[ \mathcal{S}_v^{^\dagger  c'd'} \mathcal{S}_n^{^\dagger c'e'}  \chi_{\bar{\mbf{p}}}^\dagger \left( \Gamma \cdot \mathcal K \right)_{[n]}^{\dagger \mu \alpha} T^{d'} \psi_{\mbf{p}}\right](b_\perp)\big|J/\psi,X_{s}\rangle\\
 &\times\langle J/\psi, X_{s}\big|\left[\mathcal{S}_v^{cd} \mathcal{S}_n^{ce} \psi_{\mbf p}^\dagger  \left( \Gamma \cdot \mathcal K \right)_{[n]}^{\nu \alpha} T^d \chi_{\bar{\mbf{p}}}\right](0)\big|0\rangle +\mathcal{O}(\lambda)\;.
\end{aligned}
\label{eq:hadronictensor4}
\end{equation}
Performing the integral over $b^+$ yields the delta function $\delta(q^-\!+\!P^-_N\!-\!P_{\psi}^-\!-\!P_X^-)$. However, since $P_N^\mu \!\sim\! Q_H(1,\lambda^2,\lambda)$, $P_{\psi}^{\mu}\!\sim\!Q_H(1,1,\lambda)$, and $P_X\!=\!P_{X_n}\!+\!P_{X_s}$ with $P_{X_n}^\mu \!\sim\! Q_H(1,\lambda^2,\lambda)$ and $P_{X_s}^\mu \!\sim\! Q_H(\lambda,\lambda,\lambda)$, one obtains:
\begin{equation}
    \begin{aligned}
       \delta(q^-\!+\!P^-_N\!-\!P_{\psi}^-\!-\!P_X^-)&=\delta(q^-\!-\!P_{\psi}^-)+\mathcal{O}(\lambda)\\
       &=\frac{x_B P^+_N}{Q^2}\Big(1\!-\!\frac{x_B^2 M_N^2 M_{\psi \perp}^2}{\gamma_q^2 Q^4}\Big)\delta\Big(\!z\!-1\!-\!\frac{2 x_B^2 M_N^2 (M_{\psi \perp}^2+Q^2_\perp)}{\gamma_q Q^4}\Big)
       + \mathcal O(\lambda)\;.
    \end{aligned}
\end{equation}
Finally, using the completeness relations for the undetected hadron states:
\begin{equation}
\begin{aligned}
 \sum_{\,\,X_{n}}\!\!\!\!\!\!\!\!\int \ket{X_n} \bra{X_n}   =  1 \quad\mathrm{and}\quad
\sum_{\,\,X_{s}}\!\!\!\!\!\!\!\!\int \ket{J/\psi, X_s} \bra{J/\psi, X_s }   =
a_\psi^\dagger \sum_{X_s} \ket{ X_s} \bra{ X_s } a_\psi =
{\mathcal N}_\psi \; ,
\end{aligned}
\end{equation}
where we have defined the number operator as ${\mathcal N}_\psi \equiv a_\psi^\dagger a_\psi$, we obtain:
\begin{equation}
\begin{aligned}
W^{\mu\nu} 
& =\frac{x_B P^+_N}{Q^2}\Big(1\!-\!\frac{x_B^2 M_N^2 M_{\psi \perp}^2}{\gamma_q^2 Q^4}\Big)\delta\Big(\!z\!-1\!-\!\frac{2 x_B^2 M_N^2 (M_{\psi \perp}^2+Q^2_\perp)}{\gamma_q Q^4}\Big)\\
& \times \int\frac{\mathrm{d}^{2}b_\perp}{(2\pi)^{2}}e^{ib_\perp\cdot q_{\perp}} J_{n, \alpha\alpha'}^{(0)}( b_\perp)
\Bigg\{
\big|C_{^1S_0^{[8]}}\big|^{2} \, \Gamma_{^1S_0^{[8]}}^{\dagger \mu \alpha'} \Gamma_{^1S_0^{[8]}}^{\nu\alpha}S^{(0)}_{^1S_0^{[8]} \to J/\psi}(b_\perp)\\
& + \big|C_{^3P_0^{[8]}}\big|^{2} \,  \Gamma_{^3P_{0}^{[8]}}^{\dagger \mu \alpha'} \Gamma_{^3P_{0}^{[8]}}^{\nu \alpha}  S^{(0)}_{^3P_0^{[8]} \to J/\psi}(b_\perp) + \big|C_{^3P_1^{[8]}}\big|^{2} \,  \Gamma_{ ^3P_{1}^{[8]}}^{\dagger \mu \alpha',k'} \Gamma_{ ^3P_{1}^{[8]}}^{\nu \alpha, k}  S^{(0) \, k' k}_{^3P_1^{[8]} \to J/\psi} (b_\perp)\\
& + \big|C_{^3P_2^{[8]}}\big|^{2} \,  \Gamma_{ ^3P_{2}^{[8]}}^{\dagger \mu \alpha', i'j'} \Gamma_{ ^3P_{2}^{[8]}}^{\nu \alpha, ij}  S^{(0) \, i'j' i j}_{^3P_{2}^{[8]} \to J/\psi} (b_\perp)
\Bigg\}\;.
\end{aligned}
\label{eq:hadronictensorfinal}
\end{equation} 
In a hadron frame, all the measured transverse momentum is due to the virtual photon, and hence is contained in a simple phase factor. In the above result, we have introduced the collinear gluon matrix element:
\begin{equation}
\begin{aligned}
J_{n, \alpha\alpha'}^{(0)}(b_\perp) & =
\frac{1}{2}\int\frac{db^-}{2\pi}e^{\frac{1}{2}ib^-q^+}
\langle N\big|\mathcal B_{n\perp\alpha'}^{\dagger a}(b^-,b_\perp) \mathcal B_{n\perp\alpha}^{a}(0)\big|N\rangle\;,
\end{aligned}
\label{eq:gluonTMDPDF0}
\end{equation}
while the TMD shape functions (TMDShF) $S_{[n] \to J/\psi}$ are given by:
\begin{equation}
\begin{aligned} \label{eq:TMDSHF}
S^{(0)}_{^1S_0^{[8]} \to J/\psi} (b _\perp) & = \frac{1}{N_c^2-1} \\\times\text{tr}_c \sandwich{0}{ & \big[ (\mathcal{S}_v \mathcal{S}_n)^\dagger \chi_{\bar{\mbf{p}}}^\dagger  T^a \psi_{\mbf{p}} \big]\!(b _\perp)  \mathcal N _\psi \big[ \mathcal{S}_v \mathcal{S}_n \psi_{\mbf p}^\dagger  T^a \chi_{\bar{\mbf{p}}} \big]\!(0)}{0} \;, \\
S^{(0)}_{^3P_0^{[8]} \to J/\psi} (b _\perp) & = \frac{1}{M^2(N_c^2-1)} \\
\times \text{tr}_c \sandwich{0}{& \big[ (\mathcal{S}_v \mathcal{S}_n)^\dagger \chi_{\bar{\mbf{p}}}^\dagger \, \mbf q \cdot \boldsymbol{\sigma} \, T^a \psi_{\mbf{p}} \big]\!(b _\perp)  \mathcal N _\psi \big[ \mathcal{S}_v \mathcal{S}_n \psi_{\mbf p}^\dagger \, \mbf q \cdot \boldsymbol{\sigma} \, T^a \chi_{\bar{\mbf{p}}} \big]\!(0)}{0} \; , \\
S^{(0) \, k' k}_{^3P_1^{[8]} \to J/\psi} (b _\perp) & = \frac{1}{M^2(N_c^2-1)} \\
\times  \text{tr}_c \sandwich{0}{& \big[ (\mathcal{S}_v \mathcal{S}_n)^\dagger \chi_{\bar{\mbf{p}}}^\dagger \, (\mbf q \times \boldsymbol{\sigma})^{k'} \, T^a \psi_{\mbf{p}} \big]\!(b _\perp)  \mathcal N _\psi \big[ \mathcal{S}_v \mathcal{S}_n \psi_{\mbf p}^\dagger \, (\mbf q \times \boldsymbol{\sigma})^k \, T^a \chi_{\bar{\mbf{p}}} \big]\!(0)}{0} ,  \\
S^{(0) \, i'j' ij}_{^3P_2^{[8]} \to J/\psi} (b _\perp) & = \frac{1}{M^2(N_c^2-1)} \\
\times  \text{tr}_c \sandwich{0}{& \big[ (\mathcal{S}_v \mathcal{S}_n)^\dagger \chi_{\bar{\mbf{p}}}^\dagger \, q^{(i'} \sigma^{j')} \, T^a \psi_{\mbf{p}} \big]\!(b _\perp)  \mathcal N _\psi \big[ \mathcal{S}_v \mathcal{S}_n \psi_{\mbf p}^\dagger \, q^{(i} \sigma^{j)} \, T^a \chi_{\bar{\mbf{p}}} \big]\!(0)}{0} \; .
\end{aligned}
\end{equation}

In summary, we have factorized the hadronic tensor in terms of the effective $[n]$-operators defined in the vNRQCD+SCET framework.
The factorization features two different matrix elements: $J_n$ describes the $n$-collinear incoming gluon from the proton, while $S_{[n] \to J/\psi}$ encodes the quarkonium formation in a particular $[n]$-configuration together with the (u)soft radiation of the entire process.
In the above equations, the superscript $(0)$ denotes the so-called pure matrix element in the momentum region in which it is defined, i.e., there is no overlap with the other momentum regions. 
However, when one performs the perturbative calculations of the collinear matrix elements one needs to deal with the double counting coming from the overlap of the collinear momentum region and the (u)soft one.
Therefore, one needs to subtract the contribution of (u)soft modes from the naively calculated collinear matrix element $J_{n}$ (the so-called zero-bin in SCET~\cite{Manohar:2006nz}), obtaining the pure collinear matrix elements $J^{(0)}_n$.
Since the perturbative calculation depends on the rapidity regulator that is used, the subtraction of the zero-bin does as well.
For the $\delta$ regulator (see the Wilson line definition in~\eq{eq:SoftFunctions}), it can be shown order by order in perturbative theory that the subtraction of the zero-bin is equivalent to defining the pure collinear matrix elements as follows:
\begin{equation}
J^{(0)}_{n, \alpha \alpha'} (b_\perp) \equiv \frac{J_{n, \alpha \alpha'} (b_\perp)}{S (b _\perp)} \; ,
\end{equation}
with the so-called soft function $S$ defined as:
\begin{equation}
S(b _\perp) = \frac{1}{N_c^2 -1} \text{tr}_c \sandwich{0}{\big[ \mathcal S_n^\dagger \mathcal S_{\bar n} \big] (b _\perp) \big[ \mathcal{S}_{\bar{n}}^\dagger \mathcal{S}_n \big] (0)}{0} \; .
\end{equation}
The matrix elements defined above contain rapidity divergences, which cancel in the cross section~\eqref{eq:hadronictensorfinal}. However, in order to obtain well-defined hadronic quantities, one can remove these divergences on the level of the individual matrix elements with the help of the soft function. 
In particular, we define the gluon TMDPDF as follows:
\begin{equation} \label{eq:renTMDPDF}
G^{\alpha \alpha'}_{g/N} (b _{ \perp})  \equiv 
\frac{J_n^{\alpha \alpha'} (b_\perp)}{\sqrt{ S(b _\perp) }} =  
J_n^{\alpha \alpha' (0)} (b_\perp) \sqrt{ S(b _\perp) } \; ,
\end{equation}
as well as the rapidity-subtracted TMDShF: 
\begin{equation} \label{eq:renShF}
S_{[n]\to J/\psi} (b _\perp)  =  \frac{S^{(0)}_{[n]\to J/\psi} (b _\perp)}{ \sqrt{S(b _\perp)} } \; .
\end{equation}
The hadronic tensor in terms of those quantities is the following:
\begin{equation}
\begin{aligned} \label{eq:hadronictensoralmostfinal}
W^{\mu\nu} 
& =\frac{x_B P^+_N}{Q^2}\Big(1\!-\!\frac{x_B^2 M_N^2 M_{\psi \perp}^2}{\gamma_q^2 Q^4}\Big)\delta\Big(\!z\!-1\!-\!\frac{2 x_B^2 M_N^2 (M_{\psi \perp}^2+Q^2_\perp)}{\gamma_q Q^4}\Big)\\
& \times \int\frac{\mathrm{d}^{2}b_\perp}{(2\pi)^{2}}e^{ib_\perp\cdot q_{\perp}} G_{g/N , \alpha\alpha'}( b_\perp)
\Bigg\{
\big|C_{^1S_0^{[8]}}\big|^{2} \, \Gamma_{^1S_0^{[8]}}^{\dagger \mu \alpha'} \Gamma_{^1S_0^{[8]}}^{\nu\alpha}S_{^1S_0^{[8]} \to J/\psi}(b_\perp)\\
& + \big|C_{^3P_0^{[8]}}\big|^{2} \,  \Gamma_{^3P_{0}^{[8]}}^{\dagger \mu \alpha'} \Gamma_{^3P_{0}^{[8]}}^{\nu \alpha}  S_{^3P_0^{[8]} \to J/\psi}(b_\perp)  + \big|C_{^3P_1^{[8]}}\big|^{2} \,  \Gamma_{ ^3P_{1}^{[8]}}^{\dagger \mu \alpha',k'} \Gamma_{ ^3P_{1}^{[8]}}^{\nu \alpha, k}  S^{ k' k}_{^3P_1^{[8]} \to J/\psi} (b_\perp)\\
& + \big|C_{^3P_2^{[8]}}\big|^{2} \,  \Gamma_{ ^3P_{2}^{[8]}}^{\dagger \mu \alpha', i'j'} \Gamma_{ ^3P_{2}^{[8]}}^{\nu \alpha, ij}  S^{ i'j' i j}_{^3P_{2}^{[8]} \to J/\psi} (b_\perp)
\Bigg\} \;.
\end{aligned}
\end{equation}
In the result above, as well as in its derivation, it is understood that the hard functions, the gluon TMDPDF, and the TMDShFs all depend on an ultraviolet (UV) renormalization scale $\mu$. Moreover, after the rapidity subtractions in eqs.~\eqref{eq:renTMDPDF} and~\eqref{eq:renShF}, the gluon TMDPDF and the different TMDShFs become dependent on the so-called Collins-Soper scales $\zeta_A$ and $\zeta_B$, respectively, where $\zeta_A \zeta_B = Q_H^2$ (see \hyperref[sec:ShNLO]{appendix C.2}). As usual, $G_{g/N}^{\alpha \alpha'}$ can be decomposed into a sum of different TMDPDFs, each encoding the different correlations that may exist between the polarization of the incoming gluon and the proton (e.g., see \cite{Mulders:2000sh, Meissner:2007rx, Boer:2016xqr}).

Our factorization theorem~\eqref{eq:hadronictensoralmostfinal} can be simplified considerably by summing over the helicity $\lambda$ of the $J/\psi$ in the final state, thus considering only unpolarized production.\footnote{Recent studies of polarized $J/\psi$ production can be found in, e.g., refs.~\cite{Ma:2018qvc,DAlesio:2021yws,DAlesio:2023qyo,Copeland:2023wbu,Copeland:2023qed,Kato:2024vzt}.} Note that $\lambda$ is implicitly included in the definition of the shape functions in \eq{eq:TMDSHF}, whereas the Latin indices in $S_{[n] \to J/\psi}$ are related to the polarization of the intermediate states.
To perform this summation over helicity, we first make use of the Wigner-Eckart theorem, which states that, if $\mathcal O$ is a generic operator irreducible under spatial rotations, it fulfills the following equation \cite{Brambilla:2017kgw}:
\begin{equation} \label{eq:WEtheorem}
    \sandwich{0}{\mathcal O^{\dagger i_1, i_2, \hdots, i_J}}{J/\psi(\lambda)} = \mathcal N(\mathcal O_J) \ve^{i_1, i_2, \hdots, i_J}_\lambda  \; .
\end{equation}
In the above identity, $\ve^{i_1, i_2, \hdots, i_J}_\lambda$ is the polarization tensor corresponding to a state with total spin $J$ and helicity $\lambda$, and is normalized as $\ve^{i_1, i_2, \hdots, i_J}_\lambda \ve^{i_1, i_2, \hdots, i_J}_{\lambda'} =\delta_{\lambda \lambda'}$ with $\delta_{\lambda \lambda}=2J+1$. Using these properties, the constant $\mathcal N( \mathcal O_J)$ in~\eqref{eq:WEtheorem} can be determined from:
\begin{equation}
    \mathcal N^*(\mathcal O_J)\mathcal N(\mathcal O_J) = \frac{1}{2J+1} \sum_{\lambda'} \sandwich{0}{\mathcal O^{\dagger i_1, i_2, \hdots, i_J}}{J/\psi(\lambda')} \sandwich{J/\psi(\lambda')}{\mathcal O^{i_1, i_2, \hdots, i_J}}{0} \; .
\end{equation}
Making the $J/\psi$ helicity explicit in the notation of the TMDShFs and in the number operator $\mathcal N_{\psi(\lambda)}$, the theorem~\eqref{eq:WEtheorem} can be applied to the definitions in~\eq{eq:TMDSHF} for the $J=1,2$ P-waves (for the $J=0$ this trivially amounts to adding a summation over the helicity states), yielding:
\begin{equation}
\begin{aligned} \label{eq:TMDSHF-afterWE}
S^{(0) \, k' k}_{^3P_1^{[8]} \to J/\psi(\lambda)} (b _\perp) & = \frac{\ve_\lambda^{k'} \ve_\lambda^{k}}{3 M^2(N_c^2-1)}  \\
\times \sum_{\lambda'} \text{tr}_c \sandwich{0}{& \big[ (\mathcal{S}_v \mathcal{S}_n)^\dagger \chi_{\bar{\mbf{p}}}^\dagger \, (\mbf q \times \boldsymbol{\sigma})^{m} \, T^a \psi_{\mbf{p}} \big]\!(b _\perp)  \mathcal N _{\psi(\lambda')} \big[ \mathcal{S}_v \mathcal{S}_n \psi_{\mbf p}^\dagger \, (\mbf q \times \boldsymbol{\sigma})^m \, T^a \chi_{\bar{\mbf{p}}} \big]\!(0)}{0} ,  \\
S^{(0) \, i'j' ij}_{^3P_2^{[8]} \to J/\psi(\lambda)} (b _\perp) & = \frac{\ve_\lambda^{i'j'} \ve_\lambda^{ij}}{5 M^2(N_c^2-1)} \\
\times \sum_{\lambda'}  \text{tr}_c \sandwich{0}{& \big[ (\mathcal{S}_v \mathcal{S}_n)^\dagger \chi_{\bar{\mbf{p}}}^\dagger \, q^{(l} \sigma^{m)} \, T^a \psi_{\mbf{p}} \big]\!(b _\perp)  \mathcal N _{\psi(\lambda')} \big[ \mathcal{S}_v \mathcal{S}_n \psi_{\mbf p}^\dagger \, q^{(l} \sigma^{m)} \, T^a \chi_{\bar{\mbf{p}}} \big]\!(0)}{0} \; .
\end{aligned}
\end{equation}
Using the following polarization sums (see table I in~\cite{10.1063/1.1665190}):
\begin{equation}
\begin{aligned} \label{eq:NaturalProjection}
\sum_{\lambda=-1}^1 \ve_\lambda^{k'} \ve_\lambda^{k} & = \delta^{k' k} \; ,\\
\sum_{\lambda=-2}^2 \ve_\lambda^{i'j'} \ve_\lambda^{ij} & = \frac{1}{2} \left( \delta^{i' i} \delta^{j' j} + \delta^{j' i} \delta^{i' j}  \right) - \frac{1}{3} \delta^{i'j'} \delta^{ij}  \; ,
\end{aligned}
\end{equation}
allows us, therefore, to define the unpolarized unsubtracted TMDShFs, $S_{[n] \to J/\psi}^{(0)}$:
\begin{equation}
\begin{aligned}
\sum_\lambda S^{(0) \, k' k}_{^3P_1^{[8]} \to J/\psi(\lambda)} (b _\perp)  & = \delta^{k' k} S^{(0)}_{^3P_1^{[8]} \to J/\psi} (b_\perp)  \; ,\\
\sum_\lambda S^{(0) \, i'j' ij}_{^3P_2^{[8]} \to J/\psi(\lambda)} (b _\perp) & = \left[ \frac{1}{2} \left( \delta^{i' i} \delta^{j' j} + \delta^{j' i} \delta^{i' j}  \right) - \frac{1}{3} \delta^{i'j'} \delta^{ij} \right] S_{^3P_2^{[8]} \to J/\psi}^{(0)} (b_\perp) 
\; .
\end{aligned}
\end{equation}
After summing over polarizations,  we remark that the three unpolarized TMDShFs for the P-waves with $J=0,1,2$ can be presented in the following compact way:
\begin{equation}
\begin{aligned}
    S^{(0)}_{^3P_J^{[8]} \to J/\psi} & = \frac{\Delta_J^{i' j' i j}}{(2J+1) M^2 (N_c^2 -1)} \\
    & \times \sum_{\lambda'} \text{tr}_c \sandwich{0}{ \big[ (\mathcal{S}_v \mathcal{S}_n)^\dagger \chi_{\bar{\mbf{p}}}^\dagger \, (q^{i'} \sigma^{j'}) \, T^a \psi_{\mbf{p}} \big]\!(b _\perp)  \mathcal N _{\psi(\lambda')} \big[ \mathcal{S}_v \mathcal{S}_n \psi_{\mbf p}^\dagger \, (q^{i} \sigma^{j}) \, T^a \chi_{\bar{\mbf{p}}} \big]\!(0)}{0} \; ,
\end{aligned}
\end{equation}
where we have defined the following projectors:\footnote{
Since the Pauli matrices and the relative momentum play the same role as the spin and angular momentum polarization vectors in QCD, respectively, polarization tensors $\Delta_J$ defined above can be seen as the sum over polarizations of those polarization vectors in QCD. Indeed, $\Delta_J$ are equivalent to the tensors obtained after sum over polarizations in \cite{Maltoni:1997pt} (note that $\Pi_{\alpha \beta} \Lambda^{\alpha i} \Lambda^{\beta j} = \delta^{ij}$).}
\begin{equation}
\begin{aligned}
\Delta^{i'j' ij}_0  & = \frac{1}{3} \delta^{i'j'} \delta^{ij}  \; ,\\
\Delta^{i'j' ij}_1  & = \frac{1}{2} \left( \delta^{i i'} \delta^{j j'} - \delta^{i j'} \delta^{i' j} \right) \; ,\\
\Delta^{i'j'ij}_2  & = \frac{1}{2} \left( \delta^{i i'} \delta^{j j'} + \delta^{i j'} \delta^{i' j} \right) - \frac{1}{3} \delta^{i'j'} \delta^{ij} \; .
\end{aligned}
\end{equation}

Finally, subtracting the above unpolarized TMDShFs according to~\eqref{eq:renShF}, we obtain a greatly simplified expression for the hadronic tensor~\eqref{eq:hadronictensorfinal} for unpolarized quarkonium production:
\begin{equation}
\begin{aligned}
W^{\mu\nu} 
& =\frac{x_B P^+_N}{Q^2}\Big(1\!-\!\frac{x_B^2 M_N^2 M_{\psi \perp}^2}{\gamma_q^2 Q^4}\Big)\delta\Big(\!z\!-1\!-\!\frac{2 x_B^2 M_N^2 (M_{\psi \perp}^2+Q^2_\perp)}{\gamma_q Q^4}\Big)\\
& \times \sum_{[n]} \big|C_{[n]}\big|^{2} \Gamma_{[n]}^{\dagger \mu \alpha'} \Gamma_{[n]}^{\nu\alpha} \int\frac{\mathrm{d}^{2}b_\perp}{(2\pi)^{2}}e^{ib_\perp\cdot q_{\perp}} G_{g/N , \alpha\alpha'}( b_\perp) \, S_{[n] \to J/\psi}(b_\perp) \;.
\end{aligned}
\label{eq:Unpol-hadronictensorfinal}
\end{equation}
In the above result, the matching tensors for the $J = 1,2$ P-waves are defined as the contractions of the tensors specified in~\eq{eq:EffOpLO} with the completeness relations in~\eq{eq:NaturalProjection}:
\begin{equation}
\begin{aligned}
    \Gamma_{^3P_1^{[8]}}^{\mu \alpha'} \Gamma_{ ^3P_1^{[8]}}^{\nu \alpha} & \equiv \Gamma_{ ^3P_1^{[8]},k'}^{\mu \alpha'} \Gamma_{ ^3P_1^{[8]},k}^{\nu \alpha} \delta^{k' k} \\
    & = \left( \frac{i \, g e}{M} \right)^2 \left( \frac{4 Q^4 (\gamma_q M^2 + (\gamma_q - 1) Q^2)}{M^2 (M^2 + Q^2)^2} \right) \epsilon_\perp^{\mu \alpha'} \epsilon_\perp^{\nu \alpha} \; ,\\
    \Gamma_{^3P_2^{[8]}}^{\mu \alpha'} \Gamma_{ ^3P_2^{[8]}}^{\nu \alpha}  & \equiv \Gamma_{ ^3P_2^{[8]},i'j'}^{\mu \alpha'} \Gamma_{ ^3P_2^{[8]}, i j}^{\nu \alpha} \left[ \frac{1}{2} \left( \delta^{i' i} \delta^{j' j} + \delta^{i' j} \delta^{i j'} \right) - \frac{1}{3} \delta^{i'j'} \delta^{ij} \right] 
    \\ & = \frac{2}{3} \left( \frac{- i2 \, g e}{M} \right)^2 \gamma_q^4 \,  g_\perp^{\mu \alpha'} g_\perp^{\nu \alpha} \; .
\end{aligned}
\end{equation}

For completeness, we present the results for the hadronic tensor explicitly in Fourier space, since this is where the matching onto QCD (see~\hyperref[sec:A-calculation]{appendix C}) is performed:
\begin{equation}
\begin{aligned}
W^{\mu\nu} 
& =\frac{x_B P^+_N}{Q^2}\Big(1\!-\!\frac{x_B^2 M_N^2 M_{\psi \perp}^2}{\gamma_q^2 Q^4}\Big)\delta\Big(\!z\!-1\!-\!\frac{2 x_B^2 M_N^2 (M_{\psi \perp}^2+Q^2_\perp)}{\gamma_q Q^4}\Big)
\sum_{[n]} \big|C_{[n]}\big|^{2} \Gamma_{[n]}^{\dagger \mu \alpha'} \Gamma_{[n]}^{\nu\alpha}
\\
& \times  
\int\mathrm{d}^{2}k_{n\perp}\,\mathrm{d}^{2}k_{s\perp}\,
\d^{2}(q_\perp + k_{n\perp} - k_{s\perp})\,
G_{g/N , \alpha\alpha'}(x,k_{n\perp}) \, 
S_{[n] \to J/\psi}(k_{s\perp}) 
\;,
\end{aligned}
\label{eq:Unpol-hadronictensorfinal2}
\end{equation}
where
\begin{equation}
\begin{aligned} \label{eq:GndS-momspace}
G_{g/N} (x, k_{n\perp}) & = 
\int \frac{d^2 b_\perp}{(2 \pi)^2} e^{i b_\perp \cdot k_{n\perp}} G_{g/N} (x,b _{\perp}) \; , \\
S_{[n] \to J/\psi} (k_{s\perp}) & = \int \frac{d^2 b_\perp}{(2 \pi)^2} e^{-i b_\perp \cdot k_{s\perp}} S_{[n] \to J/\psi} (b_\perp) \; .
\end{aligned}
\end{equation}

\section{TMDShF: evolution and matching}
\label{sec:3}
In this section, we introduce the $S$-state and $P$-states LDMEs and TMDShFs at NLO in $\alpha_s$.
We discuss their RG evolution and derive the matching coefficients between the TMDShFs and LDMEs at large transverse momentum.

The calculation is performed in \hyperref[sec:A-calculation]{appendix C}, using the EFT approach defined in the \hyperref[sec:A-vNRQCD]{appendix A}. We use the vNRQCD Lagrangian and its Feynman rules for the contributions of the diagrams involved at order $\alpha_s$, which are shown in figures \hyperref[fig:TMDShF-LOandCoulomb]{3}, \hyperref[fig:TMDShF-ChroElectric]{4}, \hyperref[fig:TMDShF-WilQuarks]{5}, \hyperref[fig:TMDShF-SoftExchange]{6}, and \hyperref[fig:TMDShF-uSoftExchange]{7}.
As shown in~\eq{eq:SoftFunctions}, we use the $\delta$-regularization for the rapidity divergences.
On the other hand, we use the \textoverline{MS} renormalization scheme, which is implemented by rescaling the renormalization scale $\mu^2 \to \mu^2 e^{\gamma_E}/4 \pi$ when using dimensional regularization for regularizing integrals in the evaluation of Feynman integrals.
As we will see below, the distinction between UV and IR poles will be important to obtain the mixing between channels at NLO.
Moreover, at this order in perturbative QCD, mass renormalization is not necessary.


\subsection{TMDShF at NLO}
A NLO calculation in vNRQCD, which we relegate to~\hyperref[sec:A-calculation]{appendix C}, yields the following results for the LDMEs of the $^1S_0^{[8]}$ and $^3P_J^{[8]}$ states:
\begin{equation}
\begin{aligned} \label{eq:LDMENLO-S}
\braket{^1S_0^{[8]}}
& = \left( 1 +   \left( C_F - C_A/2 \right)\frac{\pi \alpha_s}{2 \tv}  \right)\braket{^1S_0^{[8]}}^{\text{LO}} \\
& + \frac{4 \alpha_s}{3 \pi m_c^2}   \left( C_F \braket{^1P_1^{[1]}}^{\text{LO}} + B_F \braket{^1P_1^{[8]}}^{\text{LO}} \right) \left( \frac{1}{\veuv} - \frac{1}{\veir} \right) +\mathcal{O}(\alpha_s^2) \; ,
\end{aligned}
\end{equation}
and
\begin{equation}
\begin{aligned} \label{eq:LDMENLO-P}
\braket{^3P_J^{[8]}}
& = \left( 1 +   \left( C_F - C_A/2 \right)\frac{\pi \alpha_s}{2 \tv}  \right)\braket{^3P_J^{[8]}}^{\text{LO}} \\
& + \frac{4 \alpha_s}{3 \pi m_c^2}   \left( C_F \braket{^3D_{J+1}^{[1]}}^{\text{LO}} + B_F \braket{^3D_{J+1}^{[8]}}^{\text{LO}} \right) \left( \frac{1}{\veuv} - \frac{1}{\veir} \right) +\mathcal{O}(\alpha_s^2) \; .
\end{aligned}
\end{equation}
Here $C_F= (N_c^2-1)/2N_c$ and $C_A= N_c$ are the Casimir operators in the fundamental and adjoint representation, respectively, and where $B_F = (N_c^2-4)/4N_c$.\footnote{$B_F$ is defined as $\sum_{bc} d^{abc} d^{e bc} = 4 B_F \delta^{ae}$ where the $d$-coefficients are symmetric structure constants defined via the anticommutator of the generators of the Lie algebra: $d^{abc} = 2 \, \text{Tr} (\{ T^a, T^b\} T^c)$.}
In the above equation, the LDMEs at LO, $\braket{\mathcal O}^{\text{LO}}$, consist of the heavy-quark spinors and the object $\mathcal K$. The latter is a combination of Pauli matrices and the quark's relative momentum, describing the particular configuration $[n]$ of the bound state (see \hyperref[sec:A-calculation]{appendix C}).
The result contains so-called Coulomb singularities of the type $1/\tv$, which arise due to the long-range Coulomb interaction between the heavy quark and antiquark.
Moreover, the LDMEs for all channels turn out to mix with channels of higher angular momentum at NLO.
This can be traced back to the chromo-electric dipole transition by usoft radiation (the first line in the interaction vNRQCD Lagrangian in \eq{eq:Lagrangian}).

The results~\eqref{eq:LDMENLO-S} and~\eqref{eq:LDMENLO-P} were already known in the literature, see, e.g.,~\cite{Petrelli:1997ge}. In~\hyperref[sec:ShNLO]{appendix C.2}, however, we present the first calculation of the TMDShFs at NLO in SIDIS, of which the results read:\footnote{The TMDShFs turn out to depend only on the module of $b_\perp$, i.e., $b_T$.}
\begin{equation}
\begin{aligned} \label{eq:ShTotal-S}
& S_{^1S_0^{[8]} \to J/\psi} (b_T; \mu, \zeta_B) = 
 \braket{^1S_0^{[8]}}^{\text{LO}} + 
\frac{\alpha_s}{2 \pi}
 \Bigg[  \frac{C_A}{\veuv} \left( 1 - \ln \, \zeta_B  \right)
 \braket{^1S_0^{[8]}}^{\text{LO}} 
 \\ 
&  \quad
+ C_A L_T \left( 1 - \ln \, \zeta_B \right)
\braket{^1S_0^{[8]}}^{\text{LO}}
- \frac{8}{3m_c^2} L_T \left(  C_F \braket{^1P_1^{[1]}}^{\text{LO}} + B_F \braket{^1P_1^{[8]}}^{\text{LO}} \right)   \\
& \quad
+ \frac{\pi^2}{\tv} (C_F - C_A/2)
\braket{^1S_0^{[8]}}^{\text{LO}}
-  \frac{8}{3 m_c^2}  \frac{1}{\veir} \left(  C_F \braket{^1P_1^{[1]}}^{\text{LO}} + B_F \braket{^1P_1^{[8]}}^{\text{LO}} \right) \Bigg]  \; ,
\end{aligned}
\end{equation}
and
\begin{equation}
\begin{aligned} \label{eq:ShTotal-P}
& S_{^3P_J^{[8]} \to J/\psi} (b_T; \mu, \zeta_B) = 
\frac{1}{N^{(J)}_{pol}} \left\{ \braket{^3P_J^{[8]}}^{\text{LO}} + 
\frac{\alpha_s}{2 \pi}
\Bigg[  \frac{C_A}{\veuv} \left( 1 - \ln \, \zeta_B  \right)
\braket{^3P_J^{[8]}}^{\text{LO}} \right.
\\ 
& \left. \quad
+ C_A L_T \left( 1 - \ln \, \zeta_B \right)
\braket{^3P_J^{[8]}}^{\text{LO}}
-  \frac{8}{3m_c^2} L_T \left(  C_F \braket{^3D_{J+1}^{[1]}}^{\text{LO}} + B_F \braket{^3D_{J+1}^{[8]}}^{\text{LO}} \right) \right.  \\
& \left.\quad
+ \frac{\pi^2}{\tv} (C_F - C_A/2)
\braket{^3P_J^{[8]}}^{\text{LO}}
-  \frac{8}{3 m_c^2}  \frac{1}{\veir} \left(  C_F \braket{^3D_{J+1}^{[1]}}^{\text{LO}} + B_F \braket{^3D_{J+1}^{[8]}}^{\text{LO}} \right) \Bigg] \right\} \; ,
\end{aligned}
\end{equation}
with $L_T = \ln \left( \mu^2 b_T^2 e^{2 \gamma_E}/4 \right)$ and $N_{pol}^{(J)} = 2J+1$.
Here the results are renormalized with respect to the rapidity divergences, where $\zeta_B$ is the scale of rapidity subtraction. 
Similarly to the LDMEs, the TMDShFs contain Coulomb singularities and feature mixing between channels. Moreover, we obtain additional contributions from soft gluon exchanges between the soft Wilson lines, see~\hyperref[fig:TMDShF-SoftExchange]{figure 6}. As we explain in the \hyperref[sec:A-calculation]{appendix C}, the contribution from the diagrams in \hyperref[fig:TMDShF-WilQuarks]{figure 5} vanishes and we do not consider the contribution from~\hyperref[fig:TMDShF-uSoftExchange]{figure 7} to avoid double-counting.


\subsection{Evolution of LDMEs and TMDShFs}

In this section, we renormalize the bare LDMEs and TMDShFs obtained in the previous section (note there is some abuse of notation: we use the same symbols for the unrenormalized and renormalized LDMEs and TMDShFs).
Moreover, we address both the virtuality and rapidity evolution equations.
For simplicity, as usual, we work with the TMDShFs in the $b_\perp$-space.
Given that the procedure is the same for P-states, we focus on the S-state and write the result for the P-states at the end of the section.

As we can see in the NLO calculation for the S-state LDME in eq.~\eqref{eq:LDMENLO-S}, gluon radiation generates an UV divergence proportional to the $^1P_1^{[1,8]}$ LDMEs at LO. This divergence is removed through UV renormalization, leading to the following renormalization group equation (RGE):
\begin{equation}
\frac{d}{d \, \ln \mu} \braket{\mathcal O^n(\mu)} = \sum_m \gamma_{\mathcal O}^{nm} (\mu) \braket{\mathcal O^m(\mu)} \; .
\end{equation}
Here we use lowercase letters $n$ and $m$ to refer to the configuration $^{2S+1}L_J^{[col.]}$ of the heavy-quark pair (in the rest of the paper we use the notation $[n]$). 
Since there is a mixing between LDMEs at NLO, the anomalous dimension is actually a matrix $\gamma^{nm}_\mathcal{O}$.
Taking the renormalized LDME, $\braket{\mathcal{O}^m(\mu)}$, as
\begin{equation}
\begin{aligned}
\braket{ \mathcal{O}^n} = Z^{nm}_{\mathcal{O}} (\mu) \braket{\mathcal{O}^m(\mu)} \; ,
\end{aligned}
\end{equation}
with $Z_\mathcal{O}^{nm}$ the renormalization matrix, the anomalous dimension matrix is defined as
\begin{equation} \label{eq:AnomDim}
\gamma_\mathcal{O}^{nk}   \equiv - (Z^{-1}_\mathcal{O})^{nm} \frac{d}{d \, \ln \mu} Z_\mathcal{O}^{mk} \; .
\end{equation}
The renormalization matrix is a $3 \times 3$ matrix due to the three involved channels in the NLO calculation of the $^1S_0^{[8]}$ LDME.
In the last equation, the $\mu$-dependence is suppressed on both sides.
Moreover, from~\eq{eq:LDMENLO-S} we see that the countermatrix for the LDME is the following:
\begin{equation}
\begin{aligned}
    Z_{^1S_0^{[8]}} (\mu) & = 
    \begin{pmatrix}
    1 & 0 & 0\\
    0 & 1 & 0 \\
    0 & 0 & 1
    \end{pmatrix} + \frac{4 \alpha_s(\mu)}{3 \pi m_c^2} \frac{1}{\veuv} 
    \begin{pmatrix}
    0 & C_F & B_F\\
    0 & 0 & 0 \\
    0 & 0 & 0
    \end{pmatrix} \; .
\end{aligned}
\end{equation}
Combining this result with the definition of the anomalous dimension matrix in~\eq{eq:AnomDim}, we get the anomalous dimension for the $^1S_0^{[8]}$ LDME:
\begin{equation}
\begin{aligned}
\gamma_{^1S_0^{[8]}} (\mu) & = \frac{8 \alpha_s(\mu)}{3 \pi m_c^2} \begin{pmatrix}
    0 & C_F & B_F\\
    0 & 0 & 0 \\
    0 & 0 & 0
    \end{pmatrix} \; .
\end{aligned}
\end{equation}
Therefore, the RGE satisfied by the $^1S_0^{[8]}$ LDME is
\begin{equation}
\begin{aligned}
\frac{d}{d \, \ln \mu} \braket{^1S_0^{[8]} (\mu) } & = \sum_m \gamma_{^1S_0^{[8]}}^{1m} (\mu)  \braket{\mathcal{O}^m (\mu)} \\
& = \frac{8 \alpha_s(\mu)}{3 \pi m_c^2} \left( C_F \braket{^1P_1^{[1]} (\mu)} + B_F \braket{^1P_1^{[8]} (\mu)} \right) \; .
\end{aligned}
\end{equation}
Here the superscript one of the anomalous dimension refers to $^1S_0^{[8]}$ state.
We now can solve the RGE and get the $\mu$-evolution of the LDME as follows:
\begin{equation}
\begin{aligned}
    \braket{^1S_0^{[8]}(\mu)} =\braket{^1S_0^{[8]}(\mu_f)}+\omega_1(\mu,\mu_f)\braket{^1P_1^{[1]}(\mu_f)}+\omega_8(\mu,\mu_f)\braket{^1P_1^{[8]}(\mu_f)}\; .
\end{aligned}
\end{equation}
Here, $\mu_f$ terms the scale at which the LDME is extracted. The most common choice for phenomenological analyses is $\mu_f = M_\psi$. The evolution kernels are given by:
\begin{equation}
\omega_1(\mu,\mu_f) = \exp\Big(\!\!-\frac{16 C_F}{3 m_c^2 \beta_0} \ln \frac{\alpha_s(\mu)}{\alpha_s(\mu_f)}\Big)\;, \quad\quad \omega_8(\mu,\mu_f) = \exp\Big(\!\!-\frac{16 B_F}{3 m_c^2 \beta_0} \ln \frac{\alpha_s(\mu)}{\alpha_s(\mu_f)}\Big) \; ,
\end{equation}
with $\beta_0 = 11 C_A/3 - 2 n_f/3 $. 

It is clear from~\eq{eq:LDMENLO-S} and~\eq{eq:LDMENLO-P} that the UV behavior of the P-waves is the same as for the S-wave. Hence, we obtain the same evolution equations for the $^3P_J^{[8]}$ LDMEs:
\begin{equation}
\begin{aligned}
    \braket{^3P_J^{[8]}(\mu)} =\braket{^3P_J^{[8]}(\mu_f)}+\omega_1(\mu,\mu_f)\braket{^3D_{J+1}^{[1]}(\mu_f)}+\omega_8(\mu,\mu_f)\braket{^3D_{J+1}^{[8]}(\mu_f)}\; .
\end{aligned}
\end{equation}

We now turn to compute the anomalous dimension of the S-state TMDShF to order $\alpha_s$. From~\eq{eq:ShTotal-S} the complete renormalization factor for the UV divergence is the following
\begin{equation} \label{eq:RenoFactSh}
\begin{aligned}
Z_{\text{Sh}}(\mu) = 1 + \frac{\alpha_s(\mu) C_A}{2 \pi} \frac{1}{\veuv} \left( 1 - \ln \, \zeta_B \right) \; ,
\end{aligned}
\end{equation}
and the corresponding anomalous dimension, which gives the evolution in the renormalization scale $\mu$ to order $\alpha_s$, is  
\begin{equation} \label{eq:ADsh}
\begin{aligned}
\gamma_{\text{Sh}}  & = - \frac{d}{d \, \ln \mu} \ln \, Z_{\text{Sh}} = - \frac{1}{Z_{\text{Sh}}} \frac{\partial  Z_{\text{Sh}}}{\partial \ln \mu} - \frac{1}{Z_{\text{Sh}}} \frac{\partial Z_{\text{Sh}}}{\partial \alpha_s} \left( -2 \veuv \alpha_s + \mathcal O (\alpha_s^2) \right)  \\
& = \frac{\alpha_s C_A}{\pi} \left( 1- \ln \, \zeta_B \right) \; ,
\end{aligned}
\end{equation}
where we use that $d \alpha_s(\mu)/d \ln \mu = -2 \veuv \alpha_s + \beta(\alpha_s)$.\footnote{$\beta(\alpha_s)$ is the standard QCD beta function written in terms of $\alpha_s$: $\beta(\alpha_s) = - 2 \alpha_s \sum_{n= 1}^\infty \beta_{n-1} (\alpha_s/4 \pi)^n$ in the \textoverline{MS} renormalization scheme.}
Similarly to the LDMEs, the UV behavior of the TMDShFs of the S-waves and the P-waves, see eqs.~\eqref{eq:ShTotal-S} and~\eqref{eq:ShTotal-P}, is the same. Therefore, the renormalization factor~\eqref{eq:RenoFactSh} and the anomalous dimension~\eqref{eq:ADsh} equally apply to the $^3P_J^{[8]}$ TMDShF.
Considering that, we write down our final results for the renormalized LDMEs and TMDShFs at NLO. The former are:
\begin{equation}
\begin{aligned} \label{eq:renLDMENLO-S}
\braket{^1S_0^{[8]}(\mu)}
& = \left( 1 +   \left( C_F - C_A/2 \right)\frac{\pi \alpha_s}{2 \tv}  \right)\braket{^1S_0^{[8]}(\mu)}^{\text{LO}} \\
& - \frac{1}{\veir} \frac{4 \alpha_s}{3 \pi m_c^2}   \left( C_F \braket{^1P_1^{[1]}(\mu)}^{\text{LO}} + B_F \braket{^1P_1^{[8]}(\mu)}^{\text{LO}} \right)   \; ,
\end{aligned}
\end{equation}
and
\begin{equation}
\begin{aligned} \label{eq:renLDMENLO-P}
\braket{^3P_J^{[8]}(\mu)}
& = \left( 1 +   \left( C_F - C_A/2 \right)\frac{\pi \alpha_s}{2 \tv}  \right)\braket{^3P_J^{[8]}(\mu)}^{\text{LO}} \\
& - \frac{1}{\veir} \frac{4 \alpha_s}{3 \pi m_c^2}   \left( C_F \braket{^3D_{J+1}^{[1]}(\mu)}^{\text{LO}} + B_F \braket{^3D_{J+1}^{[8]}(\mu)}^{\text{LO}} \right)    \; .
\end{aligned}
\end{equation}
Likewise, the renormalized TMDShFs read:
\begin{equation} \label{eq:renSh-S}
\begin{aligned}
& S_{^1S_0^{[8]} \to J/\psi} (b_T; \mu, \zeta_B) = 
 \braket{^1S_0^{[8]}(\mu)}^{\text{LO}} + 
\frac{\alpha_s}{2 \pi}
 \Bigg[   C_A L_T \left( 1 - \ln \, \zeta_B \right)
\braket{^1S_0^{[8]}(\mu)}^{\text{LO}} 
 \\ 
&  \quad
-  \frac{8}{3m_c^2} L_T \left(  C_F \braket{^1P_1^{[1]}(\mu)}^{\text{LO}} + B_F \braket{^1P_1^{[8]}(\mu)}^{\text{LO}} \right) + \frac{\pi^2}{\tv} (C_F - C_A/2)
\braket{^1S_0^{[8]}(\mu)}^{\text{LO}}   \\
& \quad
-  \frac{8}{3 m_c^2}  \frac{1}{\veir} \left(  C_F \braket{^1P_1^{[1]}(\mu)}^{\text{LO}} + B_F \braket{^1P_1^{[8]}(\mu)}^{\text{LO}} \right) \Bigg]   \; ,
\end{aligned}
\end{equation}
and
\begin{equation} \label{eq:renSh-P}
\begin{aligned}
& S_{^3P_J^{[8]} \to J/\psi} (b_T; \mu, \zeta_B) = 
\frac{1}{N^{(J)}_{pol}} \left\{ \braket{^3P_J^{[8]}(\mu)}^{\text{LO}} + 
\frac{\alpha_s}{2 \pi}
 \Bigg[   C_A L_T \left( 1 - \ln \, \zeta_B \right)
\braket{^3P_J^{[8]}(\mu)}^{\text{LO}} \right.
 \\ 
& \left. \quad
-  \frac{8}{3m_c^2} L_T \left(  C_F \braket{^3D_{J+1}^{[1]}(\mu)}^{\text{LO}} + B_F \braket{^3D_{J+1}^{[8]}(\mu)}^{\text{LO}} \right) + \frac{\pi^2}{\tv} (C_F - C_A/2)
\braket{^1S_0^{[8]}(\mu)}^{\text{LO}} \right.  \\
& \left.\quad
-  \frac{8}{3 m_c^2}  \frac{1}{\veir} \left(  C_F \braket{^3D_{J+1}^{[1]}(\mu)}^{\text{LO}} + B_F \braket{^3D_{J+1}^{[8]}(\mu)}^{\text{LO}} \right) \Bigg] \right\}   \; .
\end{aligned}
\end{equation}
In refs.~\cite{Boer:2020bbd,Boer:2023zit}, the authors have devised an indirect way to infer the value of the TMDShFs at large transverse momentum (see also footnote~\ref{footnote:matching}). 
For a specific scale choice $\zeta_B = Q_H^2/M_\psi^2$, their results seem to agree with our calculation. 
Although a sensible choice, it will require a very detailed analysis to clarify why their method requires such a specific value of $\zeta_B$, and we leave this for future work.

We end this section by showing the $\zeta_B$-evolution of the TMDShF. 
The TMD evolution equation in $\zeta_B$ is the following: \cite{Echevarria:2012pw}
\begin{equation} \label{eq:AnomDimSh}
\begin{aligned}
\frac{d}{d \, \ln \zeta_B}  \ln\, S_{^1S_0^{[8]} \to J/\psi}(b_T; \mu,\zeta_B ) = -\mathcal{D}_g (b_T; \mu)\; ,
\end{aligned}
\end{equation}
where $\mathcal{D}_g (b_T; \mu)$ is called the rapidity anomalous dimension (RAD) or Collins-Soper (CS) kernel. 
In the above equation, it is understood that the TMDShF is renormalized. 
From \eq{eq:renSh-S}, we obtain the following result for the RAD:
\begin{equation}
\begin{aligned} \label{eq:RAD}
    \mathcal{D}_g(b_T;\mu) = \frac{\alpha_s C_A}{2 \pi} L_T +\mathcal{O}(\alpha_s^2) \; .
\end{aligned}
\end{equation} 
Once again, since they share the same dependence on $\zeta_B$, the TMDShFs for the S-wave and for the P-waves is the same.
We note that the above result is the same as for the TMDPDF (see, e.g., for the gluon TMDPDF in Higgs production \cite{Echevarria:2015uaa,Echevarria:2015byo}). This is to be expected, since the structure of the rapidity divergences in TMDPDFs and TMDShF is the same, in both cases stemming from the soft function.

Therefore, the complete evolution kernel for the TMDShFs is the same for the $^1S_0^{[8]}$ state and for the $^3P_J^{[8]}$ states:
\begin{equation}
\begin{aligned}
S_{^1S_0^{[8]} \to J/\psi} (b_T;\mu,\zeta_B) & = 
\exp \left[ \int \left( \gamma_{\text{Sh}}(\mu,\zeta_B) \frac{d \mu}{\mu} - \mathcal{D}_g (b_T; \mu) \frac{d \zeta_B}{\zeta_B} \right) \right] S_{^1S_0^{[8]} \to J/\psi} (b_T; \mu_f,\zeta_f) \; , \\
S_{^3P_J^{[8]} \to J/\psi} (b_T;\mu,\zeta_B) & = 
\exp \left[ \int \left( \gamma_{\text{Sh}}(\mu,\zeta_B) \frac{d \mu}{\mu} - \mathcal{D}_g (b_T; \mu) \frac{d \zeta_B}{\zeta_B   } \right) \right] S_{^3P_J^{[8]} \to J/\psi} (b_T; \mu_f,\zeta_f) \; .
\end{aligned}
\end{equation}


\subsection{Matching onto LDMEs}

In the limit of large transverse momentum, TMDShFs can be matched onto LDMEs, similar to how TMDPDFs can be matched onto collinear PDFs. This is done by performing operator product expansion (OPE) and then extracting the transverse dependence in terms of the Wilson coefficients:
\begin{equation}
\begin{aligned} \label{eq:OPE}
S_{[n] \to J/\psi}  (b_T ; \mu, \zeta_B) = \sum_{[m]} C^{[n]}_{[m]} (b_T; \mu, \zeta_B) \times \frac{\braket{\mathcal O ^{[m]}}}{N^{(J)}_{pol}} + \mathcal{O}(b_T \Lambda_{\text{QCD}}) \; ,
\end{aligned}
\end{equation}
where both matrix elements, LDMEs and TMDShF, are understood to be renormalized as in the previous section.
Here $C^{[n]}_{[m]}$ terms the matching coefficient between the $[n] = \, ^{2S+1} L_J^{[col.]}$ TMDShF and the $[m] = \,^{2S+1}L_J^{[col.]}$ LDME at low $b_T$. 
Note that $[m]$ can be different from $[n]$. 
We simplify the notation by naming the $S$-state TMDShF as $[n] =S$ and the $P$-states TMDShF as $[n]=P$. 
Doing this, we have, e.g., that  $C^P_{^1S_0^{[8]}}$ stands for the matching coefficient of $^3P_J^{[8]}$ TMDShF onto $^1S_0^{[8]}$ LDME. 
Following this notation and getting back to the LDMEs in \eq{eq:renLDMENLO-S} and the TMDShF in \eq{eq:renSh-S} obtained above, we get the following matching coefficients for the $^1S_0^{[8]}$ channel:
\begin{equation}
\begin{aligned} \label{eq:MatchCoefsS}
C^S_{^1S_0^{[8]}} (b_T; \mu, \zeta_B) & = 1 + \frac{\alpha_s C_A}{2 \pi}  L_T \left( 1 - \ln \, \zeta_B \right)   \; ,\\
C^S_{^1P_1^{[1]}} (b_T; \mu) & = -  \frac{\alpha_s}{2 \pi} \frac{8 \, C_F}{3 m_c^2} L_T \; , \\
C^S_{^1P_1^{[8]}} (b_T; \mu) & = -  \frac{\alpha_s}{2 \pi} \frac{8 \, B_F}{3 m_c^2} L_T \;  .
\end{aligned}
\end{equation}
Moreover, by using the results in \eq{eq:renLDMENLO-P} and in \eq{eq:renSh-P} for the $^3P_{J=0,1,2}^{[8]}$ channel, we obtain:
\begin{equation}
\begin{aligned} \label{eq:MatchCoefsP}
C^P_{^3P_J^{[8]}} (b_T; \mu, \zeta_B) & = 1 + \frac{\alpha_s C_A}{2 \pi}  L_T \left( 1 - \ln \, \zeta_B \right)   \; ,\\
C^P_{^3D_{J+1}^{[1]}} (b_T; \mu) & = -  \frac{\alpha_s}{2 \pi} \frac{8 \, C_F}{3 m_c^2} L_T \; , \\
C^P_{^3D_{J+1}^{[8]}} (b_T; \mu) & = -  \frac{\alpha_s}{2 \pi} \frac{8 \, B_F}{3 m_c^2} L_T 
\;.
\end{aligned}
\end{equation}


\subsection{Cross-check}

We would like to conclude this section by discussing two points regarding the results obtained above.
Firstly, we match the QCD cross section onto the SCET one at NLO. 
As mentioned in \hyperref[sec:2]{section 2}, the total cross section is the sum of all the cross sections for each channel, resulting in four hard functions, one for each channel.
Since the hard function originates from the virtual contribution, we can refer to the result for photoproduction in: \cite{Maltoni:1997pt} 
\begin{equation} \label{eq:VirtualCS}
\begin{aligned}
\sigma^{(b)}_{[n]} & = \delta(1-x) (\sigma_{[n]}^{0 (b)} + \sigma_{[n]}^{V(b)})   \; , \\
\sigma_{[n]}^{V(b)} & = \sigma_{[n]}^{0 (b)} \frac{\alpha^{(b)}_s}{2 \pi} f_\ve(Q_H^2) \\
& \times \left[ \frac{\beta_0}{2}\left( \frac{1}{\veuv} - \frac{1}{\veir}\right) + \left( C_F - \frac{C_A}{2} \right)  \frac{\pi^2}{\tv}  - C_A \left( \frac{1}{\veir^2} + \frac{1}{\veir} \right)  + \mathcal D_{[n]} \right] \; ,
\end{aligned}
\end{equation}
with $Q_H$ the hard scale where $Q_H = M_\psi$ for photoproduction and $Q_H = f(Q,M_\psi)$ for leptoproduction.
In the above equation, the superscript $(b)$ explicitly indicates that this result should be renormalized, $\sigma^{0(b)}_{[n]}$ is the Born cross section for the channel $[n]$, $\mathcal D_{[n]}$ are finite terms for each channel, and
\begin{equation}
\begin{aligned}
f_\ve(Q_H^2) \left. \right|_{\overline{\text{MS}}} & = \left( \frac{\mu^2 e^{\gamma_E}}{Q_H^2} \right)^{\ve} \Gamma(1+\ve) = 1+ \ve\, \ln \frac{\mu^2}{Q_H^2} + \frac{\ve^2}{12} \left( \pi^2 + \ln^2 \frac{\mu^2}{Q_H^2} \right) + \mathcal O (\ve^3) \; .
\end{aligned}
\end{equation}
Since the Born cross section depends on $\alpha^{(b)}_s (\mu)$, we renormalize the cross section by using the renormalization of the  QCD strong coupling.
By doing this, we remove the UV divergence.
Then we insert the expanded $f_\ve(Q_H^2)$ in $\sigma^V_{[n]}$, obtaining the following renormalized virtual contribution to the cross section for the channel $[n]$:
\begin{equation}
\begin{aligned} \label{eq:renorCS}
\sigma^V_{[n]} & = \sigma^0_{[n]} \left[ 1 + \frac{\alpha_s}{2 \pi} \left\{ \left[ \left( C_F - C_A/2 \right)  \frac{\pi^2}{\tv} - \frac{C_A}{\veir} \right] - \left[ \frac{C_A}{\veir^2} + \frac{1}{\veir} \left( \frac{\beta_0}{2} + C_A \, \ln \frac{\mu^2}{Q_H^2} \right) \right] \right. \right.\\
& \left. \left. - C_A \left( \frac{\beta_0}{2C_A} \, \ln \frac{\mu^2}{Q_H^2} + \ln \frac{\mu^2}{Q_H^2} + \frac{1}{2} \,  \ln^2 \frac{\mu^2}{Q_H^2}  + \frac{\pi^2}{12} \right) + \mathcal D_{[n]} \right\} \right] \; .
\end{aligned}
\end{equation}
Here $\sigma^0_{[n]}$ is the renormalized Born cross section which is a function of the renormalized strong coupling, $\alpha_s$.
Although this cross section is not specifically for leptoproduction, its UV and IR behavior is the same, allowing us to use the first line of \eq{eq:renorCS} to verify the virtual contribution of the TMDShF obtained in this work. 
In fact, the first square of the first line corresponds to the virtual contribution of the TMDShF, as described in \eq{eq:ShKperp}, while the second term corresponds to the virtual part of the gluon TMDPDF \cite{Echevarria:2015uaa}.
The remainder of the result will be determined by the hard function; however, the NLO calculation of the cross section for leptoproduction has not yet been completed.
We will present the results in the future, accompanied by a phenomenological analysis.

In addition to the previous discussion, we provide an analysis on the anomalous dimensions.
We calculate the anomalous dimension of the Hard function, $\gamma_H$, from the second line of~\eq{eq:renorCS}:
\begin{equation}
\begin{aligned}
\gamma_H \equiv \frac{1}{H} \frac{d \,H}{d \, \ln \mu} = - \frac{\alpha_s(\mu) C_A}{ \pi} \left( 1 +  \ln \frac{\mu^2}{Q_H^2} \right) \; .
\end{aligned}
\end{equation}
Since the hadronic tensor does not depend on the factorization scale, we can verify the result obtained for the TMDShF anomalous dimension in~\eq{eq:ADsh}. To do so, we consider that:
\begin{equation}
\begin{aligned}
\frac{1}{\sigma} \frac{d \, \sigma}{d \, \ln \mu} & = 0 \; , \\
\frac{1}{\sigma_0} \frac{d \, \sigma_0}{d \, \ln \mu} + \gamma_H + \gamma_{f_{1g}} + \gamma_{\text{Sh}} & = 0 \; .
\end{aligned}
\end{equation}
The evolution of the renormalized Born cross section is as follows:
\begin{equation}
\begin{aligned}
\frac{1}{\sigma_0} \frac{d \, \sigma_0}{d \, \ln \mu} & = - \frac{\alpha_s(\mu)}{\pi} b_0 \; .
\end{aligned}
\end{equation}
Moreover, we take the TMDPDF anomalous dimension, denoted by $\gamma_{f_{1g}}$ from, e.g., \cite{Echevarria:2015uaa}:
\begin{equation}
\begin{aligned}
\gamma_{f_{1g}} & = \frac{\alpha_s(\mu)}{\pi} \left( b_0 - C_A \, \ln \frac{\zeta_A}{\mu^2} \right) \; .
\end{aligned}
\end{equation}
Therefore, the TMDShF anomalous dimension should be:
\begin{equation}
\begin{aligned} \label{eq:ADsh2}
\gamma_{\text{Sh}} = \frac{\alpha_s(\mu) C_A}{\pi} \left( 1 - \ln \, \zeta_B \right) \; ,
\end{aligned}
\end{equation}
because $\zeta_A \zeta_B = Q_H^2$.
Indeed, \eq{eq:ADsh2} coincides with the result obtained in~\eq{eq:ADsh}, confirming that the TMDShF anomalous dimension derived in this work is as expected.


\section{Conclusions}
\label{sec:4}
In this work, we have studied $J/\psi$ production at small transverse momentum $P_{\psi\perp}\!\ll\! Q_H$ in SIDIS. To do so, we have used an effective field theory approach which combines vNRQCD with SCET. At leading power in $\lambda \equiv P_{\psi\perp}/Q_H$, the $J/\psi$ is produced in gluon-photon fusion through the intermediate color-octet states $^1S_{0}^{[8]}$, $^3S_{1}^{[8]}$, $^3P_{J\!=\!0,1,2}^{[8]}$. We have provided a precise definition of the relevant vNRQCD+SCET operators and matched the corresponding coefficients onto the tree-level QCD calculation. At the leading power in $\lambda$ and to all orders in $\alpha_s$, the $^3S_{1}^{[8]}$ contributions are observed to vanish, as was earlier observed in refs.~\cite{Mukherjee:2016qxa,Bacchetta:2018ivt}.

The main result of our paper is the first leading-power factorization theorem for this process in terms of the gluon TMDPDF as well as different TMDShFs. Every color-octet state has a corresponding TMDShF which parameterizes the hadronization of the bound state into the outgoing $J/\psi$ at low transverse momentum. We have calculated the TMDShFs and their collinear counterparts, the LDMEs, at next-to-leading order in perturbation theory. Moreover, we have obtained the Wilson coefficients to match the TMDShFs onto the LDMEs in the limit of large transverse momentum. Finally, we have established the evolution equations for the TMDShFs from a renormalization-group analysis. Intermediate results of our study, namely the LDMEs at NLO, agree with earlier results in the literature, see~\cite{Maltoni:1997pt}. 

At the scales under consideration, we expect the color-singlet contribution to be relevant as well. Indeed, although suppressed in powers of $\alpha_s$ and $\lambda$, the color singlet is enhanced in the NRQCD parameter $\tv$ with respect to the color-octet states. We plan to address this channel in future work. 


\acknowledgments
We are grateful to S.~Fleming, T.~Mehen, and C.~Pisano for helpful discussions.
This project is also supported by the State Agency for Research of the Spanish Ministry of Science and Innovation through the grants PCI2022-132984, PID2022-136510NB-C33 and CNS2022-135186, by the Basque Government through the grant IT1628-22, as well as by the European Union Horizon 2020 research and innovation program under grant agreement No.~824093 (STRONG-2020). The work of PT is supported by a postdoctoral fellowship fundamental research of the Research Foundation Flanders (FWO) No.~1233422N.


\appendix

\section{vNRQCD} \label{sec:A-vNRQCD}

In this section we establish the formalism we have used in the LDME and TMDShF calculations.
In vNRQCD \cite{Luke:1999kz, Rothstein:2018dzq} there are two hard scales: the heavy quark mass, $m$, and the hard scale of the process, $Q_H$. In the low energy theory there are two scales: the heavy quark momentum, $m \tv$, and the kinetic energy of the heavy quark, $m \tv^2$, inside quarkonium. At the matching scale $m$ we integrate out the hard modes with momentum $p^\mu \sim m$ and the off-shell ones including gluons with energy $p^0 \sim m \tv$ and momentum $\mbf p \sim m \tv^2$.

By using the time-like vector $v^\mu$ such that $v^2 = 1$, we can split the four-momentum of the heavy quark $Q$ into
\begin{equation}
\begin{aligned}
p^\mu_Q = m v^\mu + k^\mu \; ,
\end{aligned}
\end{equation}
where $k^0$ is the kinetic energy and $\mathbf{k}$ is the three-momenta of the heavy quark. We consider the heavy quark is on-shell, so $p_Q^2 = m^2$.
In the nonrelativistic limit, the three-momentum $\mathbf{k}$ is small compared to the mass, so $\mathbf{k} \sim m \tv$ and the only solution for the on-shell equation is for $k^0$ to be the kinetic energy, $k^0 \sim m \tv^2$, so we set the $k$-scaling from the on-shell condition:
\begin{equation}
m^2 + 2 m k^0 + (k^0)^2 - \mbf k^2  = m^2 \quad \mathrm{such\;that}\quad
k \sim m (\tv^2, \tv, \tv, \tv)  \; .
\end{equation}
To distinguish the soft and ultrasoft modes in the $k^\mu$ definition we use the label-momentum formalism, in which we split the $k^\mu$ momentum into a large label (soft) part, $\ell^\mu$, and a small residual (ultrasoft) part, $r^\mu$:
\begin{equation}
k^\mu  =   \ell^\mu + r^\mu \quad \mathrm{with}\quad \ell^\mu \sim m (0,\tv,\tv,\tv) \quad \mathrm{and} \quad r^\mu \sim m (\tv^2,\tv^2,\tv^2,\tv^2) \; .
\end{equation}
Therefore, denoting the soft gluons and massless quark fields at the scale $m \tv$ as $A_{\boldsymbol \ell}^\mu$ and $\psi_{\boldsymbol \ell}$, and denoting the ultrasoft gluons and massless quark fields at the scale $m \tv^2$ as $A_{us}^\mu$ and $\psi_{us}$, the vNRQCD Lagrangian is constructed:
\begin{equation}
\mathcal L _{\text{vNRQCD}} =  \mathcal L _p + \mathcal L _s + \mathcal L _{us} \;.
\end{equation}
The potential vNRQCD Lagrangian, $\mathcal L_{p}$, describes the interaction between heavy (anti) quark fields $\psi_{\boldsymbol \ell}$ resp. $\chi_{-\boldsymbol \ell}$ and ultrasoft gluon fields $ A_{us}^\mu$:
\begin{equation} \label{eq:LagrangianvNRQCD}
\begin{aligned}
\mathcal{L}_{p} & =  \sum_{\boldsymbol \ell} \psi^\dagger_{\boldsymbol \ell} \left\{ i D^0 - \frac{(\mbf{P} - i \mbf{D})^2}{2m} \right\} \psi_{\boldsymbol \ell} + (\psi \rightarrow \chi, \, T \to \bar T) \\
& + \sum_{\boldsymbol \ell, \boldsymbol \ell '} \frac{4 \pi \alpha_s}{(\boldsymbol \ell - \boldsymbol \ell ')^2} \, \psi_{\boldsymbol \ell}^\dagger T^a \psi_{\boldsymbol \ell '} \, \chi_{- \boldsymbol \ell}^\dagger \bar T^a \chi_{- \boldsymbol \ell '}  + \mathcal{O}(\tv^6) \; .
\end{aligned}
\end{equation}
In the above equation, the covariant derivative is defined as $i D^\mu = i \partial^\mu - g A_{us}^\mu$.\footnote{Note that this definition is not consistent with the one provided in section 2. In the current section, we use the definition of the covariant derivative so that the vNRQCD framework is in accordance with the previous works.} The full derivative operator has been decomposed as follows: $i \partial_\mu  \rightarrow \mbf P + i \partial_\mu$ with the label operator $\mbf P$ acting on the label momentum as $\mbf P \, \psi_{\boldsymbol \ell} = \boldsymbol \ell \, \psi_{\boldsymbol \ell}$. The potential heavy-quark fields scale as $\psi\sim\chi\sim \tv^{3/2}$ and the covariant derivative as $D^\mu \sim \tv^2$. Thus the potential vNRQCD Lagrangian scales as $\mathcal L_p \sim \tv^5 + \alpha_s \tv^4 + \mathcal{O}(\tv^6)$. 

The soft vNRQCD Lagrangian, $\mathcal L_{s}$, involves a pure soft fields part and interaction terms involving both potential heavy quarks, $\psi_{\boldsymbol \ell}$, and soft fields. At leading power in v, the only term in the interaction Lagrangian is the following:
\begin{equation}
\begin{aligned} \label{eq:DoubleEmission}
\mathcal L^{\text{int}}_s = - 4 \pi \alpha_s \sum_{q,q', \boldsymbol \ell, \boldsymbol \ell '} \frac{1}{q^0} \, \psi^\dagger_{\boldsymbol \ell} \left[ A^0_{q} , A^0_{q'} \right] \psi_{\boldsymbol \ell} + (\psi \rightarrow \chi , \, T \to \bar T) + \hdots \;.
\end{aligned}
\end{equation}
As we can see, at this order there is not interaction between heavy (anti)quarks and a single soft gluon. Moreover, the double soft gluon emission described by \eq{eq:DoubleEmission} is not relevant for the calculation at NLO, so in the NLO calculation of the LDME and TMDShF in \hyperref[sec:A-calculation]{appendix C}, we will only have interactions between the quarks and the usoft gluons.

The ultrasoft piece of the vNRQCD Lagrangian, $\mathcal L_{us}$, involves only pure ultrasoft field terms.
Note that the ultrasoft field strength scales as $G^{\mu \nu}_{us} \sim \tv^4$. The leading dimension term would be $\psi^\dagger_{us} \, i \Dslash \, \psi_{us} \sim \tv^5$ and the next-to-leading contribution $G^{\mu \nu}_{us} G_{us, \mu \nu} \sim \tv^8$.

The first term in the potential vNRQCD Lagrangian, see (\ref{eq:LagrangianvNRQCD}) and denoted as $\mathcal L_p^{(1)}$ for the following, is the only one in which the ultrasoft gluon couples to the heavy quarks at leading order in $\alpha_s$. However, the ultrasoft gluon can be decoupled from the heavy quarks using a BPS field redefinition:
\begin{equation}
\label{eq:BPS}
\psi_{\boldsymbol \ell} (x) \rightarrow  \, Y_u(x) \psi_{\boldsymbol \ell}(x) \quad\mathrm{with}\quad
Y_u(x) = P \exp \left( - i g \int_{- \infty}^0 dt A_{us}^0 (x+t,\mbf{x}) \right) \;.
\end{equation}
The term $(\mbf P - i \mbf D)^2$ in $\mathcal L_p^{(1)}$ can be expanded into $\mbf P^2 - \mbf D^2 - i 2 \mbf D \cdot \mbf P \sim \tv^2 - \tv^4 - \tv^3$, so $\mbf D^2$ is suppressed. The operator $i D^0 - \mbf P^2/2m$ is the kinetic term of the heavy-quark fields. As we will see later, the operator $ \mbf D \cdot \mbf P$, which must be treated as a perturbation, will turn out to play an important role since it is responsible for the mixing of the LDMEs at NLO. After the BPS field redefinition, the next-to-leading contribution $ \mbf D \cdot \mbf P$ becomes:
\begin{equation}
\begin{aligned}
\mathcal L_p^{(1)} = - \sum_{\boldsymbol \ell} \psi^\dagger_{\boldsymbol \ell}(x) \frac{-i \mathbf{D} \cdot \mathbf{P}}{m} \psi_{\boldsymbol \ell}(x) \rightarrow - g \sum_{\boldsymbol \ell} \psi^\dagger_{\boldsymbol \ell}(x) \frac{\mbf{B}_{us} \cdot \mbf{P}}{m} \psi_{\boldsymbol \ell}(x) \;,\\
\mbf B_{us}(x) = - \frac{1}{g} Y_u^\dagger(x) \, i \mbf D \, Y_u(x) \;,
\end{aligned}
\end{equation}
where $\mbf B_{us}$ is called the ultrasoft gluon building block.

Finally, the leading and next-to-leading contributions of the vNRQCD interaction Lagrangian are given by:
\begin{equation}
\begin{aligned} \label{eq:Lagrangian}
\mathcal L^{\text{int.}}_{\text{vNRQCD}} & = - g \sum_{\boldsymbol \ell} \psi_{\boldsymbol \ell}^\dagger \left( \frac{\mbf B_{us} \cdot \mbf P}{m} \right) \psi_{\boldsymbol \ell}(x) + (\psi \rightarrow \chi, \, T \to \bar T)\\
& + \sum_{\boldsymbol \ell, \boldsymbol \ell '} \frac{4 \pi \alpha_s}{(\boldsymbol \ell - \boldsymbol \ell ')^2} \, \psi_{\boldsymbol \ell}^\dagger T^a \psi_{\boldsymbol \ell '} \, \chi_{- \boldsymbol \ell}^\dagger \bar T^a \chi_{- \boldsymbol \ell '} \\
& - 4 \pi \alpha_s \sum_{q,q', \boldsymbol \ell, \boldsymbol \ell '} \frac{1}{q^0} \, \psi^\dagger_{\boldsymbol \ell} \left[ A^0_{q} , A^0_{q'} \right] \psi_{\boldsymbol \ell} + (\psi \rightarrow \chi, \, T \to \bar T) + \mathcal{O}(\tv^6) \; ,
\end{aligned}
\end{equation}
which scales as $\mathcal L_{\text{vNRQCD}} \sim  -  \tv^5 + \alpha_s \tv^4 - \alpha_s \tv^4 + \mathcal{O}(\tv^6)$.


\section{Matching onto QCD} \label{sec:A-MatTensors}

In this section, we compute the QCD amplitude for the partonic process:
\begin{equation}
\gamma^* (k_1) + g (k_2) \to c (p) \bar c (\bar p) \; .
\end{equation}
Notice we have changed the notation in this section for the virtual photon ($q \to k_1$) and the incoming gluon ($p_g \to k_2$).
After the QCD calculation, we compute the same amplitude using the effective $[n]$-operators, and compare both results to obtain the matching tensors $\Gamma_{[n]}$.

The heavy-quark fields in the $[n]$-operators are described by the vNRQCD framework (see \hyperref[sec:A-vNRQCD]{appendix A}), and their momenta, $p$ and $\bar p$, are defined in the rest frame of the $J/\psi$.
In this frame, in which the three-momenta of the heavy quarks satisfy $\mbf p = - \bar{\mbf p}$, their relative momentum $\mbf q $, which is defined as $\mbf q \equiv ( \mbf p - \bar{\mbf p} )/2 $, is small w.r.t. the heavy-quark mass $m_c$.
We perform the QCD calculation in this scenario, which is the so-called nonrelativistic limit of QCD. As shown in this section, one needs to Taylor expand the amplitude of QCD around $\mbf q$, and each order of the expansion gives the contribution of an particular angular momentum configuration denoted by the quantum number $L$. Therefore, we perform the matching onto the effective amplitude term by term, i.e., the S-state matching tensor will be given by the matching at order $\mbf q^0$, while the P-state one by the matching at order $\mbf q^1$.

Moreover, the QCD calculation in the nonrelativistic limit is easier in the frame in which the total three-momentum of the quark and antiquark is equal to the three-momentum $\mbf P$.
As usual, the heavy-quark momenta are defined as follows: $p = P/2 + \tilde{q}$ and $\bar p = P/2 - \tilde{q}$, where $P$ is the total momentum of the heavy-quark pair defined as $P = p + \bar p$, and $\tilde{q} = (p - \bar p)/2$ is the boosted relative momentum $\mbf q$ from the $J/\psi$ rest frame to this frame: $\tilde{q}^\mu \equiv (\boldsymbol \Lambda \cdot \mbf q)^\mu$. The boost matrix, $\Lambda_i^\mu$, is defined as follows~\cite{Braaten:1996rp}:
\begin{equation} \label{eq:BoostMatrixDef}
\Lambda^0_i = \frac{P_i}{2 E_q} 
\quad \text{and} \quad
\Lambda^j_i = \delta_i^{j} + \left( \frac{P^0}{2 E_q} - 1 \right) \frac{P_i P^j}{\mbf P^2} \; ,
\end{equation}
where $\mbf P = \mbf p + \bar{\mbf p}$, $E_q = \sqrt{m_c^2 + \mbf q^2}$ and $P^2 = 4 E_q^2$.
As we will see later, an important consequence of how the boost matrix is defined is that:
\begin{equation} \label{eq:Freedom}
\left( \Lambda \cdot P \right)_i = 0  
\quad \text{and} \quad
\left( \Lambda \cdot k_1 \right)_i =  - \left( \Lambda \cdot k_2 \right)_i \; .
\end{equation}

\begin{figure}
    \centering  \includegraphics[width=0.65\linewidth]{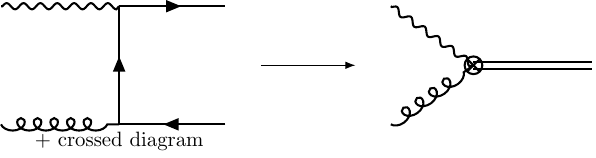}
    \caption{Feynman diagram for the QCD calculation (left) of the process $\gamma^* g \to c \bar c$ and for the amplitude calculation (right) by using the $[n]$-effective operators.}
    \label{fig:A-PgFusion}
\end{figure}

We now are ready to start with the calculation of the QCD amplitude. The process we are interested is shown in the left picture of the \hyperref[fig:A-PgFusion]{figure 2}. 
The amplitude is given by:
\begin{equation}
\begin{aligned}
\mathcal A (\gamma^* g \to c \bar c) & = \frac{i g e}{2} \epsilon_\mu(k_1) \epsilon_\nu(k_2) \\
& \times
\bar u(p) T^a 
\left[ 
\frac{\gamma^\mu (\Slash p - \Slash k_1 + m_c ) \gamma^\nu}{k_1^2 - 2 p \cdot k_1 } +
\frac{\gamma^\nu (- \Slash{ \bar p} + \Slash k_1 + m_c) \gamma^\mu }{k_1^2 - 2 \bar p \cdot k_1} 
\right]
v(\bar p)\\
& = 
\frac{i g e}{2} \epsilon_\mu(k_1) \epsilon_\nu(k_2)
\left( \tilde s -  \frac{4  (\tilde{q} \cdot k_1)^2}{\tilde s}  \right)^{-1} 
\bar u(p) T^a 
\left[
D^{\mu \nu}_+ -  \frac{2 \, \tilde{q} \cdot k_1}{\tilde s} \, D^{\mu \nu}_-
\right]
v(\bar p) \; ,
\end{aligned}
\end{equation}
where we have used that the gluon is on-shell $k_2^2 = 0$. Here, the center-of-mass energy is $s = (k_1+k_2)^2$, which can be written as $s = P^2$ by momentum conservation, and we define $\tilde s$ as follows: $\tilde s \equiv (P^2-k_1^2)/2$. Moreover, $e$ stands for the charge of the electron such that $e^2 = 4 \pi \alpha$, with $\alpha$ the fine-structure constant in electrodynamics and $g^2 = 4 \pi \alpha_s$ in QCD. The polarization vectors of the virtual photon and the incoming gluon are denoted by $\epsilon_\mu(k_1)$ and $\epsilon_\nu(k_2)$, respectively, while the quantities $D_\pm$ are defined as follows:
\begin{equation}
\begin{aligned}
D^{\mu \nu}_\pm & \equiv
\gamma^\mu (\Slash p - \Slash k_1 + m_c ) \gamma^\nu 
\pm
\gamma^\nu (- \Slash{ \bar p} + \Slash k_1 + m_c) \gamma^\mu \\
& =
- (\Slash p - m_c) \gamma^\mu \gamma^\nu \pm \gamma^\nu \gamma^\mu (\Slash{\bar p} + m_c) + 2 (p \mp \bar p)^\mu \gamma^\nu - ( \gamma^\mu \Slash k _1 \gamma^\nu \mp \gamma^\nu \Slash k _1 \gamma^\mu ) \; .
\end{aligned}
\end{equation}
We now use the Dirac equation, getting the following formula for the amplitude:
\begin{equation}
\begin{aligned}
\mathcal A (\gamma^* g \to c \bar c) & =
\frac{i g e }{ 2} \epsilon_\mu(k_1) \epsilon_\nu(k_2)
\left( \tilde s -  \frac{4  (\tilde{q} \cdot k_1)^2}{\tilde s}  \right)^{-1} 
\left[ 
\mathcal D _+^{\mu \nu,a} -   \frac{2 \, \tilde{q} \cdot k_1}{\tilde s} \, \mathcal D _-^{\mu \nu,a} 
\right] \; ,
\end{aligned}
\end{equation}
with
\begin{equation}
\begin{aligned}
\mathcal D _+^{\mu \nu,a}  \equiv
\bar u(p) T^a D_+^{\mu \nu} v(\bar p) & = 
\bar u(p) T^a \left[ 4 \, \tilde{q}^\mu \gamma^\nu - \left( \gamma^\mu \Slash k _1 \gamma^\nu - \gamma^\nu \Slash k _1 \gamma^\mu \right) \right] v(\bar p) \; , \\
\mathcal D _-^{\mu \nu,a}  \equiv
\bar u(p) T^a D_-^{\mu \nu} v(\bar p) & = 
\bar u(p) T^a \left[ 2 \, P^\mu \gamma^\nu - \left( \gamma^\mu \Slash k _1 \gamma^\nu + \gamma^\nu \Slash k _1 \gamma^\mu \right) \right] v(\bar p) \\
& = \bar u (p) T^a 
\left[ 
2 k_2^\mu \gamma^\nu - 2 k_1^\nu \gamma^\mu + 2 g^{\mu \nu} \Slash k _1 
\right] 
v(\bar p) \; .
\end{aligned}
\end{equation}
Note that the Dirac structures in $\mathcal D_\pm$ determine the configuration of spin in which the heavy-quark pair is produced.
In the boosted frame, described by the following relations: $\mbf p + \bar{\mbf{p}} = \mbf P$ and $\mbf p - \bar{\mbf{p}} = 2 \mbf q$, the heavy-quark spinors are the following:
\begin{equation} \label{eq:DiracSpinors}
u(p) = \frac{1}{\mathcal N} (2 E_q + \Slash P \gamma_0)
\begin{pmatrix}
    (E_q + m_c) \, \xi \\
    \mbf q \cdot \boldsymbol \sigma \, \xi 
\end{pmatrix} \; , 
\quad v(\bar p) = \frac{1}{\mathcal N} (2 E_q + \Slash P \gamma_0)
\begin{pmatrix}
    - \mbf q \cdot \boldsymbol \sigma \, \eta \\
    (E_q + m_c) \, \eta 
\end{pmatrix} \; .
\end{equation}
Here, $\mathcal N = \sqrt{4 E_q(P^0 + 2 E_q)(E_q + m_c)}$ and $\xi$ and $\eta$ are Pauli spinors. 
Taking the results from the appendix of \cite{Braaten:1996rp}, we reduce the Dirac bilinears as a functions of Pauli spinors, Pauli matrices and relative momentum. 
Thus, we obtain the objects $\mathcal D_\pm$ as a combination of the spin-singlet and -triplet contributions described by the following spin factors: $\xi^\dagger T^a \eta$ and $\xi^\dagger T^a (\boldsymbol \Lambda \cdot \boldsymbol \sigma) \eta$, respectively.
As mentioned above, the angular momentum configuration will be described by the order in the $\mbf q$-expansion.
Then we expand the result obtained after using the expressions of \cite{Braaten:1996rp} for the bilinears, getting the following expressions for $\mathcal D_\pm$ up to order $\mbf q$:
\begin{equation}
\begin{aligned}
\mathcal D _+^{\mu \nu,a} & = 
i 4 m_c \epsilon^{i j k} \Lambda_i^\mu ( \Lambda \cdot k_1 )_j \Lambda^\nu_k \, \xi^\dagger T^a \eta
+ \left[ 8 m_c \tilde{q}^\mu g^{\nu \rho} + \frac{2 k_{1,\beta}}{m_c} \left( g^{\nu \rho} (P^\mu \tilde{q}^\beta - P^\beta \tilde{q}^\mu ) \right. \right. \\
& + \left. \left.  g^{\mu \rho} (P^\beta \tilde{q}^\nu - P^\nu \tilde{q}^\beta) + g^{\beta \rho} (P^\nu \tilde{q}^\mu - P^\mu \tilde{q}^\nu)  \right) \right]
\xi^\dagger T^a \left( \boldsymbol \Lambda \cdot \boldsymbol \sigma \right)_\rho \eta
\; , \\
\mathcal D _-^{\mu \nu,a} & =
4 m_c \left( 
k_2^\mu g^{\nu \rho} -  k_1^\nu g^{\mu \rho} + k_1^{\rho} g^{\mu \nu} 
\right) \xi^\dagger T^a \left( \boldsymbol \Lambda \cdot \boldsymbol \sigma \right)_\rho \eta \; .
\end{aligned}
\end{equation}
From those equations, we see that the spin-singlet contribution comes from $\mathcal D_+$, whereas the spin-triplet contribution arises from a combination of $\mathcal D_+$ and $\mathcal D_-$.
We now put these results in the amplitude and expand around $\mbf q$:
\begin{equation} \label{eq:Amp}
\begin{aligned}
\mathcal A (\gamma^* g \to c \bar c) & =
\frac{i g e }{2} \epsilon_\mu(k_1) \epsilon_\nu(k_2) \frac{1}{\tilde s^2}
\left[ \tilde s \, \mathcal D_+^{\mu \nu ,a} - 2 \, \tilde{q} \cdot k_1 \mathcal D_-^{\mu \nu,a} \right] \\
& = \frac{i g e}{2} \epsilon_\mu(k_1) \epsilon_\nu(k_2) \\
& \times \frac{1}{\tilde s^2} \left\{ 
\tilde s \left[ i 2  \,  \epsilon^{\mu \nu \alpha \beta} k_{1,\alpha} P_\beta  \right] \times \left( \xi^\dagger T^a \eta \right) \right. \\
 & \left. + 4 \left[
 \tilde s \, g^{\mu \rho} g^{\nu \sigma} \left( \frac{P^2 + k_1^2}{2} \right)  + \tilde s^2 g^{\mu \sigma} g^{\nu \rho}  - P^2 \, k_1^\sigma k_1^\rho g^{\mu \nu} \right. \right. \\
& \left. \left.
+ g^{\nu \rho} \left(  \tilde s \, k_1^\mu k_1^\sigma - \frac{P^2+k_1^2}{2} k_2^\mu  k_1^\sigma  \right) - g^{\mu \rho} \left( \tilde s \,  k_2^\nu k_1^\sigma  - \frac{P^2+k_1^2}{2} k_1^\nu k_1^\sigma \right) \right. \right. \\
& \left. \left. + \tilde s \left( g^{\mu \sigma} P^\nu k_1^\rho - g^{\nu \sigma} P^\mu k_1^\rho
\right) \right] \times \left( \frac{(\boldsymbol \Lambda \cdot \mbf q)_\sigma}{M} \, \xi^\dagger T^a \left( \boldsymbol \Lambda \cdot \boldsymbol \sigma \right)_\rho \eta \right)
\right\} \; ,
\end{aligned}
\end{equation}
where we have used that $\epsilon^{ijk} \Lambda_i^\mu \Lambda_j^\alpha \Lambda_k^\nu = \epsilon^{\beta \mu \alpha \nu} P_\beta/\sqrt{P^2}$. 
We have split the result into the S-state contribution, given by the third line of the last equation, and P-states contribution, given by the last three lines.
Moreover, we would like to note a tricky aspect in \eq{eq:Amp} (which is not relevant for the further discussion): since the boost matrices are defined as \eq{eq:BoostMatrixDef}, one can make the replacement $k_1^{\sigma (\rho)} \Lambda_{\sigma(\rho)}^i  \leftrightarrow - k_2^{\sigma(\rho)} \Lambda_{\sigma (\rho)}^i $ in \eq{eq:Amp}, resulting in different consistent expressions of the matching tensor.

We now write the amplitude with the kinematics of the process in a photon frame considering mass corrections, which were established in \hyperref[sec:2]{section 2}.
Moreover, it is convenient to decompose the particles momenta by using the following vectors: $\kappa_+^\mu = \frac{P_N^+}{2} n^\mu $ and $\kappa_-^\mu = \frac{2}{P_N^+} \bar n^\mu$, such that $\kappa^2_+ = \kappa^2_- = 0$ and $\kappa_+ \cdot \kappa_+ = 2$. Thus the virtual photon momentum reads:
\begin{equation}
    k_1^\mu =\frac{-x_{B}P_{N}^{+}}{\gamma_{q}}\frac{n^{\mu}}{2}+\frac{Q^{2}\gamma_{q}}{x_{B}P_{N}^{+}}\frac{\bar{n}^{\mu}}{2} 
    = \frac{- x_B}{\gamma_q} \kappa_+^\mu + \frac{Q^2 \gamma_q}{4 x_B} \kappa_-^\mu \;,
\end{equation}
with $Q^2 = - k_1^2$ and $\gamma_q$ denotes the mass corrections which is defined in \eq{eq:targetmass}; note we choose a photon frame, so $Q_\perp = Q$ ($\mbf q_\perp = \mbf k_{1\perp} = 0$) in the definition of $\gamma_q$. The Bjorken variable $x_B$ can be written in terms of the variable defined above, $\tilde s \equiv (P^2 - k_1^2)/2 = (M^2 + Q^2 )/2$:
\begin{equation}
x_B = \frac{Q^2 }{2 P_N \cdot k_1} =  x_S \frac{Q^2 }{2 \,  \tilde s} \; ,
\end{equation}
where $x_S$ is the collinear momentum fraction carried by the gluon from the proton which includes kinematic power corrections \cite{Scimemi:2019cmh}. Thus the momenta involved in the amplitude above are the following:
\begin{equation}
    \begin{aligned} \label{eq:kinpartonic}
        k_1^\mu &  = \frac{- x_S Q^2}{\gamma_q \,  \tilde s} \frac{\kappa_+^\mu}{2} + \frac{ \gamma_q \, \tilde s}{ x_S} \frac{\kappa_-^\mu}{2} \; , \\
        k_2^\mu & =  x_S \kappa_+^\mu \; ,
    \end{aligned}
\end{equation}
with $P^\mu = k_1^\mu + k_2^\mu$ by momentum conservation.

As mentioned in \hyperref[sec:2]{section 2}, given that the effective $[n]$-operators in \eq{eq:operators} scale as $\lambda$, the effective amplitude scales as $\lambda$, so in order to perform the matching we need to expand the gluon polarization vector to order $\lambda$:
\begin{equation} \label{eq:polvector}
\epsilon_\nu = \epsilon_{\perp \nu } - \frac{k_{2 \perp \nu}}{\bar n \cdot k_2} \bar n \cdot \epsilon \; .
\end{equation}
Therefore, inserting \eq{eq:kinpartonic} and \eq{eq:polvector} into \eq{eq:Amp} yields:
\begin{equation}
\begin{aligned}
\mathcal A (\gamma^* g \to c \bar c) & = 
\frac{ig e}{2} \epsilon_\mu(k_1) \,  \epsilon_{\perp \nu }(k_2)  \\
& \times \left\{ \frac{\gamma_q}{2} i 2 \,  \epsilon^{\mu \nu \alpha \beta} \kappa_{- \alpha} \kappa_{+ \beta}  \times \left( \xi^\dagger T^a \eta \right) \right.  \\
& \left. + 4 \left[ 
g_\perp^{\nu \sigma} \left( g^{\mu \rho} \frac{M^2 - Q^2}{M^2 + Q^2}  +  \frac{ x_S^2}{M^2 + Q^2} \left( 1 - \frac{2 \,x_S Q^2}{\gamma_q (M^2+Q^2)} \right) \kappa_+^\mu \kappa_+^\rho + \frac{\gamma_q}{2} \, \kappa_-^\mu \kappa_+^\rho  \right) \right. \right. \\
& \left. \left. + g_\perp^{\nu \rho} \left( g^{\mu \sigma} + \frac{2 x_S^2}{ (M^2 + Q^2)^2} \left( M^2 + Q^2 \frac{1-\gamma_q}{\gamma_q} \right) \kappa_+^\mu \kappa_+^\sigma -  \frac{\gamma_q}{2} \kappa_-^\mu \kappa_+^\sigma \right) \right. \right. \\
& \left. \left. - g^{\nu \mu}_\perp \frac{ 4 \, x_S^2 M^2 }{(M^2 + Q^2)^2} \, \kappa_+^\sigma \kappa_+^\rho
\right] \times \left( \frac{(\boldsymbol \Lambda \cdot \mbf q)_\sigma}{M} \,  \xi^\dagger T^a \left( \boldsymbol \Lambda \cdot \boldsymbol \sigma \right)_\rho \eta \right) 
\right\} \; ,
\end{aligned}
\end{equation}
with $k_{2\perp \nu} = 0 $.
In the light-cone basis, built by $n$ and $\bar n$, the amplitude is given by:
\begin{equation} \label{eq:QCDamplitude}
\begin{aligned}
\mathcal A (\gamma^* g \to c \bar c) & = 
\frac{ig e}{2} \epsilon_\mu(k_1) \,  \epsilon_{\perp \nu }(k_2) \\
\times & \left\{ -i 4 \,  \epsilon^{\mu \nu}_\perp \frac{\gamma_q}{2}   \times \left( \xi^\dagger T^a \eta \right) \right.  \\
+ & \, 4 \left. \left[ 
g_\perp^{\nu \sigma} \left( g^{\mu \rho} \frac{M^2 - Q^2}{M^2 + Q^2}  +  \frac{ x_S^2 P_N^{+2} }{4(M^2 + Q^2)} \left( 1 - \frac{2 \,x_S Q^2}{\gamma_q (M^2+Q^2)} \right) n^\mu n^\rho + \frac{\gamma_q}{2} \, \bar n^\mu n^\rho  \right) \right. \right. \\
+ & \left. \left. g_\perp^{\nu \rho} \left( g^{\mu \sigma} + \frac{x_S^2 P_N^{+2} }{2 (M^2 + Q^2)^2} \left(  M^2 + Q^2 \frac{1-\gamma_q}{\gamma_q} \right) n^\mu n^\sigma -  \frac{\gamma_q}{2} \bar n^\mu n^\sigma \right) \right. \right. \\
- & \left. \left. g^{\nu \mu}_\perp \frac{x_S^2 P_N^{+2}  M^2 }{(M^2 + Q^2)^2} \, n^\sigma n^\rho
\right] \times \left( \frac{(\boldsymbol \Lambda \cdot \mbf q)_\sigma}{M} \,  \xi^\dagger T^a \left( \boldsymbol \Lambda \cdot \boldsymbol \sigma \right)_\rho \eta \right) 
\right\} \; ,
\end{aligned}
\end{equation}
with $\epsilon_\perp^{\mu \nu} = \tensor{\epsilon}{^\mu^\nu_\alpha_\beta} \bar n^\alpha n^\beta/2$ and $g_\perp^{\mu \nu} = g^{\mu \nu} - (n^\mu \bar n ^\nu + \bar n^\mu n^\nu)/2$. For easy comparison with the effective-theory side of the matching equation, we rewrite this result as follows:
\begin{equation}
\begin{aligned}
\mathcal A (\gamma^* g \to c \bar c) = \epsilon_\mu (k_1) \left( \mathcal A_{^1S_0^{[8]}}^\mu + \mathcal A_{^3P_J^{[8]}}^\mu \right) \; ,
\end{aligned}
\end{equation}
with
\begin{equation}
\begin{aligned} \label{eq:QCDmatching}
\mathcal A_{^1S_0^{[8]}}^\mu & = 2 \, g e \, \epsilon_\perp^{\mu \nu} \frac{\gamma_q}{2}  \,  \epsilon_{\perp \nu }(k_2) \times \left( \xi^\dagger T^a \eta \right)  \; , \\
\mathcal A_{^3P_J^{[8]}}^\mu  &  = i 2 \, g e 
\left\{ 
g_\perp^{\nu \sigma} \left[ g^{\mu \rho} \frac{M^2 - Q^2}{M^2 + Q^2}  +  \frac{ x_S^2 P_N^{+2} }{4(M^2 + Q^2)} \left( 1 - \frac{2 \,x_S Q^2}{\gamma_q (M^2+Q^2)} \right) n^\mu n^\rho + \frac{\gamma_q}{2} \, \bar n^\mu n^\rho  \right] \right.   \\
& + g_\perp^{\nu \rho}  \left[ \left. g^{\mu \sigma} + \frac{x_S^2 P_N^{+2} }{2 (M^2 + Q^2)^2} \left( M^2 + Q^2 \frac{1-\gamma_q}{\gamma_q} \right) n^\mu n^\sigma -  \frac{\gamma_q}{2} \bar n^\mu n^\sigma \right] 
\right. \\
& \left. - g^{\nu \mu}_\perp \frac{x_S^2 P_N^{+2}  M^2 }{(M^2 + Q^2)^2} \, n^\sigma n^\rho
\right\} \times  \left( \epsilon_{\perp \nu } - \frac{k_{2 \perp \nu}}{\bar n \cdot k_2} \bar n \cdot \epsilon \right) \times \left( \frac{(\boldsymbol \Lambda \cdot \mbf q)_\sigma}{M} \,  \xi^\dagger T^a \left( \boldsymbol \Lambda \cdot \boldsymbol \sigma \right)_\rho \eta \right) \; .
\end{aligned}
\end{equation}
We now use that the product $q^i \sigma^j$ can be decomposed into the following three tensors:
\begin{equation} \label{eq:Decomposition}
q^i \sigma^j = \frac{\delta^{ij}}{3} \mbf q \cdot \boldsymbol \sigma + \frac{\epsilon^{ijk}}{2} (\mbf q \times \boldsymbol \sigma)^k + q^{(i} \sigma^{j)} \; ,
\end{equation}
with $q^{(i} \sigma^{j)} = (q^i \sigma^j + q^j \sigma^i)/2 - \mbf q \cdot \boldsymbol \sigma \delta^{ij}/3$.
We obtain the tesnorial structures for $J=0,1,2$ from this decomposition.
In fact, the scalar term is describing the state $J= 0$, the vector one is describing the state $J= 1$ and the symmetric-traceless one is describing the state $J= 2$.
Thus we apply (\ref{eq:Decomposition}) to the P-wave contribution of the QCD amplitude (\ref{eq:QCDmatching}):
\begin{equation} 
\begin{aligned} \label{eq:QCDmatchingJs}
\mathcal A^\mu_{^3P_0^{[8]}} & =  \Gamma^{\mu \nu \sigma \rho} \left( -g_{\sigma \rho} + \frac{P_\sigma P_\rho}{P^2} \right)  \,  \epsilon_{\perp \nu } \times \frac{1}{3} \left( \xi^\dagger_{\mbf q} (\mbf q \cdot \boldsymbol \sigma) \, T^a \eta_{- \mbf q} \right) \; , \\
\mathcal A^\mu_{^3P_1^{[8]}} & =  \Gamma^{\mu \nu \sigma \rho} \epsilon_{\alpha \beta \sigma \rho} \Lambda^\alpha_k \frac{P^\beta}{M}  \,  \epsilon_{\perp \nu } \times \frac{1}{2} \left( \xi^\dagger_{\mbf q}  (\mbf q \times \boldsymbol \sigma)^k \, T^a \eta_{- \mbf q} \right) \; , \\
\mathcal A^\mu_{^3P_2^{[8]}} & =
\Gamma^{\mu \nu \sigma \rho} \Lambda^i_\sigma \Lambda^j_\rho  \,  \epsilon_{\perp \nu } \times \left( \xi^\dagger_{\mbf q} \, ( q_{(i} \sigma_{j)}) \, T^a \eta_{- \mbf q} \right) \; ,
\end{aligned}
\end{equation}
where we have used that $\Lambda^\alpha_i \Lambda^\beta_i = - g^{\alpha \beta} + P^\alpha P^\beta/P^2$ and $\epsilon^{ijk} \Lambda_j^\alpha \Lambda_k^\beta = \epsilon^{\mu \nu \alpha \beta} \Lambda_\mu^i P_\nu/\sqrt{P^2}$ and the tensor $\Gamma^{\mu \nu \sigma \rho}$ is defined as:
\begin{equation}
\begin{aligned}
    \Gamma^{\mu \nu \sigma \rho} & = \frac{i 2 g e }{M} \left\{ g_\perp^{\nu \sigma} \left[ g^{\mu \rho} \frac{M^2 - Q^2}{M^2+ Q^2} + \frac{M^2 + Q^2}{4 Q^4} x_B^2 P_N^{+2} \left( 1 - \frac{2 x_B}{\gamma_q} \right) n^{\mu} n^\rho + \frac{\gamma_q}{2} \bar n^\mu n^\rho \right] \right. \\ 
    + g_\perp^{\nu \rho} & \left. \left[ g^{\mu \sigma} + \frac{x_B^2 P_N^{+2}}{2 Q^4} \left( M^2 + Q^2 \frac{(1-\gamma_q)}{\gamma_q} \right) n^\mu n^\sigma - \frac{\gamma_q}{2} \bar n^\mu n^\sigma   \right]  - g_\perp^{\nu \mu} \frac{x_B^2 P_N^{+2}  M^2 }{Q^4} n^\sigma n^\rho \right\} \; .
\end{aligned}
\end{equation}
where we use that $x_S = x_B (M^2 + Q^2)/Q^2$.

So far, we have calculated the amplitude for the S and P -states in the nonrelativistic limit of QCD.
To obtain the matching tensors, we need to deal with the calculation of the amplitude by using the $[n]$-operators defined in~\eq{eq:EffOpLO}.
Note that the expansion of the gluon building block of SCET can be written in terms of gluon fields as follows:
\begin{equation}
\begin{aligned}
    \mathcal B_{n \perp}^\mu = A_{n \perp}^\mu - \frac{k_{2\perp}^\mu}{\bar n \cdot q} \bar n \cdot A_{n,k_2} + \hdots \; ,
\end{aligned}
\end{equation}
where the dots denote terms with more collinear gluon fields, i.e., higher orders in the $\lambda$-expansion. 
Since the components of the collinear gluon field scales in the same way as the components of the collinear momentum, the transverse component of a collinear gluon field $A_{n \perp}$ scales as $\lambda$. 
In addition, we can see from the vNRQCD Lagrangian that the heavy-quark fields scale as $\lambda^0$.
Thus the effective $[n]$-operators scale as $\lambda$.
Therefore, since we have expanded the QCD amplitude up to order $\lambda$, we are able to do the matching between both sides.
The effective amplitude is given by:
\begin{equation} \label{eq:EffAmplitude}
\begin{aligned}
\mathcal{A}^{\mu, \text{eff}} & = \mathcal{A}^{\mu,\text{eff}}_{^1 S_0^{[8]}} + \mathcal{A}^{\mu, \text{eff}}_{^3 P_0^{[8]}} + \mathcal{A}^{\mu, \text{eff}}_{^3 P_1^{[8]}} \; + \mathcal{A}^{\mu, \text{eff}}_{^3 P_2^{[8]}} \; ,
\end{aligned}
\end{equation}
with
\begin{equation} \label{eq:QCDamplitude2}
\begin{aligned}
\mathcal{A}^{\mu, \text{eff}}_{^1 S_0^{[8]}} & = \Gamma^{\mu \nu}_{^1S_0^{[8]}}
\sandwich{c \bar c}
{\psi^\dagger_{\mbf{p}} \left( A_{n \perp \nu} - \frac{k_{2 \perp \nu}}{\bar n \cdot k_2} \bar n \cdot A_{n,k_2} \right)  \chi_{\bar{\mbf p}}  }
{g} \\
& = M \, \Gamma^{\mu \nu}_{^1S_0^{[8]}} 
\,  \epsilon_{\perp \nu }  
\times \left( \xi_{\mbf p}^\dagger T^a \eta_{\bar{\mbf p}} \right) \; , \\
\mathcal{A}^{\mu, \text{eff}}_{^3 P_0^{[8]}} & = M \, \Gamma_{^3P_0^{[8]}}^{\mu \nu} 
\,  \epsilon_{\perp \nu }
\times 
\left( \frac{1}{M} \,  
\xi^\dagger_{\mbf p} \, \left( \mbf q \cdot \boldsymbol{\sigma} \right) \, T^a   \eta_{\bar{\mbf p}} \right)  \; ,\\
\mathcal{A}^{\mu, \text{eff}}_{^3 P_1^{[8]}} & = M \, \Gamma_{^3P_1^{[8]}}^{\mu \nu, k} 
\,  \epsilon_{\perp \nu }
\times 
\left( \frac{1}{M} \,  
\xi^\dagger_{\mbf p} \, \left( \mbf q \times \boldsymbol{\sigma} \right)^k \, T^a   \eta_{\bar{\mbf p}} \right)  \; ,\\
\mathcal{A}^{\mu, \text{eff}}_{^3 P_2^{[8]}} & = M \, \Gamma_{^3P_2^{[8]}}^{\mu \nu, ij} 
\,  \epsilon_{\perp \nu }
\times 
\left( \frac{1}{M} \,  
\xi^\dagger_{\mbf p} \, \left( q^{(i} \sigma^{j)}  \right) \, T^a   \eta_{\bar{\mbf p}} \right)  \; .
\end{aligned}
\end{equation}
where we have used that $k_{2\perp}^\nu = 0$, $ \sandwich{c \bar c}{\psi^\dagger \chi}{0} = M \, \xi^\dagger \eta $ in perturbative NRQCD. Equating the QCD amplitude (\ref{eq:QCDamplitude}) with the effective amplitude (\ref{eq:EffAmplitude}), we obtain the $^1S_0^{[8]}$ and $^3P_{J=0,1,2}^{[8]}$ matching tensors:
\begin{equation}
\begin{aligned}
\Gamma^{\mu \nu}_{^3 P_0^{[8]}} & = - \frac{i 2 \, ge }{3M} \left( \frac{M^2 (\gamma_q^2 + 2) + Q^2 \gamma_q^2}{M^2 + Q^2} \right) g_\perp^{\mu \nu} \; , \\
\Gamma^{\mu \nu, k}_{ ^3 P_1^{[8]}} & =  \frac{i  \, ge}{M} \left( \frac{x_B^2 P_N^{+2} \left(M^2 \gamma_q + Q^2 \left(\gamma_q - 1 \right)\right) (n \cdot \Lambda)_k - \gamma _q^2 Q^4 ( \bar{n} \cdot \Lambda)_k}{\gamma_q x_B P_N^+ M \left(M^2+Q^2\right)} \right) \epsilon_\perp^{\mu \nu} \; ,\\
\Gamma^{\mu \nu, ij}_{ ^3 P_2^{[8]}} & =  - \frac{i2\, ge}{M} \left( \frac{x_B^2 P_N^{+2} M^2}{Q^4} \right) (n \cdot \Lambda)_i (n \cdot \Lambda)_j \, g_\perp^{\mu \nu} \; ,
\end{aligned}
\end{equation}
with
\begin{equation}
\begin{aligned}
    (n \cdot \Lambda)^i = \delta^{3i} + P^i \left( \frac{1}{2 M} + \frac{1}{P^3} \left( \frac{P^0}{M} -1 \right) \right) \; , \\
    (\bar n \cdot \Lambda)^i = - \delta^{3i} + P^i \left( \frac{1}{2 M} - \frac{1}{P^3} \left( \frac{P^0}{M} -1 \right) \right) \; .
\end{aligned}
\end{equation}
Considering that $n^\mu = (1,0,0,1)$ and $\bar n^\mu = (1,0,0,-1)$, we can write $P^0$ and $P^3$ as follows:
\begin{equation}
\begin{aligned}
    P^0 & = \frac{x_B P_N^+}{2} \left( \frac{M^2 + Q^2}{Q^2} - \frac{1}{\gamma_q} \right) + \frac{\gamma_q Q^2}{2 x_B P_N^+} \; ,\\
    P^3 & = \frac{x_B P_N^+}{2} \left( \frac{M^2 + Q^2}{Q^2} - \frac{1}{\gamma_q} \right) - \frac{\gamma_q Q^2}{2 x_B P_N^+} \; .
\end{aligned}
\end{equation}


\section{Details of the calculation} \label{sec:A-calculation}

In this section, we compute the $^1S_0^{[8]}$ LDME and TMDShF at NLO in the vNRQCD framework, which is defined in \hyperref[sec:A-vNRQCD]{appendix A}.
Given the close similarity in the computation of the P-wave contributions, we have delineated the distinctions in \hyperref[sec:A-Pwaves]{appendix D} compared to the present section. 

\subsection{NLO calculation of the LDME}

\begin{figure} \centering
    \includegraphics[scale= 0.6]{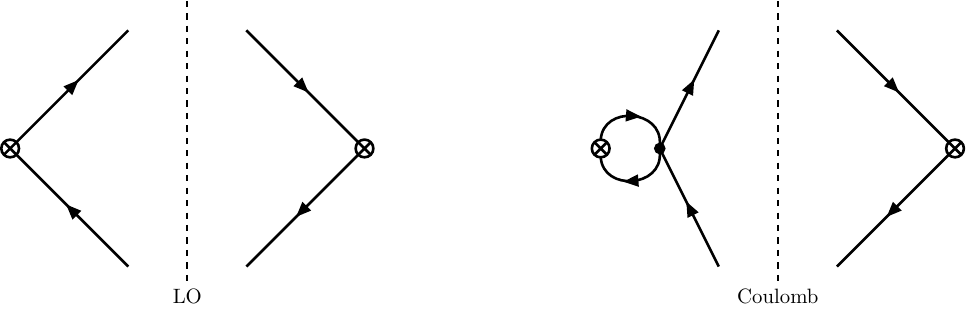}
    \caption{Diagrams showing the LO and the Coulomb interaction. Hermitian conjugate of right diagram is not shown.}
    \label{fig:TMDShF-LOandCoulomb}
\end{figure}

\begin{figure}
    \centering
    \includegraphics[scale=0.6]{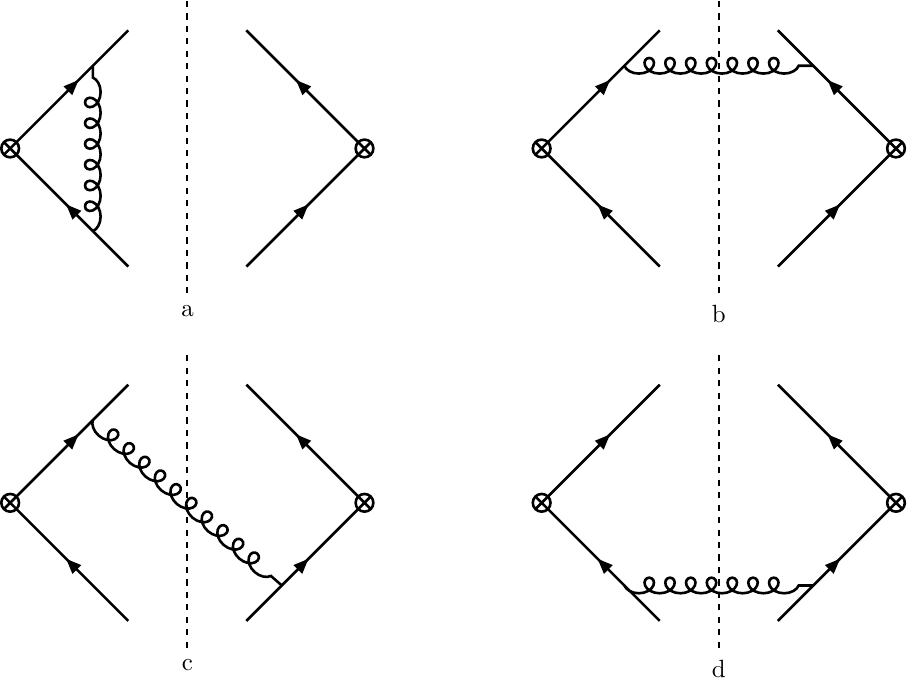}
    \caption{Diagrams showing the ultrasoft gluon exchanges between heavy quarks and (anti)quarks contributing to NLO. Hermitian conjugates of diagrams a and c are not shown. The soft and ultrasoft Wilson lines are not shown.}
    \label{fig:TMDShF-ChroElectric}
\end{figure}

From \cite{Bodwin:1994jh}, we know that the operator for state $^1S_0^{[8]}$, in terms of the heavy-quark annihilation field $\psi$ and the heavy-antiquark creation field $\chi$, is as follows
\begin{equation} \label{eq:Soperator}
\mathcal{O}_8  \left( ^1S_0 \right) = \chi^\dagger T^a \psi \, \mathcal{N}_{\psi} \, \psi^\dagger T^a \chi \;,
\end{equation}
where $\mathcal N_\psi \equiv a_\psi^\dagger a_\psi$ is defined above as the number operator for the state $J/\psi$.
Consequently, the $^1S_0^{[8]}$ LDME will be the vacuum expectation value of that operator. 
From the figures presented in this appendix, it can be seen how the LDME can be illustrated as a crossed circle and two quark lines. The crossed circle represents the state in which the heavy-quark pair is produced or decays through the bilinear formed by the fields and the color and angular momentum structure, e.g., for the state we are concerned with, the left crossed circle will be $\psi^\dagger T^a \chi$.
In particular, in the \hyperref[fig:TMDShF-LOandCoulomb]{figure 3} and \hyperref[fig:TMDShF-ChroElectric]{figure 4} the Feynman diagrams for the LDME NLO contribution are shown.

At LO, the $^1S_0^{[8]}$ LDME looks like this
\begin{equation} \label{eq:LOldme}
\braket{ 0| \mathcal O_8 (^1S_0) |0 }^{\text{LO}} \equiv \braket{^1S_0^{[8]}}^{\text{LO}} = M^2 \, \eta^\dagger T^a \xi \times \xi^\dagger T^a \eta \; .
\end{equation}
Here $\eta$ and $\xi$ are the Pauli 2-component spinors. The Dirac 4-component spinors $u(p)$ and $v(\bar p)$ are defined in terms of the Pauli spinors in the standard nonrelativistic form (see~\eq{eq:DiracSpinors}, where we explicitly show the relation between both). Expanding the Dirac spinors in the velocity in perturbative NRQCD, one finds $\sandwich{c \bar c}{ \psi^\dagger \chi}{0} = M \xi^\dagger \eta$, with $M$ the mass of the heavy-quark state, and where we have implicitly summed over the helicity of the bound state.

The diagrams which contribute to the NLO are called Coulomb interaction and chromo-electric dipole transition.
The Coulomb interaction contribution arises from the term of the second line term of vNRQCD interaction Lagrangian in~\eq{eq:Lagrangian}:
\begin{equation}
\begin{aligned}
\mathcal L_{\text{vNRQCD}}^{\text{Coul.}} = \sum_{\mbf p, \mbf p '} \frac{4 \pi \alpha_s}{(\mbf p - \mbf p ')^2} \, \psi_{\mbf p}^\dagger T^a \psi_{\mbf p '} \, \chi_{- \mbf p}^\dagger \bar T^a \chi_{- \mbf p '} \; .
\end{aligned}
\end{equation}
The leading order Lagrangian gives the intermediate fermion propagator for the below equation $p^0 + k^0 - (\mbf p + \mbf k)^2/2m$. Additionally, considering that $\mbf p - \mbf p' = \mbf k$, we obtain
\begin{equation}
\begin{aligned}
\braket{^1S_0^{[8]}}^{\text{Coul.}} & = \braket{^1S_0^{[8]}}^{\text{LO}} (-i 4\pi \alpha_s) \left( C_F - C_A/2 \right) \int \frac{d^4 k}{(2 \pi)^4} \frac{1}{\mbf{k}^2} \frac{1}{[p^0 + k^0 - (\mbf{p}+\mbf{k})^2/2m + i \epsilon]} \\
& \times  \frac{1}{[p^0 - k^0 - (\mbf{p}+\mbf{k})^2/2m+ i \epsilon]} + h.c. \; .
\end{aligned}
\end{equation}
Here we compute the integral on $k^0$ by contour integration, i.e., closing the contour in the lower half of the complex plane and picking up the pole $k^0 = - p^0 + (\mbf{p}+\mbf{k})^2/2m - i \epsilon$. We find that the contribution reduces to the following integral on the three-momentum:
\begin{equation} \label{eq:Coulomb}
\begin{aligned}
\braket{^1S_0^{[8]}}^{\text{Coul.}} & = \braket{^1S_0^{[8]}}^{\text{LO}} (4 \pi \alpha_s m_c) \left( C_F - C_A/2 \right) \int \frac{d^{3} k}{(2\pi)^{3}} \frac{1}{\mbf{k}^2} \frac{1}{\mbf{k}^2 + 2 \mbf{p} \cdot \mbf{k} - i \epsilon} + h.c. \\
& = \braket{^1S_0^{[8]}}^{\text{LO}} \left( C_F - C_A/2 \right) \frac{\pi \alpha_s}{2 \tv} \; ,
\end{aligned}
\end{equation}
where we have used that $\mbf{v} = \mbf{p}/m_c$ and $\tv = |\mbf{v}|$.

The chromo-electric effect contribution arises from the term of the first line of the vNRQCD interaction Lagrangian:
\begin{equation}
\begin{aligned} \label{eq:Chromo}
\mathcal L_{\text{vNRQCD}}^{\text{Chro.}} = - g \sum_{\mbf p} \psi_{\mbf p}^\dagger \left( \frac{\mbf B_{us} \cdot \mbf P}{m_c} \right) \psi_{\mbf p}(x) + (\psi \rightarrow \chi, \, T \to \bar T) \; .
\end{aligned}
\end{equation}
Let us look at this term in more detail. As mentioned before, after BPS redefinition in~\eq{eq:BPS}, the usoft Wilson lines arise. In the rest of the paper, we assume they are implicit in the soft Wilson lines but here we need write explicitly (at NLO):
\begin{equation}
\begin{aligned} \label{eq:usoftWilsonLine}
Y_v(x) & = 1 - i g \int_{-\infty}^0 dt' A_{us}^0 (t+t',\mbf{x}) + \mathcal{O}(g^2)\\
& = 1 + \int \frac{d^d k}{(2\pi)^d} e^{-i k \cdot x} \left( \frac{g \, T^a }{k^0} \right) A_{us}^{0,a} (k) + \mathcal{O}(g^2) \; ,
\end{aligned}
\end{equation}
where we use that $v=(1,\mbf 0)$ in the rest frame of the $J/\psi$.
Inserting the expanded usoft Wilson line (\ref{eq:usoftWilsonLine}) in~\eq{eq:Chromo} through the product $\mbf B _{us} \cdot \mbf P$, we obtain
\begin{equation}
\begin{aligned}
\mbf{B}_{us} \cdot \mbf{P} & = \left(\frac{-1}{g} Y_u^\dagger (i D^i) Y_u \right) P^i\\
& =  \int \frac{d^d k}{(2 \pi)^d} e^{-i k \cdot x} \left( \mbf{A}_{us} \cdot \mbf{P} - \frac{A_{us}^0}{k^0} \mbf{k} \cdot \mbf{P} \right) \; ,
\end{aligned}
\end{equation}
and consequently, the chromo-electric term in the Lagrangian is given by
\begin{equation}
\begin{aligned} \label{eq:LagrChromo}
\mathcal L_{\text{vNRQCD}}^{\text{Chro.}} & = \frac{-g}{m_c} \sum_{\mbf{p}} \int \frac{d^d k}{(2 \pi)^d} e^{-i k \cdot x} \psi^\dagger_{\mbf{p}} \left( \mbf{A}_{us} \cdot \mbf{p} - \frac{A_{us}^0}{k^0} \mbf{k}\cdot \mbf{p} \right)  \psi_{\mbf{p}} + (\psi \rightarrow \chi, \, T \to \bar T) \; .
\end{aligned}
\end{equation}
Here we act with the label momentum operator on the heavy-(anti)quark fields as follows $\mbf P \,  \psi_{\mbf p} = \mbf p \,  \psi_ {\mbf p}$. 
From this form of the interaction term, the loop integral reads (e.g., here we show the contribution of the \hyperref[fig:TMDShF-ChroElectric]{figure 4}b)
\begin{equation} \label{eq:4.2b}
\begin{aligned}
    \braket{^1 S_0^{[8]}}^{\ref{fig:TMDShF-ChroElectric}\text{b}}  & =  \frac{4 \pi \alpha_s}{m_c^2}  \, I^{\ref{fig:TMDShF-ChroElectric}\text{b}} \eta^\dagger  T^a T^b \xi \times \xi^\dagger  T^b T^a \eta \; ,
\end{aligned}
\end{equation}
with
\begin{equation}
\begin{aligned}
I^{\ref{fig:TMDShF-ChroElectric}\text{b}} & = i \mu^{2 \ve} \int \frac{d^d k}{(2 \pi)^d} \frac{\mbf{p}^2 - (\mbf{p} \cdot \mbf{k})^2/\mbf{k}^2}{(k^0)^2 - \mbf{k}^2 +  i \epsilon}  \, \frac{1}{[p^0 - k^0 - (\mbf{p}+\mbf{k})^2/2m]^2} \\
& = \mu^{2 \ve} \int \frac{d^{3-2 \ve} \mbf k}{(2 \pi)^{3-2\ve}} \frac{\mbf{p}^2 - (\mbf{p} \cdot \mbf{k})^2/\mbf{k}^2}{2|\mbf{k}|} \frac{1}{- |\mbf{k}|} \, \frac{1}{- |\mbf{k}|}\\
& = \mu^{2 \ve} \frac{1}{2} \int \frac{d^{3-2\ve} \mbf k}{(2 \pi)^{3-2\ve} } \frac{\mbf{p}^2 - (\mbf{p} \cdot \mbf{k})^2/\mbf{k}^2}{|\mbf{k}|^3} = \frac{1}{2} \frac{\mbf{p}^2}{6 \pi^2} \left( \frac{1}{\veuv} - \frac{1}{\veir} \right) \; .
\end{aligned}
\end{equation}
In the second equality we have computed the contour-integral over $k^0$ around the pole $k^0 = |\mbf{k}| - i \epsilon$ and in the third equality we have expanded the integrand in powers of $\mbf{p}/m_c$ and $\mbf{k}/m_c$. To do so we need the following integral:
\begin{equation}
\begin{aligned}
\mu^{2 \ve} \int d^{3-2\ve} \mbf k \, \frac{1}{|\mbf k|^3} & = 2 \pi \left( \frac{1}{\veuv} - \frac{1}{\veir} \right) \; .
\end{aligned}
\end{equation}
Diagrams \hyperref[fig:TMDShF-ChroElectric]{figure 4}c,d give the same contribution as (\ref{eq:4.2b}), but the color configuration of the spin factor changes as follows
\begin{equation}
\begin{aligned} \label{eq:2bcd}
\braket{^1S_0^{[8]}}^{\ref{fig:TMDShF-ChroElectric} \text{c} } & = \braket{^1S_0^{[8]}}^{\ref{fig:TMDShF-ChroElectric} \text{b}} \left( T^a T^b \otimes T^b T^a \rightarrow T^a T^b \otimes T^a T^b \right) \; ,\\
\braket{^1S_0^{[8]}}^{\ref{fig:TMDShF-ChroElectric} \text{c}^* } & = \braket{^1S_0^{[8]}}^{\ref{fig:TMDShF-ChroElectric} \text{b}} \left( T^a T^b \otimes T^b T^a \rightarrow T^b T^a \otimes T^b T^a \right) \; ,\\
\braket{^1S_0^{[8]}}^{\ref{fig:TMDShF-ChroElectric} \text{d} } & = \braket{^1S_0^{[8]}}^{\ref{fig:TMDShF-ChroElectric} \text{b}} \left( T^a T^b \otimes T^b T^a \rightarrow T^b T^a \otimes T^a T^b \right) \; .
\end{aligned}
\end{equation}
Using the following relations
\begin{equation}
\begin{aligned} \label{eq:RelColor}
T^a T^b \otimes T^b T^a & = \frac{C_F}{2 N_c} \, 1 \otimes 1 + \frac{N_c^2 -2}{2N_c} \, T^a \otimes T^a  \; , \\
T^a T^b \otimes T^a T^b & = \frac{C_F}{2 N_c} \, 1 \otimes 1 - \frac{1}{N_c} \, T^a \otimes T^a \, ,
\end{aligned}
\end{equation}
and summing the contribution of all diagrams, we obtain the following
\begin{equation}
\begin{aligned}
\braket{^1S_0^{[8]}}^{\ref{fig:TMDShF-ChroElectric}\text{b,c,d}}  & =  \frac{4 \alpha_s}{3 \pi m_c^2}  \left(C_F \frac{1 \otimes 1}{2 N_c} + B_F \, T^a \otimes T^a \right)  \eta^\dagger p^k \xi \times \xi^\dagger p^k \eta \left( \frac{1}{\veuv} - \frac{1}{\veir} \right) \; ,
\end{aligned}
\end{equation}
where $B_F = (N_c^2-4)/4N_c$.
Here we do not consider the virtual diagram (\hyperref[fig:TMDShF-ChroElectric]{figure 4}a) because it gives a contribution at higher order in the relative velocity expansion:
\begin{equation}
\begin{aligned}
\braket{^1S_0^{[8]}}^{\ref{fig:TMDShF-ChroElectric}\text{a}} \propto \frac{\mbf{p}^2}{m_c^2}\,  \eta^\dagger T^b T^a T^b \xi \times \xi^\dagger T^a \eta = \frac{\mbf{p}^2}{m_c^2} (C_F- C_A/2) \eta^\dagger T^a \xi \times \xi^\dagger T^a \eta \; .
\end{aligned}
\end{equation}
Therefore the following result is the contribution for the $^1S_0^{[8]}$ LDME of the usoft gluon exchange between heavy-quarks and (anti)quarks at leading order on $\tv$
\begin{equation} \label{eq:ChroElectric}
\begin{aligned}
\braket{^1S_0^{[8]}}^{\ref{fig:TMDShF-ChroElectric}}  & = \frac{4 \alpha_s}{3 \pi m_c^2} \left( C_F \braket{^1P_1^{[1]}}^{\text{LO}} + B_F \braket{^1P_1^{[8]}}^{\text{LO}} \right) \left( \frac{1}{\veuv} - \frac{1}{\veir} \right) \; .
\end{aligned}
\end{equation}

We put together the previous results and conclude that the $^1S_0^{[8]}$ LDME at NLO is the following
\begin{equation}
\begin{aligned} \label{eq:LDMENLO}
\braket{^1S_0^{[8]}}^{\text{NLO}} & = \braket{^1S_0^{[8]}}^{\ref{fig:TMDShF-LOandCoulomb}\text{LO}} + \braket{^1S_0^{[8]}}^{\ref{fig:TMDShF-LOandCoulomb} \text{Coul.}} + \braket{^1S_0^{[8]}}^{\ref{fig:TMDShF-ChroElectric}} \\
& = \left( 1 +   \left( C_F - C_A/2 \right)\frac{\pi \alpha_s}{2 \tv}  \right)\braket{^1S_0^{[8]}}^{\text{LO}} \\
& + \frac{4 \alpha_s}{3 \pi m_c^2}   \left( C_F \braket{^1P_1^{[1]}}^{\text{LO}} + B_F \braket{^1P_1^{[8]}}^{\text{LO}} \right) \left( \frac{1}{\veuv} - \frac{1}{\veir} \right)  \; .
\end{aligned}
\end{equation}

\subsection{NLO calculation of the TMDShF} \label{sec:ShNLO}

\begin{figure} \centering
    \includegraphics[scale=0.6]{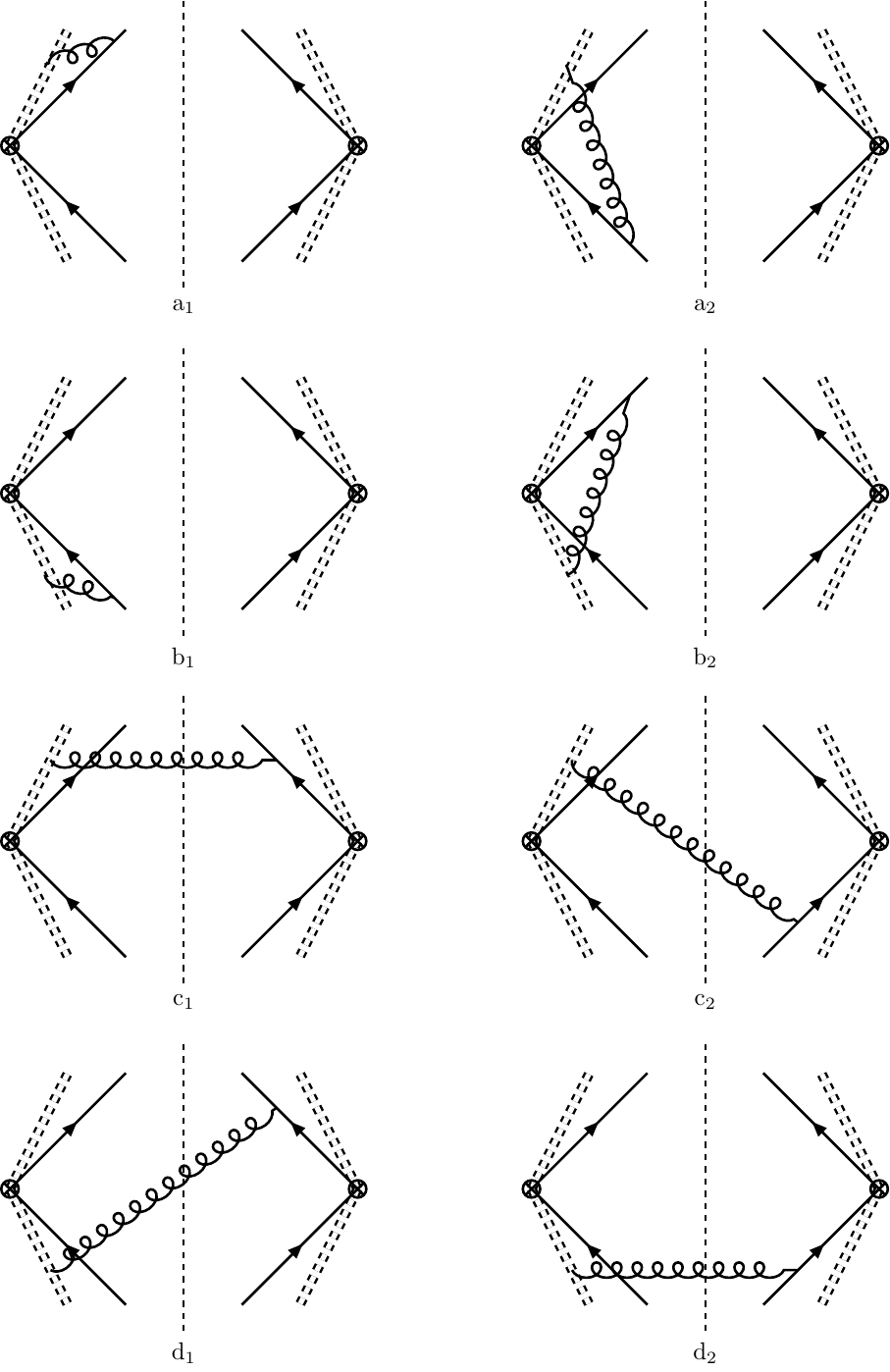}
    \caption{Diagrams showing the ultrasoft gluon exchanges between heavy-(anti)quarks and ultrasoft Wilson lines contributing to NLO. Hermitian conjugates are not shown. The dashed lines are illustrating the ultrasoft Wilson lines.}
    \label{fig:TMDShF-WilQuarks}
\end{figure}

\begin{figure} \centering
    \includegraphics[scale=0.6]{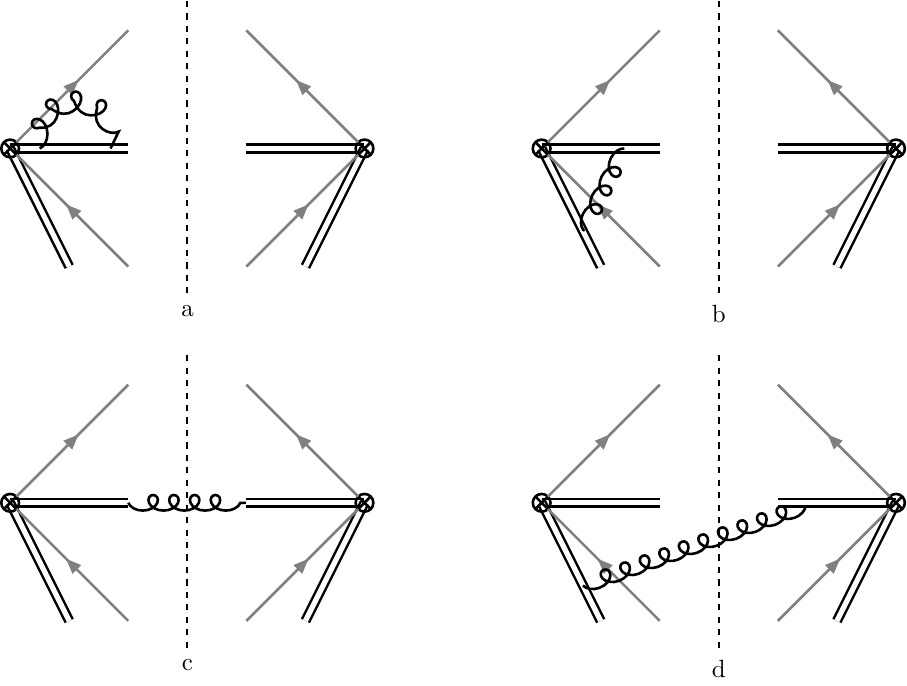}
    \caption{Non-zero diagrams showing the soft gluon exchange between soft Wilson lines contributing to NLO. Hermitian conjugates of diagrams a, b and d are not shown. The double lines are illustrating the soft Wilson lines.}
    \label{fig:TMDShF-SoftExchange}
\end{figure}

\begin{figure} \centering
    \includegraphics[scale=0.6]{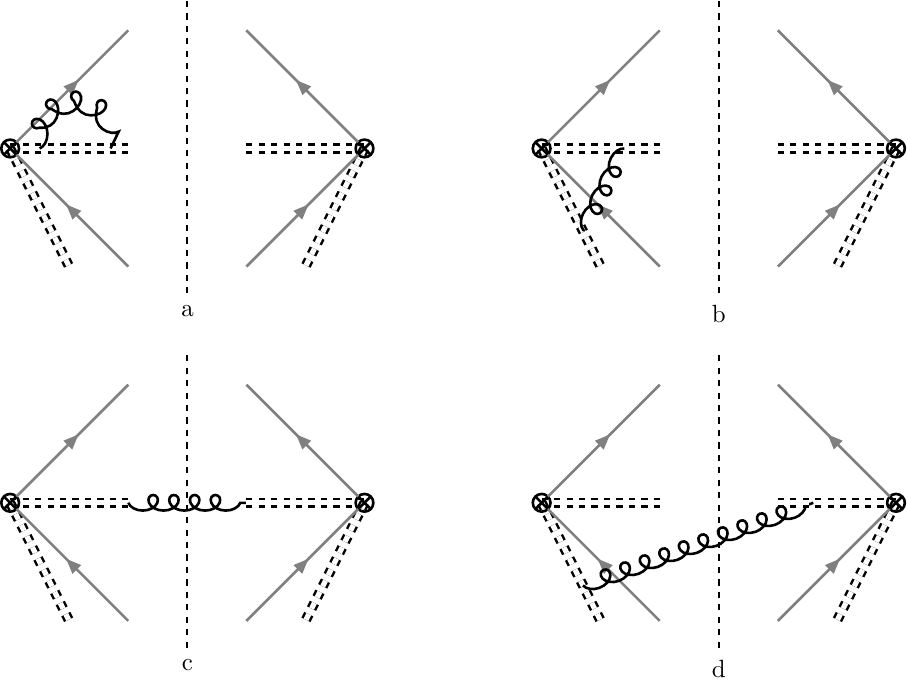}
    \caption{Non-zero diagrams showing the ultrasoft gluon exchange between ultrasoft Wilson lines contributing to NLO. Hermitian conjugates of diagrams a, b and d are not shown. The dashed lines are illustrating the ultrasoft Wilson lines.}
    \label{fig:TMDShF-uSoftExchange}
\end{figure}

We proceed to calculate the $^1S_0^{[8]}$ TMD shape function in \eq{eq:renShF} at NLO. In addition to the Coulomb interaction and chromo-electric transition, there are three additional contributions arising from the Wilson lines in the definition of the TMDShF. The Feynman diagrams illustrating these interactions are depicted in \hyperref[fig:TMDShF-WilQuarks]{figure 5}, \hyperref[fig:TMDShF-SoftExchange]{figure 6} and \hyperref[fig:TMDShF-uSoftExchange]{figure 7}.
As previously mentioned, due to rapidity divergences arising from one of the Wilson lines (the one in the $n$ collinear direction) in the TMDShF, we employ the delta rapidity regulator at the operator level in the Wilson lines definition (\ref{eq:SoftFunctions}).
For clarity, we opt to modify the notation for the shape functions in this subsection, replacing $S_{[n] \to J/\psi}$ to $S_{[n]}$, with the understanding that the final state is the $J/\psi$.

From the previous results in~\eq{eq:LOldme} and~\eq{eq:Coulomb}, it is straightforward to deduce that
\begin{equation}
\begin{aligned}
S_{^1S_0^{[8]} }^{ \ref{fig:TMDShF-LOandCoulomb} \text{LO}} (\mbf k_\perp) & =  \delta^2(\mbf{k}_\perp) \braket{^1S_0^{[8]}}^{\text{LO}} \; , \\
S_{^1S_0^{[8]} }^{ \ref{fig:TMDShF-LOandCoulomb} \text{Coul.}} (\mbf k_\perp) & =   \delta^2(\mbf k_\perp)  \left( C_F - C_A/2 \right)\frac{\pi \alpha_s}{2 \tv} \braket{^1S_0^{[8]}}^{\text{LO}} \; .
\end{aligned}
\end{equation}
In the LDME NLO calculation, as evident from the analysis, the virtual contributions for the chromo-electric dipole transition interaction manifest at higher orders in the v-expansion. Hence, we only consider the real contributions from \hyperref[fig:TMDShF-ChroElectric]{figure 4}:
\begin{equation}
\begin{aligned}
S_{^1S_0^{[8]} }^{ \ref{fig:TMDShF-ChroElectric} \text b}  (\mbf k_\perp) & =   \frac{4 \pi \alpha_s}{m_c^2}  \, I^{\ref{fig:TMDShF-ChroElectric}\text{b}} (\mbf k_\perp) \eta^\dagger  T^a T^b \xi \times \xi^\dagger  T^b T^a \eta \; , \\
I^{\ref{fig:TMDShF-ChroElectric}b}(\mbf k_\perp) & = \frac{1}{2} \int_{-\infty}^\infty \frac{d k_z}{(2 \pi)^3 } \frac{\mbf{p}^2 - (\mbf{p} \cdot \mbf{k})^2/\mbf{k}^2}{|\mbf{k}|^3} = \frac{1}{3} \int_{-\infty}^\infty \frac{d k_z}{(2 \pi)^3} \frac{\mbf p^2}{|\mbf k|^3}  = \frac{1}{12 \pi^3} \frac{\mbf p^2}{\mbf k_\perp^2} \; ,
\end{aligned}
\end{equation}
where $\mbf k= (\mbf k_\perp, k_z)$.
As in the LDME calculation, the remaining diagrams provide equivalent contributions, with the only variation being the color configuration. Using the relations (\ref{eq:2bcd}) and (\ref{eq:RelColor}), we obtain the total contribution:
\begin{equation}
\begin{aligned}
S_{^1S_0^{[8]} }^{ \ref{fig:TMDShF-ChroElectric}}  (\mbf k_\perp^2) & =  \frac{4 \alpha_s}{3 \pi^2 m_c^2} \frac{1}{\mbf k_\perp^2} \left( C_F \braket{^1P_1^{[1]}}^{\text{LO}} + B_F \braket{^1P_1^{[8]}}^{\text{LO}} \right) \; .
\end{aligned}
\end{equation}

The ultrasoft gluon exchanges between heavy (anti)quarks and ultrasoft Wilson lines, as illustrated in \hyperref[fig:TMDShF-WilQuarks]{figure 5}, originate from a single iteration of the chromo-electric term in the vNRQCD interaction Lagrangian (\ref{eq:LagrChromo}) and the insertion of an ultrasoft gluon from the ultrasoft Wilson lines. As indicated in~\eq{eq:LagrChromo}, this contribution is directly proportional to the three-momentum of the heavy (anti)quark. Given the relationship $ \bar{\mbf{p}} = - \mbf{p}$ between the 3-momenta of the heavy quark and antiquark, the sum $a1+a2$ evaluates to zero. This relationship holds true for the other pairs of diagrams as well: $b1 + b2 = 0$, and so forth.

The soft gluon exchanges between soft Wilson lines are shown in the \hyperref[fig:TMDShF-SoftExchange]{figure 6}. To obtain the contribution of that kind of interactions we need compute the following integrals:
\begin{equation}
\begin{aligned} \label{eq:Integrals-I}
I^{\ref{fig:TMDShF-SoftExchange}\text{a}} (\mathbf{k}_\perp^2; \mu) & =  i g^2 C_A \mu^{2 \ve} \delta^2(\mbf{k}_\perp) \int \frac{d^d k}{(2 \pi)^{d}} \frac{1}{(k^2 + i0)}\frac{1}{(k^0+i0)(-k^0+i0)} + h.c. \\
& = \frac{g^2 C_A \mu^{2 \ve}}{2}\delta^2(\mbf{k}_\perp)  \int \frac{d^{d-1} k}{(2 \pi)^{d-1}} \frac{1}{|\mbf{k}|^3}  + h.c. \\
& =  \frac{\alpha_s C_A}{2 \pi} \delta^2(\mbf{k}_\perp) \left( \frac{1}{\veuv} - \frac{1}{\veir} \right) \; ,\\
I^{\ref{fig:TMDShF-SoftExchange}\text{b}}(\mathbf{k}_\perp^2; \mu) & =-i g^2 C_A \mu^{2 \ve} \delta^2(\mbf{k}_\perp) \int \frac{d^d k}{(2 \pi)^d} \frac{1}{(k^2 + i0)} \frac{v \cdot \bar n}{(k^+ + i \delta^+) (\frac{ k^+ v^- + k^- v^+}{2} + i0)} + h.c. \\
& = \frac{-i g^2 C_A \mu^{2 \ve} 2 (-2 \pi i)}{2(2 \pi)^2 (2 \pi)^{2-2\ve}} \delta^2(\mbf{k}_\perp)\int_{0}^\infty \frac{dk^+ d^{d-2} \mbf{k}_\perp}{(k^+ + i\delta^+) (\mbf{k}_\perp^2 + k^{+2}(v^-/v^+) )} + h.c. \\
& = - \frac{g^2 C_A }{2 \pi} \frac{\Gamma(\ve) \mu^{2 \ve}}{(4 \pi)^{1- \ve} (v^-/v^+)^\ve} \int_0^{\infty} \frac{dk^+ (k^+)^{-2 \ve}}{k^+ + i \delta^+} + h.c. \\
& = - 2  \alpha_s C_A \frac{(i \delta^+)^{-2\ve}}{(4 \pi)^{1-\ve} (v^-/v^+)^\ve} \Gamma(\ve) \Gamma(2 \ve) \Gamma(1-2 \ve) + h.c.\\
& = - \frac{\alpha_s C_A}{2 \pi}  \delta^2(\mbf{k}_\perp) 
\left[ \frac{1}{\veuv^2} - \frac{1}{\veuv} \ln \frac{\delta^{+2} (v^-/v^+)}{\mu^2} + \frac{1}{2} \ln^2 \frac{\delta^{+2} (v^- / v^+)}{\mu^2} + \frac{\pi^2}{4} \right] \;,
\end{aligned}
\end{equation}
\begin{equation} \label{eq:Integrals-II}
\begin{aligned}
I^{\ref{fig:TMDShF-SoftExchange}\text{c}} (\mathbf{k}_\perp^2; \mu) & =   2 \pi g^2 C_A  \int \frac{d^d k}{(2 \pi)^{d}} \frac{v^2 \delta(k^2) \theta(k^+) }{(\frac{ k^+ + k^-}{2})(\frac{- k^+ - k^-}{2})} \\
& = - \alpha_s  C_A  2^{2 \epsilon} \pi^{2 \epsilon -2} \int_0^\infty    \frac{ d k^+ d^{d-2} \mbf{k}_\perp\left| k^+ \right| }{\left(\mbf{k}_\perp^2 +(k^+)^2 \right)^2}  \\
& = - \frac{\alpha_s C_A}{2 \pi^2} \frac{1}{\mbf{k}_\perp^2} \; ,\\
I^{\ref{fig:TMDShF-SoftExchange}\text{d}}(\mathbf{k}_\perp^2; \mu) & = - 2\pi g^2 C_A   \int \frac{d^d k}{(2 \pi)^d} \frac{v \cdot \bar n \, \delta(k^2) \theta(k^+) }{(k^+ + i \delta^+) (\frac{- k^+ v^- - k^- v^+}{2})} + h.c.\\ 
& = \frac{4 \pi g^2 C_A }{2 (2 \pi)^2 (2 \pi)^{2-2\ve} } \int_0^\infty \frac{d k^+  \, d^{d-2} \mathbf{k}_\perp  }{(k^+ + i \delta^+) (\mathbf{k}_\perp^2 + k^{+2} (v^-/v^+) )} + h.c.\\
& = - \frac{ \alpha_s C_A  }{2 \pi^2 } \frac{1}{\mathbf{k}_\perp^2- \delta^{+2} (v^-/v^+) } \ln \left( \frac{\delta^{+2} (v^-/v^+)}{\mathbf{k}_\perp^2 }  \right) \; .
\end{aligned}
\end{equation}
Here we only explicitly show the dependence on $v^+$ and $v^-$ in integrals \textit{b} and \textit{d} because as we will see later, it will be crucial for the renormalization of rapidity divergences. For the diagrams \textit{a} and \textit{c}, we use that $v$ is a time-like vector such that $v^2=1$ and $v^-/v^+ = 1$.
Using the previous results, the contribution to the shape function of each diagram in \hyperref[fig:TMDShF-SoftExchange]{figure 6}, denoted by $i$, is the following:
\begin{equation}
\begin{aligned}
S_{^1S_0^{[8]} }^{ \ref{fig:TMDShF-SoftExchange} i} (\mbf k_\perp^2)  =  \braket{^1S_0^{[8]}}^{\text{LO}} I^{\ref{fig:TMDShF-SoftExchange} i} (\mbf k_\perp^2), \quad i = \text{a,b,c,d} \; .
\end{aligned}
\end{equation} Putting everything together, we get the total contribution of the soft gluon exchanges between soft Wilson lines:
\begin{equation}
\begin{aligned}
S_{^1S_0^{[8]} }^{ \ref{fig:TMDShF-SoftExchange}} (\mbf k_\perp^2; \mu , \delta)  & = -  \braket{^1S_0^{[8]}}^{\text{LO}}\frac{\alpha_s C_A}{2 \pi^2} \\
\times & \left[ \pi \delta^2 (\mbf k _\perp) \left( \frac{1}{\veuv^2} - \frac{1}{\veuv} \ln \frac{\delta^{2}}{\mu^2} + \frac{1}{2} \ln^2 \frac{\delta^{2}}{\mu^2} + \frac{\pi^2}{4} \right) +
\frac{1}{\mathbf{k}_\perp^2- \delta^{2}} \ln \left( \frac{\delta^{2}}{\mathbf{k}_\perp^2 }  \right) \right. \\
& \left. +  \delta^2(\mbf{k}_\perp) \left( \frac{\pi}{\veir} - \frac{\pi}{\veuv} \right) + \frac{1}{\mbf{k}_\perp^2}
\right] \; .
\end{aligned}
\end{equation}
Here we define $\delta^2 \equiv \delta^{+2} (v^-/v^+)$ which is boost invariant.
Since there is only one soft Wilson line having rapidity divergences, only diagrams exhibiting rapidity divergences are \textit{b} and \textit{d} and the sum of their contributions is half of soft function.

The exchanges of usoft gluons between the ultrasoft Wilson lines are depicted in \hyperref[fig:TMDShF-uSoftExchange]{figure 7}. These diagrams yield scaleless integrals. For instance, the integral arising from the virtual interaction between the ultrasoft Wilson lines $n$ and $v$, after integrating over $k^0$, is as follows:
\begin{equation}
\begin{aligned}
I^{us} = \int \frac{d^4 k}{(2 \pi)^4} \frac{g^2 \, v \cdot \bar n}{(k^0)^2 [(k^0)^2 - \mbf{k}^2 + i \eps]} = i g^2  \int \frac{d^3 k}{(2 \pi)^3} \frac{v \cdot \bar n}{|\mbf{k}|^3} \; .
\end{aligned}
\end{equation}
According to distinguish between IR and UV poles, we would have to take into account the contribution of these diagrams to the TMDShF. However it will cancel against the zero-bin subtraction of the corresponding soft diagrams shown in \hyperref[fig:TMDShF-SoftExchange]{figure 6}.

Summing all the results obtained in this section, we get the TMD shape function at NLO in transverse momentum space:
\begin{equation}
\begin{aligned} \label{eq:ShKperp}
S_{^1S_0^{[8]} } (\mbf{k}_\perp; \mu, \delta) & =   \frac{\alpha_s }{2 \pi} \delta^2(\mathbf{k}_\perp) \left( (C_F-C_A/2) \frac{\pi^2}{\tv} -  \frac{C_A}{\veir} \right) \braket{^1S_0^{[8]}}^{\text{LO}}  \\
&  -   \frac{\alpha_s C_A }{2 \pi^2}  \pi \delta^2 (\mbf k _\perp) \left( \frac{1}{\veuv^2} - \frac{1}{\veuv} \ln \frac{\delta^{2}}{\mu^2} + \frac{1}{2} \ln^2 \frac{\delta^{2}}{\mu^2} + \frac{\pi^2}{4} \right)   \braket{^1S_0^{[8]}}^{\text{LO}}  \\
&  -   \frac{\alpha_s C_A }{2 \pi^2} 
\frac{1}{\mathbf{k}_\perp^2- \delta^{2}} \ln \left( \frac{\delta^{2}}{\mathbf{k}_\perp^2 }  \right)  \braket{^1S_0^{[8]}}^{\text{LO}}  \\
&   + \frac{\alpha_s C_A}{2 \pi^2} \left( \frac{\pi }{\veuv} \delta^2(\mbf{k}_\perp) - \frac{1}{\mbf{k}_\perp^2} \right)\braket{^1S_0^{[8]}}^{\text{LO}}   \\
&  + \frac{4 \alpha_s}{3 \pi^2 m_c^2} \frac{1}{\mbf k_\perp^2} \left( C_F \braket{^1P_1^{[1]}}^{\text{LO}} + B_F \braket{^1P_1^{[8]}}^{\text{LO}} \right)   \; .
\end{aligned}
\end{equation}
Before anything else, it is convenient to perform the Fourier transform of the shape function such that:
\begin{equation}
\begin{aligned} \label{eq:SofExchange}
S_{^1S_0^{[8]} } (b_T; \mu, \delta) & =  \braket{^1S_0^{[8]}}^{\text{LO}} \left[
\frac{\alpha_s C_A}{2 \pi} \left( \frac{1}{\veuv} + L_T \right) + \frac{\pi \alpha_s}{2 \tv} \left( C_F - C_A/2 \right) \right.  \\
&  \left. + \frac{\alpha_s C_A}{2 \pi} \left( - \frac{1}{\veuv^2} + \frac{1}{\veuv} \ln \frac{\delta^{2}}{\mu^2} + \frac{L_T^2}{2} + L_T \ln \frac{\delta^{2}}{\mu^2} + \frac{\pi^2}{12}  \right)
\right]  \\
&  - \frac{4 \alpha_s}{3 \pi m_c^2} \left( \frac{1}{\veir} + L_T \right) \left(  C_F \braket{^1P_1^{[1]}}^{\text{LO}} + B_F \braket{^1P_1^{[8]}}^{\text{LO}} \right)
 \; ,
\end{aligned}
\end{equation}
with $L_T = \ln(\mu^2 b_T^2 e^{2 \gamma_E}/4)$.
Note we employ the same symbols to denote functions in $k_T$-space and in $b_T$-space. To perform the Fourier transform, we need the following integrals:
\begin{equation}
\begin{aligned}
\int d^{2-2\ve} \mbf{k}_\perp \frac{e^{i \mbf{k}_\perp \cdot \mbf{b}_\perp}}{\mbf{k}_\perp^2} & = 4^{-\ve} \pi^{1-\ve} b_T^{2 \ve}  \Gamma(-\ve)  \; , \\
\int d^{2-2 \ve} \mbf k _\perp \frac{e^{i \mbf k_\perp \cdot \mbf b _\perp}}{k_T^2 - \Lambda^2} \, \ln \frac{\Lambda^2}{k_T^2} & = - \pi \left( \frac{1}{2} \ln^2 \frac{4 e^{-2 \gamma_E}}{\Lambda^2 b_T^2} + \frac{\pi^2}{3} \right) \; .
\end{aligned}
\end{equation}
We proceed to renormalize the rapidity divergences by redefining the shape function as in~\eq{eq:renShF}.
The soft function was calculated, e.g., in \cite{Echevarria:2015uaa}, its Fourier-transformed result at NLO is
\begin{equation}
S(b_T; \mu, \delta^+, \delta^-) = \frac{\alpha_s C_A}{2 \pi} \left[ - \frac{2}{\veuv^2} + \frac{2}{\veuv} \ln \frac{\delta^+ \delta^-}{\mu^2} + L_T^2 + 2 L_T \ln \frac{\delta^+ \delta^-}{\mu^2} + \frac{\pi^2}{6}  \right] \; .
\end{equation}
According to the discussions in \cite{Echevarria:2012js}, the soft function splits (to all orders) into two pieces:
\begin{equation}
S(b_T; \mu, \d^+,\d^-) =
\sqrt{S(b_T; \mu, \d^+,\tfrac{\d+}{\alpha})}\,
\sqrt{S(b_T; \mu, \alpha\d^-,\d^-)}
\;,
\end{equation}
where then at NLO we have
\begin{equation} \label{eq:halfSF}
S(b_T; \mu, \delta^+,\tfrac{\d^+}{\alpha}) = \frac{\alpha_s C_A}{2 \pi} \left[ - \frac{2}{\veuv^2} + \frac{2}{\veuv} \ln \frac{\delta^{+2}}{\alpha \mu^2} + L_T^2 + 2 L_T \ln \frac{\delta^{+2}}{\alpha \mu^2} + \frac{\pi^2}{6}  \right] 
\;,
\end{equation}
where $\alpha$ is a real number which transforms under boosts like $(p^+)^2$.
Here we define this number such that $\alpha/(v^-/v^+) \equiv \zeta_B$, where the factor $\zeta_B$ is a finite boost dimensionless invariant real number.
The procedure to get $\zeta_B$ is the TMD factorization theorem, similar to the one used, e.g., in \cite{Echevarria:2012js}: in that case, $\zeta_A$ and $\zeta_B$ have dimensions, but in the case that concerns us, $\zeta_B$ is dimensionless and $\zeta_A$ has mass-squared dimensions.
Consequently, we have $\zeta_A \zeta_B = Q_H^2$ in this work. Note the similarity to the process of dijet production in DIS \cite{delCastillo:2020omr}.
Given that the shape function in the previous result includes a contribution from half of the soft function (look at second line in~\eq{eq:SofExchange}), the renormalization using the Fourier-transformed soft function in~\eq{eq:halfSF} is straightforward. Consequently, the rapidity renormalization scale $\zeta_B$ emerges, and the subtracted $^1S_0^{[8]}$ TMDShF at NLO is as follows:
\begin{equation}
\begin{aligned} \label{eq:ShTotal}
&S_{^1S_0^{[8]} } (b_T; \mu, \zeta_B) = 
 \braket{^1S_0^{[8]}}^{\text{LO}} + 
\frac{\alpha_s}{2 \pi}
 \Bigg[  \frac{C_A}{\veuv} \left( 1 - \ln \, \zeta_B  \right)
 \braket{^1S_0^{[8]}}^{\text{LO}} 
 \\ 
&  \quad
+ C_A L_T \left( 1 - \ln \, \zeta_B \right)
\braket{^1S_0^{[8]}}^{\text{LO}}
-  \frac{8}{3m_c^2} L_T \left(  C_F \braket{^1P_1^{[1]}}^{\text{LO}} + B_F \braket{^1P_1^{[8]}}^{\text{LO}} \right)   \\
& \quad
+ \frac{\pi^2}{\tv} (C_F - C_A/2)
\braket{^1S_0^{[8]}}^{\text{LO}}
-  \frac{8}{3 m_c^2}  \frac{1}{\veir} \left(  C_F \braket{^1P_1^{[1]}}^{\text{LO}} + B_F \braket{^1P_1^{[8]}}^{\text{LO}} \right) \Bigg]  \; .
\end{aligned}
\end{equation}
At NLO, we have explicitly verified, through the utilization of the $\delta$-regularization scheme, the expected cancellation of rapidity divergences.
It's noteworthy to observe an additional UV divergence in~\eq{eq:ShTotal} compared to the LDME calculation in~\eq{eq:LDMENLO}. 
This divergence, stemming from the virtual gluon self-exchanges of the $\bar c c$ state depicted in \hyperref[fig:TMDShF-SoftExchange]{figure 6}a, is relevant in the RG evolution of the shape functions, as we see in \hyperref[sec:3]{section 3}.

\section{Results for P-states} \label{sec:A-Pwaves}

As previously mentioned, the computation of the $^3P_J^{[8]}$ LDMEs and TMDShFs can be inferred from the S-wave calculation. In this section, we summarize the results for the P-wave channel.

Upon inspection of the diagrams and the vNRQCD Lagrangian interaction, it becomes evident that the methodology for computing the P-waves mirrors that of the S-wave. At LO, the LDME primarily encompasses the spin factor denoting the configuration of angular momentum, spin, and color of the heavy quarks in the final state. Analogous to our explanation in~\eq{eq:Soperator} and in~\eq{eq:LOldme}, and taking the operators in NRQCD, the LDMEs are as follows
\begin{equation}
\begin{aligned} \label{eq:PLOldme}
\braket{^3P_0^{[8]}}^{\text{LO}}  & = \frac{M^2}{3} \, \eta^\dagger \mbf q \cdot \boldsymbol{\sigma} T^a \xi \times \xi^\dagger \mbf q \cdot  \boldsymbol{\sigma} T^a \xi \eta   \; ,\\
\braket{^3P_1^{[8]}}^{\text{LO}}  & = \frac{M^2}{2} \, \eta^\dagger (\mbf q \times \boldsymbol{\sigma})^k T^a \xi \times \xi^\dagger (\mbf q \times \boldsymbol{\sigma})^k T^a \xi \eta   \; ,\\
\braket{^3P_2^{[8]}}^{\text{LO}}  & = M^2 \, \eta^\dagger  q^{(i}  \sigma^{j)} T^a \xi \times \xi^\dagger q^{(i}  \sigma^{j)} T^a \xi \eta   \;. \\
\end{aligned}
\end{equation}
Here we do not write the boost matrix (\ref{eq:BoostMatrixDef}) because we are considering this result in the $J/\psi$ rest frame.
Consequently, the exchange of soft and ultrasoft gluons originating from the vNRQCD interaction Lagrangian yields analogous contributions for the P-wave as observed for the S-wave.
The chromo-electric term corresponds to a label operator which gives a three-momentum in the spin factor. This, in turn, corresponds to an angular momentum with a value one unit higher than that of the LO spin factor. For instance, if the LO spin factor has angular momentum $L$, the chromo-electric term provides a combination of the color singlet and color octet for an angular momentum value of $L+1$:
\begin{equation}
\braket{^{2S+1} L_J^{[8]}}^{\ref{fig:TMDShF-ChroElectric}}   = \frac{4 \alpha_s}{3 \pi m_c^2} \left( C_F \braket{^{2S+1} {L'}_{J'}^{[1]}}^{\text{LO}} + B_F \braket{^{2S+1} {L'}_{J'}^{[8]}}^{\text{LO}} \right) \left( \frac{1}{\veuv} - \frac{1}{\veir} \right) \;,
\end{equation}
with $L' = L+1$, and therefore $J' = J+1$ (the spin $S$ remains unchanged). From this result and considering that the Coulomb singularity always appears regardless of the state, we have that the LDME at NLO is
\begin{equation}
\begin{aligned}
\braket{^3P_J^{[8]}}^{\text{NLO}} & = \left( 1 +   \left( C_F - C_A/2 \right)\frac{\pi \alpha_s}{2 \tv}  \right)\braket{^3P_J^{[8]}}^{\text{LO}} \\
& + \frac{4 \alpha_s}{3 \pi m_c^2}   \left( C_F \braket{^3D_{J+1}^{[1]}}^{\text{LO}} + B_F \braket{^3D_{J+1}^{[8]}}^{\text{LO}} \right) \left( \frac{1}{\veuv} - \frac{1}{\veir} \right) \; .
\end{aligned}
\end{equation}

Regarding the shape function, gluon exchanges between the Wilson lines do not affect the state of the heavy quark pair, so the integrals in eqs. \eqref{eq:Integrals-I} and \eqref{eq:Integrals-II} remain unchanged. Consequently, the result in~\eq{eq:SofExchange} also remains the same, and therefore at the end of the day, we find that the shape function at NLO is:
\begin{equation}
\begin{aligned} 
& S_{^3P_J^{[8]} \to J/\psi} (b_T; \mu,\zeta_B) =  
\frac{1}{N^{(J)}_{pol}} \left\{ \braket{^3P_J^{[8]}}^{\text{LO}} + 
\frac{\alpha_s}{2 \pi}
 \Bigg[  \frac{C_A}{\veuv} \left( 1 - \ln \, \zeta_B  \right)
 \braket{^3P_J^{[8]}}^{\text{LO}} \right.
 \\ 
& \left. \quad
+ C_A L_T \left( 1 - \ln \, \zeta_B \right)
\braket{^3P_J^{[8]}}^{\text{LO}}
-  \frac{8}{3m_c^2} L_T \left(  C_F \braket{^3D_{J+1}^{[1]}}^{\text{LO}} + B_F \braket{^3D_{J+1}^{[8]}}^{\text{LO}} \right) \right.  \\
& \left.\quad
+ \frac{\pi^2}{\tv} (C_F - C_A/2)
\braket{^3P_J^{[8]}}^{\text{LO}}
-  \frac{8}{3 m_c^2}  \frac{1}{\veir} \left(  C_F \braket{^3D_{J+1}^{[1]}}^{\text{LO}} + B_F \braket{^3D_{J+1}^{[8]}}^{\text{LO}} \right) \Bigg] \right\} \; .
\end{aligned}
\end{equation}

\bibliographystyle{utphys}
\bibliography{references}

\end{document}